\pdfoutput=1 
\documentclass[twoside,11pt]{CUEDthesisPSnPDF}

\usepackage[english]{babel}

\usepackage{setspace}
\usepackage{amsmath}
\usepackage{graphicx}

\usepackage{hyperref}
\hypersetup{colorlinks=true,allcolors=blue}

\usepackage{siunitx}
\usepackage{textgreek}
\usepackage{caption}
\usepackage{subcaption}
\usepackage{float}

\usepackage[frozencache,cachedir=.]{minted}

\usepackage[toc,page]{appendix}
\usepackage{comment}
\usepackage{chemformula}
\usepackage{afterpage}

\usepackage{rotating}
\title{Implementing photometric stereo for SHeM to reconstruct true-to-size 3D surfaces}

\author{Aleksandar \textsc{Radi\'{c}}}

\date{\today}

\collegeordept{{Girton College}}
\university{{Cavendish Laboratory}}
\crest{\includegraphics[width=0.45\textwidth]{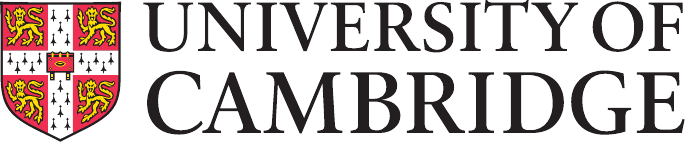}}
\spcrest{\includegraphics[width=0.45\textwidth]{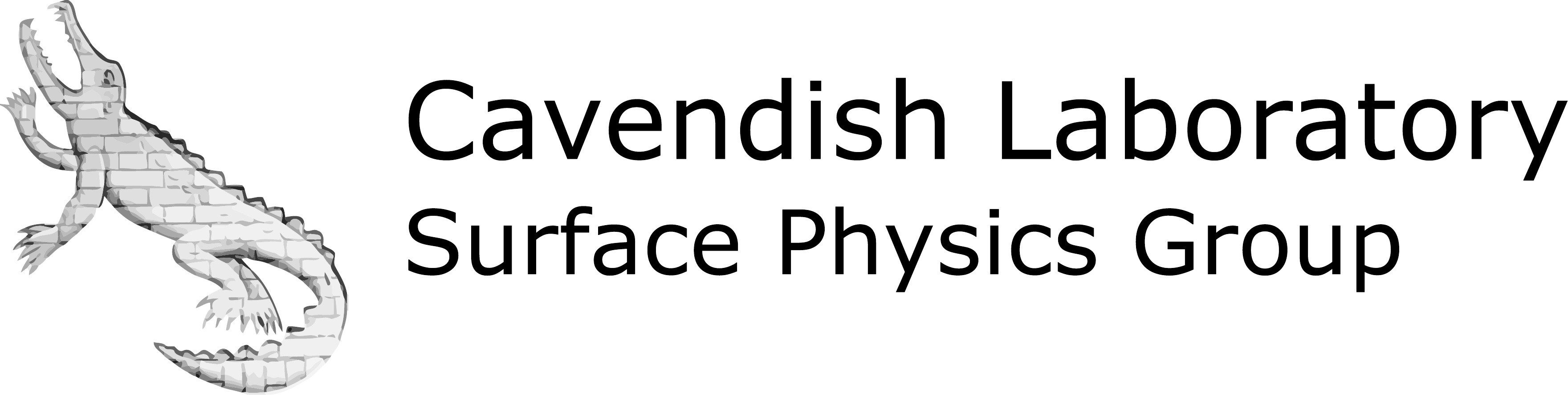}}
\degree{For the degree of Master of Philosophy}
\degreedate{August 2021}

\begin{document}
\maketitle
\newpage

\pagenumbering{gobble}
\thispagestyle{empty}
\begin{flushleft}
\section*{Acknowledgements}
I would like to thank my supervisor Andrew Jardine for his continual input during the course of my work. My work was only made possible by the support and knowledge of the SMF group and its collaborators, with particular thanks to Sam Lambrick, Matt Bergin and Nick von Jeinsen with whom I have worked closely on SHeM. I would like to show my appreciation for Dave Ward who has mentored and guided me, both academically and personally, throughout the year. Finally, I would like to acknowledge the overwhelming support and belief in me provided by my family throughout my time at university. I also acknowledge the department for their flexibility surrounding experimental work this year. 
\end{flushleft}

\newpage
\thispagestyle{empty}
\begin{center}
\section*{Declaration}
This thesis is the result of my own work and includes nothing which is the outcome of work done in collaboration except where specifically indicated in the text. It does not exceed the word limit prescribed by the Degree Committee for the Faculty of Physics and Chemistry and has not been submitted in whole or part for a degree, diploma, or other qualification at this or any other university.
\end{center}

\newpage
\thispagestyle{empty}
\section*{Publications}
Publications prepared using work performed during the course of research for the presented thesis include:
\begin{itemize}
    \item \textit{Measuring facet angles using a combined spatially and angularly resolved atom beam instrument} - Poster presentation, A. Radi\'{c}, N. A. von Jeinsen, S. M. Lambrick, A. P. Jardine, D. J. Ward, ISSC-23 (2021).
    \item \textit{Seminar on scanning helium microscopy (SHeM)}, oral presentation, University of Lincoln, UK (2021).
    \item \textit{Selected area helium diffraction for surface science on a spot}, N. A. von Jeinsen, S. M. Lambrick, M. Bergin, A. Radi\'{c}, B. Liu, D. Seremet, A. P. Jardine, D. J. Ward - in preparation
    \item \textit{3D printing vacuum components for scanning helium microscopy}, M. Bergin, T. A. Myles, A. Radi\'{c}, C. Hatchwell, D. J. Ward, A. Fahy, M. Barr, P.C. Dastoor - in preparation
    \item \textit{Vacuum compatibility of SLA 3D printed plastic for rapid prototyping}, A. Radi\'{c}, D. J. Ward - in preparation
    \item \textit{Diffuse scattering observed in scanning helium microscopy}, S. Lambrick, D. J. Ward, M. Bergin, A. Radi\'{c}, M. Barr, A. Fahy, T. Myles, PC Dastoor, J. Ellis, AP. Jardine - in preparation
\end{itemize}

\newpage
\thispagestyle{empty}
\begin{center}
\section*{Abstract}
\begin{flushleft}
Scanning Helium Microscopy (SHeM) is a unique imaging technique with both spatial and angular resolution that uses a beam of thermal energy helium-4 atoms for non-destructive imaging of delicate samples. In the present thesis, a novel 3D imaging mode for SHeM has been realised for both single and multi-detector instruments. The method allows for true-to-size 3D surface reconstruction using an adaptation of the photometric stereo algorithm, known as ``heliometric stereo'', that has been published by the group during the course of the work.
\\[12pt]
An important aspect of realising 3D image reconstruction is achieving a normally incident beam onto the sample. Limited by traditional machining techniques, stereolithography (SLA) 3D printed plastics have been explored as a novel fabrication technique for SHeM pinhole plates, the feature which defines the microscope's imaging properties. The vacuum compatibility and re-wetting properties of SLA printed FormLabs ``Clear Resin'' have been investigated. The work concludes that it is an ideal printing process and material combination for rapid prototyping of small vacuum components for use in HV to UHV pressure ranges by following a baking protocol developed and outlined here as standard operating procedure. The work presented here suggests that re-wetting of such plastics is exclusively a surface process over timescales of the order of weeks. 
\\[12pt]
3D image reconstruction using both the single and multi-detector solutions necessitated the development of a real-space point tracking method. A complete analytical model of sample motion was developed and has been applied outside of 3D image reconstruction. The method has enabled a novel imaging mode for SHeM in heliometric stereo, and enables the measurement of facet angles of multi-faceted polycrystalline materials, with particular applications in technological samples, biological crystals and synthetic substrates. The point tracking method has become part of standard operating procedure for imaging in SHeM where it is used to debug the sample manipulator.
\\[12pt]
A multi-detector solution to heliometric stereo was implemented by building \textbeta-SHeM, a second generation platform for developing scanning helium microscopy which allows imaging from different azimuthal angles simultaneously.
\end{flushleft}
\newpage
\begin{flushleft}
The heliometric stereo methods implemented in the work have motivated the development of a GPU accelerated version of the in-house Monte-Carlo based ray tracing framework, which is the de-facto standard for SHeM image analysis. GPU parallelisation was explored as a method for decreasing simulation time and enabling previously inaccessible simulations involving complex scattering distributions and high resolution, realistic sample geometries. Preliminary testing on an analogous problem yielded a potential performance increase of up to 380 times.
\end{flushleft}
\end{center}

\tableofcontents

\newpage

\pagenumbering{arabic} 
\setcounter{page}{1}
\chapter{Introduction \& Background}
\section*{Background}
Microscopy has been a staple investigative tool across all fields of science for centuries since the first optical microscope was made by Antonie van Leeuwenhoek in the late $17^{\text{th}}$ century, which he used to see microbial life for the first time \cite{VanLeeuwenhoek1677}\cite{Lane2015}. Throughout history, scientific discoveries such as this have been facilitated exclusively by advances in microscopy. The ability to probe length-scales smaller than that which the naked eye can resolve is key in gaining further understanding of the structures and processes that form the world around us. Continual development in microscopy, predominantly in optical and charged particle probe instruments, has enabled countless discoveries across science. It is crucial therefore that not only historically significant techniques continue to be developed, but novel ones too.
\\[12pt]
Current leading techniques are typically highly specialised and have significant restrictions on the types of samples they can image. Electron microscopy, for example, is both destructive and requires a conductive sample. Therefore, a compromise must be reached when imaging non-ideal samples via irreparably altering them through surface coatings or from the electron beam itself. Environmental scanning electron microscopy (ESEM) is able to image a wide range of delicate samples, but maintains the possibility of inducing permanent changes in the sample. Alternatively, there are numerous non-destructive techniques that are more versatile in terms of imaging samples without preparation, but again a compromise must be made. Most optical techniques offer such non-destructive imaging at the cost of relatively poorer maximum resolution and depth-of-field. It should be noted that it is possible to achieve some depth-of-field using confocal microscopy at the cost of a reduced field of vision \cite{Misri}. Although the array of currently available microscopes span a vast range of sample types and imaging capabilities as a collective, none can offer the unique imaging properties and versatility of scanning helium microscopy (SHeM).
\section*{The History of SHeM}
\begin{figure}[h]
    \centering
    \includegraphics[width=0.7\textwidth]{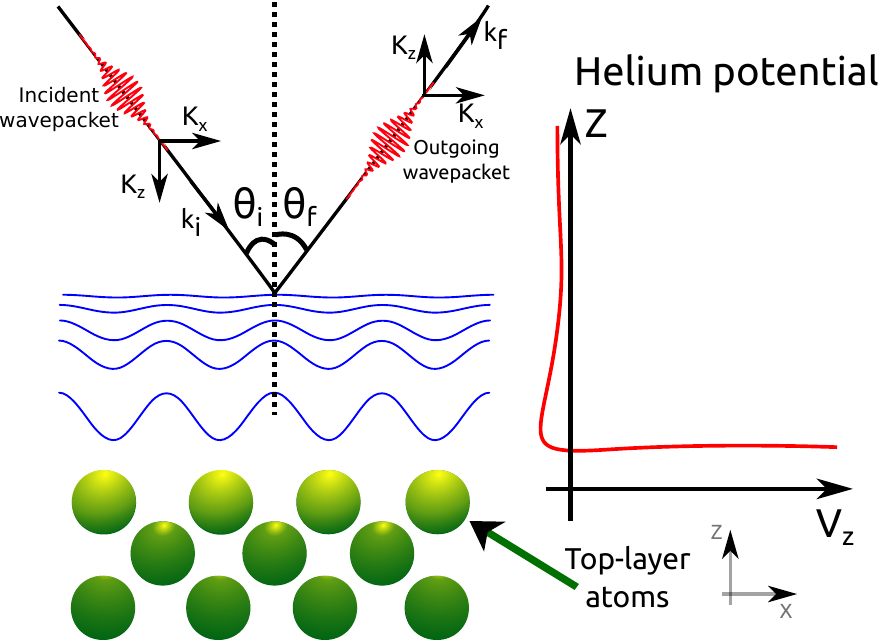}
    \caption{Diagram of the helium matter-wave scattering from the outermost electron density of a surface, with wavevectors denoted by $K$, and angle of incidence/scattering by $\theta$. Figure adapted from David J. Ward's PhD Thesis \cite{DaveThesis}}
    \label{fig:helium_scattering}
\end{figure}

Scanning helium microscopy (SHeM) \cite{Bergin2019} \cite{Bergin2020} \cite{Palau2016} uses a beam of thermal energy helium-4 atoms to probe surface topography with both spatial and angular resolution. The properties of the SHeM beam make it ideal for imaging delicate, electrically insulating and transparent samples. Owing to its thermal energy of $\approx 50\, m\,eV$, the beam has a de Broglie wavelength of only $0.05\,nm$ \cite{Bracco2013} which when coupled with being electrically net neutral and chemically inert results in a sensitive and delicate surface probe that can image a wide variety of samples. The low energy beam not only prevents sample damage but means that the beam scatters from the outermost electron density, shown in Figure \ref{fig:helium_scattering}, making SHeM a true surface probe \cite{Farias1998}, unlike typical electron and helium ion beams which have energies $3-6$ orders of magnitude greater \cite{Joy2009}.

\newpage
\begin{flushleft}
The first microscope to use a helium-4 beam and generate transmission 2D images was reported in 2008 by Koch et al. \cite{Koch2008}. However, the more traditional reflection mode imagining was first reported by Witham et al. in 2011 \cite{Witham2011}\cite{Witham2014} in parallel with the development of SHeM in Cambridge.

\end{flushleft}
\section*{The Cambridge SHeM}
\begin{figure}[h]
    \centering
    \includegraphics[width=\textwidth]{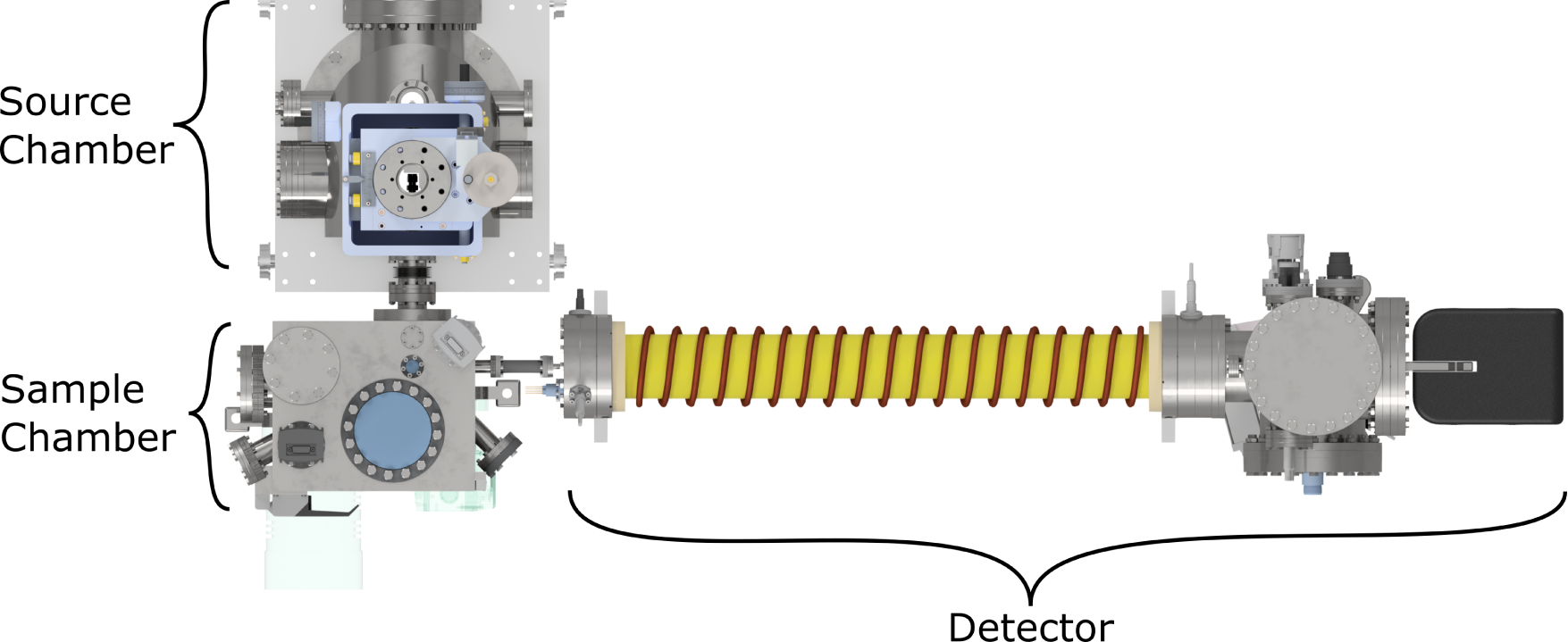}
    \caption{Top-down render of the single detector scanning helium microscope in Cambridge, \textalpha-SHeM.}
    \label{fig:ashem_top}
\end{figure}
\begin{figure}[h]
    \centering
    \includegraphics[width=\textwidth]{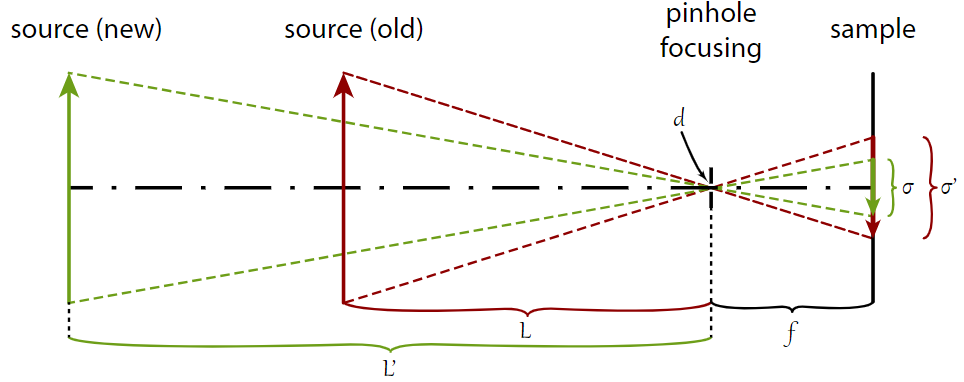}
    \caption{Schematic geometrical optics diagram of source demagnification through a pinhole where $L$ is the distance of the source from the focusing plane (pinhole aperture) and $\sigma$ is the FWHM or standard deviation of the spot which the beam illuminates on the imaging plane. Figure and caption reproduced from Nick A. von Jeinsen's MPhil thesis \cite{NickThesis}.}
    \label{fig:geometrical_optics}
\end{figure}
\begin{figure}[h]
    \centering
    \includegraphics[width=\textwidth]{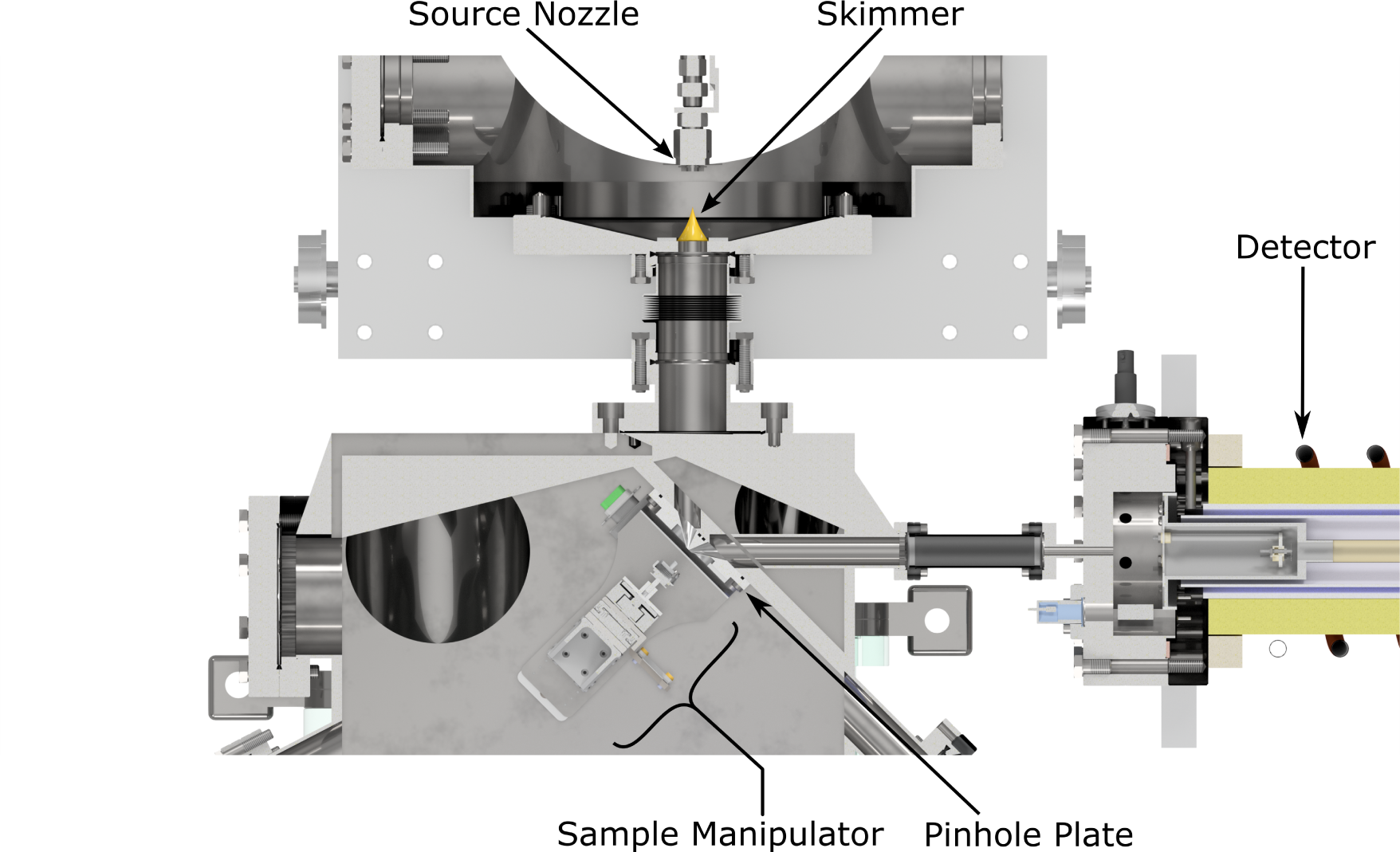}
    \caption{Diagram of \textalpha-SHeM sample chamber showing the \ang{90} total scattering angle with \ang{45} beam incidence on the sample. Helium expands from the source nozzle and is selected by the skimmer before being collimated by the pinhole aperture after which the beam scatters from the sample. The detector aperture selects a subset of the scattered atoms that are then counted and recorded.}
    \label{fig:ashem_close}
\end{figure}

\begin{flushleft}
The current iteration of helium microscope in use at the Cavendish Laboratory shown in Figures \ref{fig:ashem_top} and \ref{fig:ashem_close}, \textalpha-SHeM, was developed and built by the SMF group at the University of Cambridge in collaboration with the Centre for Organic Electronics at the University of Newcastle in Australia \cite{Barr2014}. The helium beam is formed by a supersonic free jet expansion \cite{1988Aamb}\cite{Reisinger2007} from a $10\,\mu m$ nozzle at room temperature. The centre of the expansion is selected spatially by the $\approx 100\,\mu m$ skimmer, producing a beam with a well defined velocity distribution.
\end{flushleft}
The complexity of the helium-helium interaction is confined to the area immediately surrounding the nozzle, the rest of the system can be considered, to a first order approximation, as a simple optics problem using classical optics methods. The Cambridge SHeM uses a pinhole to de-magnify the virtual source onto the imaging plane by standard rules of geometrical optics, as shown in Figure \ref{fig:geometrical_optics}. The resulting de-magnified beam has spot size, or full-width half-maximum (FWHM), as small as $350\,nm$ in the current work \cite{NickThesis}. The FWHM is a measure of the microscope's spatial resolution. 

After scattering from the sample, the signal to be detected, passes through the pinhole plate again via the detector aperture, the diameter of which defines the instrument's angular resolution. The Cambridge SHeM then uses a single custom built solenoid detector operating at efficiencies of up to $0.5\%$ \cite{Bergin2021} to detect the helium-4. The pinhole plate is the key technological requirement to build the Cambridge SHeM.
\newpage
\begin{flushleft}
In the work presented, I have enabled a new imaging mode using the single detector \textalpha-SHeM that can reconstruct a sample in true-to-size 3D using an algorithm similar to photometric stereo but applied to SHeM, as outlined by Lambrick et al. \cite{Lambrick2021}, in Chapter \ref{section:exp_heliometric}. By reconstructing samples with real-space units, referred to here as being true-to-size, one can use the technique for height profilometry. To realise the reconstruction method, a normal incidence pinhole plate \cite{Bergin2019a} was designed and fabricated using 3D printed plastics to surpass the limitations of traditional machining methods. Chapter \ref{section:polymer_plates} shows the development and characterisation process of using 3D printed plastics in low-HV to UHV pressure ranges. The novel fabrication techniques employed in the work have unlocked a vast range of potential imaging modes for SHeM through the ability to fabricate more complex pinhole plate geometries than ever before. The next instrument development platform of SHeM, \textbeta-SHeM, was developed during the course of the work and is discussed in detail in Chapter \ref{section:hardware3D}. \textbeta-SHeM has multiple physical detectors for native 3D reconstruction and significant advances across the instrument, from sample environment to source, enabling a variety of novel experiments not previously possible in \textalpha-SHeM. 
\end{flushleft}
Motivated by computationally expensive, and therefore slow, simulations of 3D surface reconstructions using the Monte-Carlo based in-house ray tracing framework \cite{Lambrick2018}, graphics card parallelisation was explored as an avenue to accelerate the simulations. Simulating surface reconstructions is valuable for exploring the strengths and limitations of the heliometric stereo algorithm, and further optimising the method in an environment where all experimental variables can be controlled precisely. The preliminary performance testing on an analogous problem in Chapter \ref{section:cuda} indicates a potential performance increase of up to 400 times under ideal conditions. Such an improvement in compute time would enable simulations using high resolution, realistic sample and pinhole plate geometries, and complex scattering distributions beyond cosine-like \cite{Feres2004}\cite{Celestini2008}\cite{Lambrick2018}\cite{OKEEFEDR1971} that were previously prohibitively computationally expensive.

\section*{3D Surface Reconstruction}
The ability to map the topography of microscopic surfaces is important across many applications in technology and research. However, on the micro- and nanoscale, the established techniques used for such imaging can be challenging. For example, implementation of 3D imaging in SEM \cite{Gontard2016} proves difficult because of secondary electron emission \cite{Postek2013} \cite{AgudoJacome2012}, and methods using scanning probes are limited by the tip profile \cite{Shen2017}. Some work has been done on 3D reconstruction using ionic helium beams \cite{Vollnhals2018}. SHeM's low energy beam allows it to scatter from the outermost electron density distribution without penetrating the sample making it an ideal instrument for true topographical mapping. There is a 3D image reconstruction method as described by Lambrick et al. \cite{Lambrick2021}, referred to as heliometric stereo. The algorithm reconstructs a true-to-size 3D surface by using images taken from different azimuthal angles of the sample to find the helium intensity normal to the surface at every pixel. One can achieve imaging from different azimuthal angles with either a single detector instrument and the ability to rotate the sample, or natively in a multiple detector instrument. Subsequently, the method integrates the intensities to find normal vectors for each pixel, from which a tiled surface is constructed. The current work details the novel instrumentation and experimental methods that have enabled the first experimental reconstruction using heliometric stereo on a single detector instrument.

\chapter{Enhancements to SHeM for Heliometric Stereo}
\label{section:enhancements_heliometric}
It is convenient to implement 3D reconstructions in existing, single detector instruments. There is a 3D image reconstruction method, as described by Lambrick et al. \cite{Lambrick2021}, akin to photometric stereo \cite{Stover2016} \cite{Woodham1980} \cite{Goldman2010}. The algorithm reconstructs a true-to-size 3D surface by using images taken from different azimuthal angles of the sample to find surface normals for each pixel, hence forming a tiled surface. The accuracy of the reconstruction with regards to being true-to-size is quantified by comparison of lengths on the sample to SEM images because topographical comparison to a CAD model is not possible. To achieve reconstruction, a method of tracking a point on the sample in real-space must be achieved so that features can be accurately mapped across images. The chapter starts by outlining the \textalpha-SHeM sample environment so that both a calibration routine to find the centre of rotation on the the sample, and a complete analytical model of the motion of said sample for predictive tracking can be developed. A statistically derived method to reduce the error in the point tracking accuracy is also discussed and implemented, with further statistical analysis also detailed in anticipation of further work providing larger data sets to give statistically significant results.
\\[12pt]
A novel SHeM imaging mode which measures facet angles on multi-faceted polycrystalline samples \cite{ISSC23} was developed in parallel as a result of the analytical sample motion model. By tracking individual crystal domains, in conjunction with an absolute measurement of the domain size, one can find the facet angle of that domain in SHeM.

\section{Point Tracking}
The invisible master geometry about which all additional geometries in SHeM are defined is the scattering plane – the region in which the imaging beam exists. The scattering plane can only be exactly defined by the beam and its interaction with the sample. With the beam line being defined as starting at the source nozzle and finishing at the point on the sample where the beam scatters, the scattering plane is defined as being perpendicular to the beam at the point of scattering.
\begin{figure}[H]
    \centering
    \begin{subfigure}{0.45\textwidth}
    \centering
    \includegraphics[width=\textwidth]{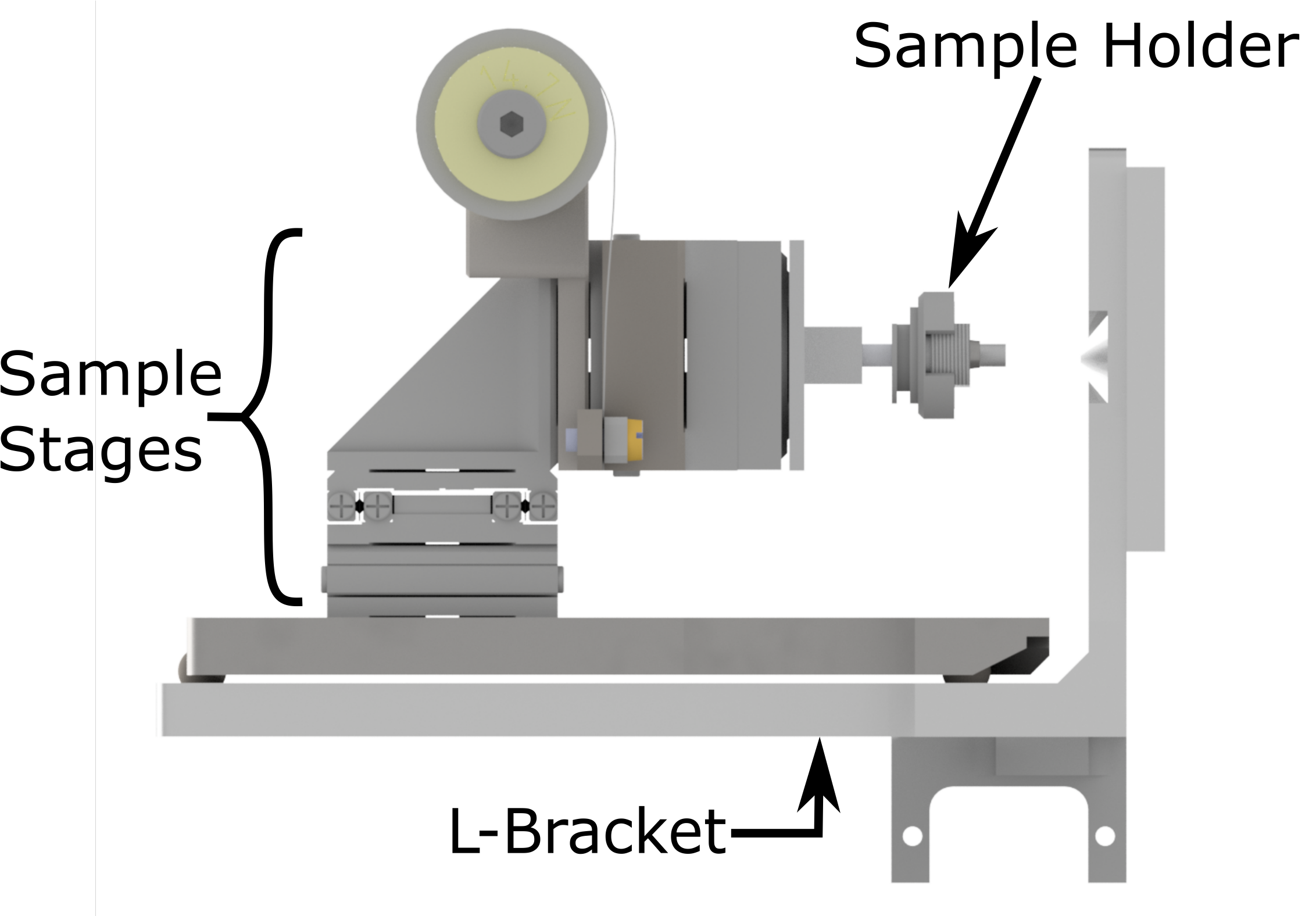}
    \end{subfigure}
    \begin{subfigure}{0.45\textwidth}
    \centering
    \includegraphics[width=\textwidth]{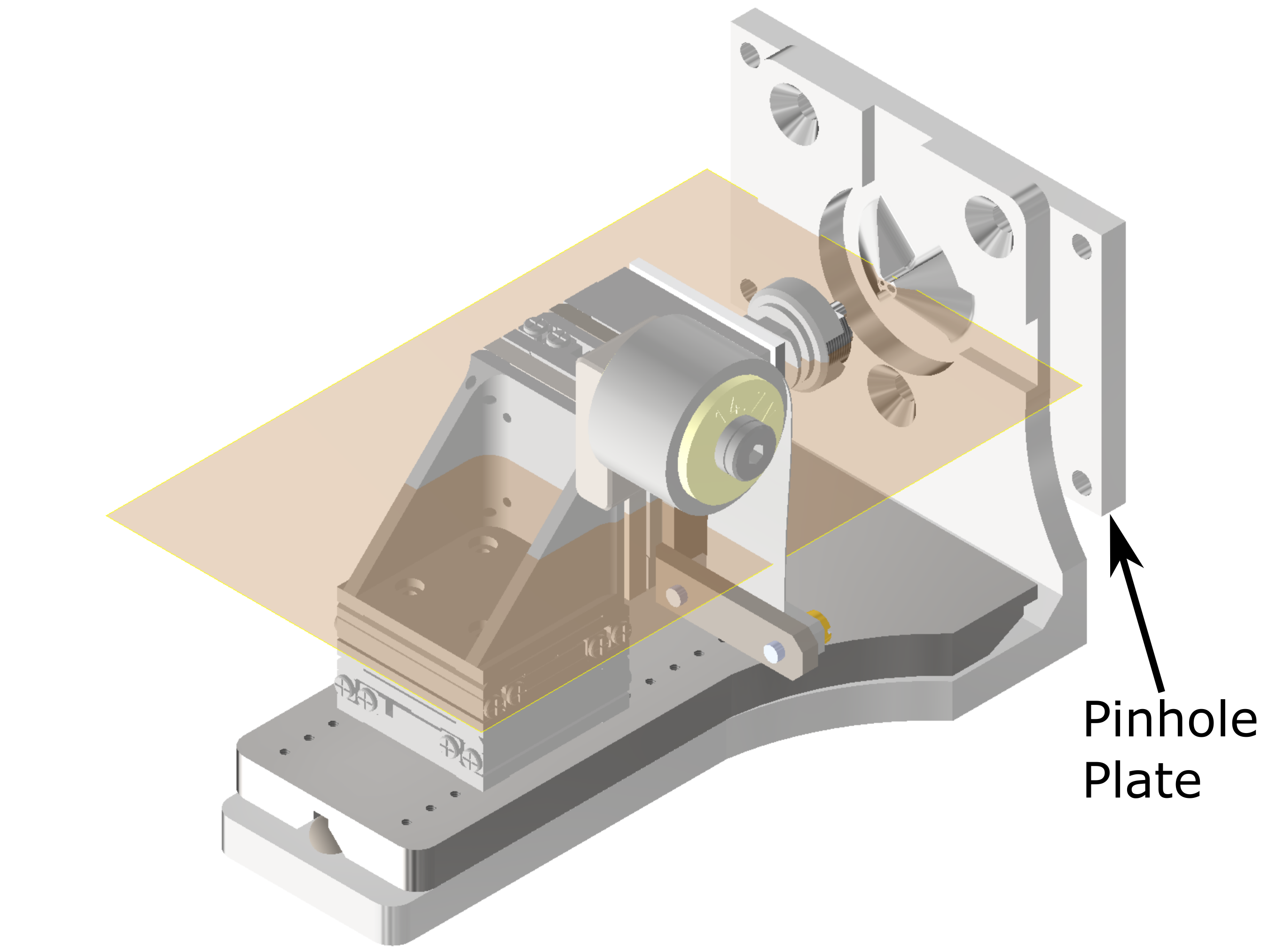}
    \end{subfigure}
    \caption{CAD model of the sample manipulator and pinhole plate assembly in Cambridge's \textalpha-SHeM. It is convenient to define the geometry of the instrument with respect to the scattering plane shown in orange in the right-hand panel.}
    \label{fig:scattering_plane_stages}
\end{figure}
Figure \ref{fig:scattering_plane_stages} shows the \textalpha-SHeM sample manipulator from Figure \ref{fig:ashem_close}. The sample manipulator assembly, consisting of three linear stages (Attocube ECS3030R), is attached to the pinhole plate via an ``L-bracket'' as shown in Figure \ref{fig:scattering_plane_stages}. The pinhole plate mounts directly to the wall of the sample chamber.
\\[12pt]
During imaging the sample rasters across the beam, coplanar to the scattering plane. The current configuration in SHeM has two linear stages required to do such rastering, ideally in the plane perpendicular to the scattering plane, with a third to position the sample at the correct working distance from the pinhole. Thus, to explore azimuthal angles the addition of a rotational stage (Attocube ECR3030), shown in Figure \ref{fig:scattering_plane_stages}, is implemented here to control the azimuthal angle. A variety of novel measurement modes like point diffraction \cite{vonJeinsen2021} and 3D image reconstruction are facilitated by the upgrade. Thus far all SHeM images have been acquired in 2D and it is useful to get a 3D perspective to acquire topographical sample information.
\\[12pt]
There are multiple ways to measure accurate sample topography by finding surface normals in SHeM. One can use a goniometer to tilt the sample, as demonstrated by the SHeM team in Newcastle, Australia \cite{Myles2019}, or by rotating the sample azimuthally as described by Lambrick et al. \cite{Lambrick2021} to apply traditional photometric stereo to helium microscopy, referred to as ``heliometric stereo''. In the presented work, the latter is explored. For the most efficient image acquisition, a multiple detector SHeM could be built, removing the necessity of sample rotation by imaging different azimuthal angles simultaneously. The current Cambridge SHeM, however, has a single detector, so the experimental realisation of heliometric stereo requires sample rotation.
\\[12pt]
The current work presents the full experimental implementation of heliometric stereo using a single detector microscope. Starting from the design and simulation of a novel pinhole plate, to the sample manipulator's hardware upgrade adding a rotational stage, and finishing with the development of a real-space point tracking method. The culmination of which enables the collection of at least three azimuthally rotated images that are given to the heliometric stereo algorithm.
\\[12pt]
To enable heliometric stereo, a method by which points on a sample can be tracked in real-space must be developed. The method must fully capture the degrees of freedom of the sample manipulator so that the position of a feature can be accurately tracked/predicted when the sample is rotated azimuthally.
\begin{figure}[h]
    \centering
    \includegraphics[width=0.7\textwidth]{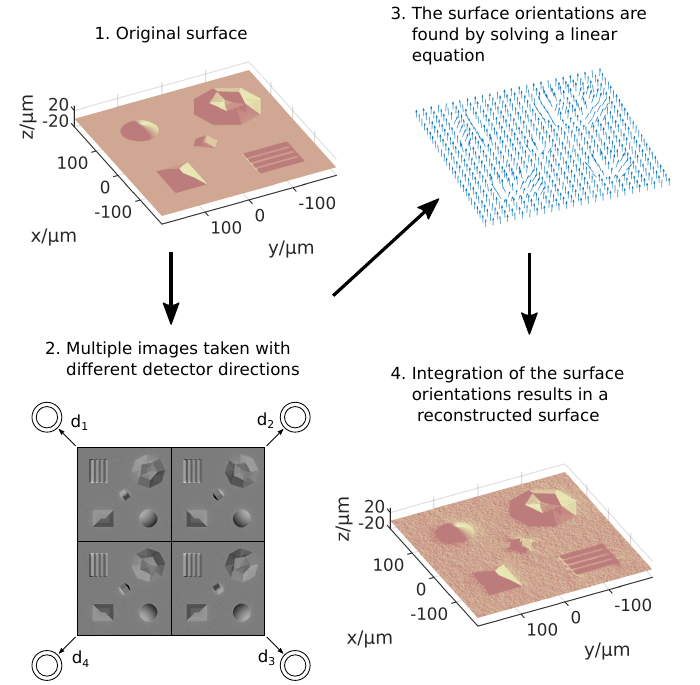}
    \caption{Overview of the basic heliometric stereo method. (1) The simulated sample. (2) The sample is imaged used multiple detectors placed in different directions to yield a series of helium images. (3) Those helium images are used in terms of the intensity vector to acquire surface normals. (4) Finally, as the normals to a surface are the gradient of that surface they may be integrated to give a reconstructed surface. Diagram reproduced from Figure 4 of Lambrick et al. \cite{Lambrick2021}}
    \label{fig:heliometricstereo_algorithm}
\end{figure}

\newpage
\section{Point Tracking Calibration}
\begin{figure}[h]
    \centering
    \includegraphics[width=\textwidth]{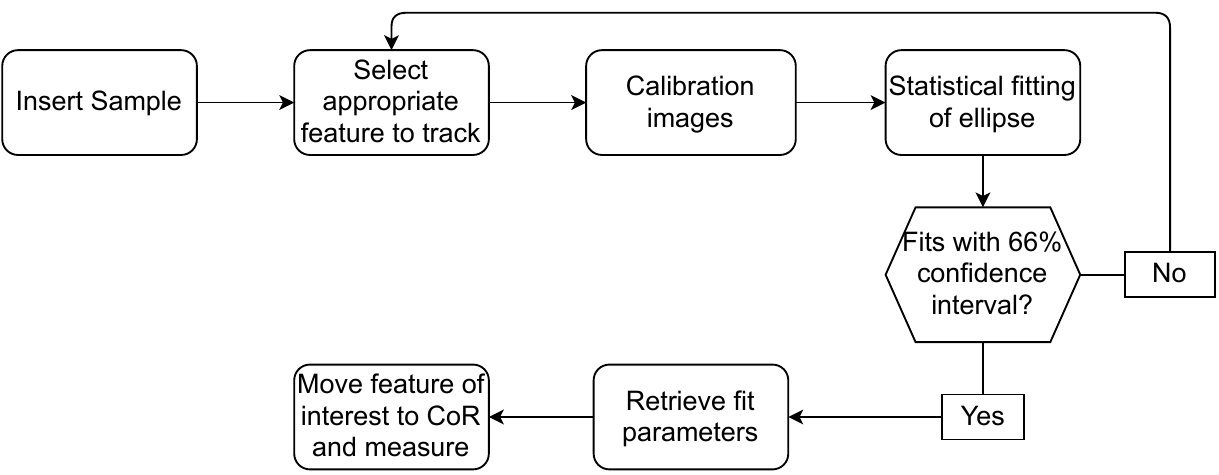}
    \caption{Flowchart outlining the point tracking calibration routine to find the centre of rotation (CoR) for a generic sample.}
    \label{fig:point_tracking_routine}
\end{figure}
\begin{flushleft}
After sample insertion, a high-resolution overview image of the entire sample was made to identify a suitable feature to track for the calibration of the centre of rotation. Ideally, multiple overview images from different rotation angles should be taken to eliminate any potential tracking candidates which become obscured due to masking or shadows, as described in the work of Fahy et al. \cite{Fahy2018}. A feature is masked when it is obscured because of the projection of the detector’s perspective onto the sample whereas shadowing is when a feature is obscured due to the projection of the beam onto the sample.
\end{flushleft}

\begin{figure}[h]
    \centering
    \begin{subfigure}{0.4\textwidth}
    \includegraphics[width=\textwidth]{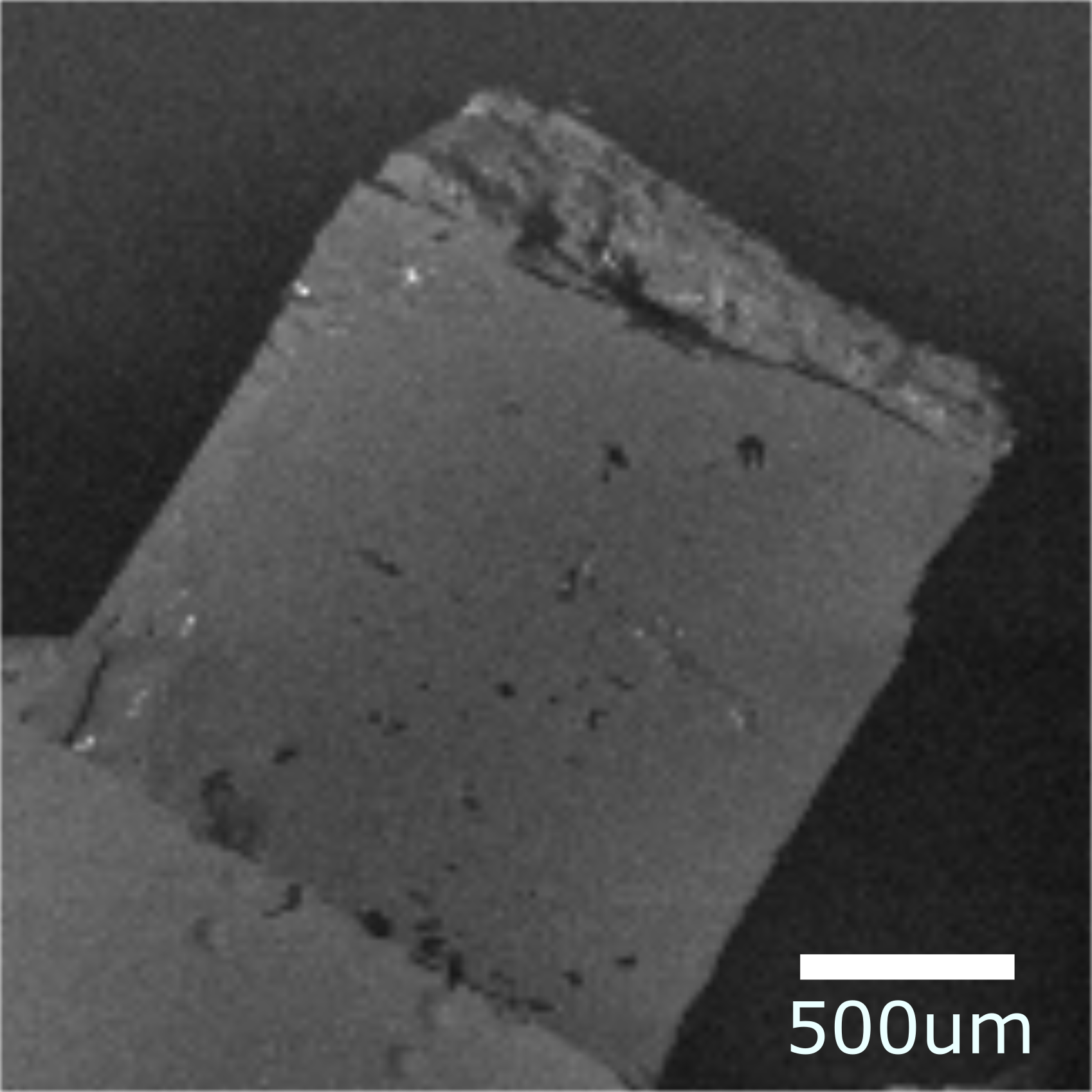}
    \subcaption[a]{}
    \label{lif_overviews_a}
    \end{subfigure}
    \begin{subfigure}{0.405\textwidth}
    \includegraphics[width=\textwidth]{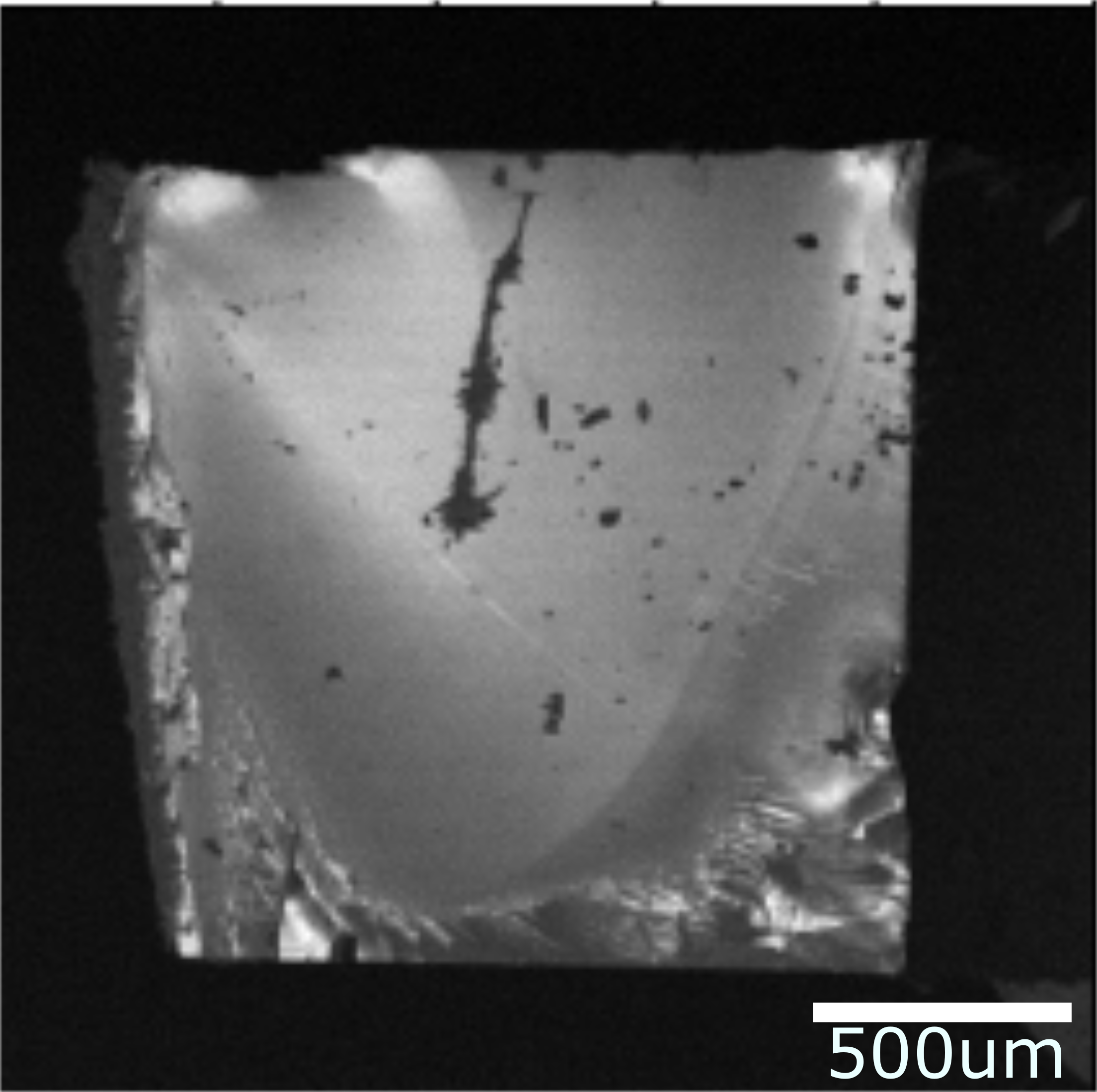}
    \subcaption[b]{}
    \label{lif_overviews_b}
    \end{subfigure}
    \caption{Reference images of entire LiF surface taken with pixel sizes of $15\,\mu m$ and $10\,\mu m$ for Panels \ref{lif_overviews_a} and \ref{lif_overviews_b}, respectively. It is clear from the reference images that the sample has multiple very sharp corners which protrude out from the edge of the sample holder, higher resolution images of these corners were chosen as the ideal feature to track.}
    \label{fig:lif_overviews}
\end{figure}
\newpage
\begin{flushleft}
The main considerations when picking a calibration point are:
\end{flushleft}
\begin{enumerate}
\setlength\itemsep{-0.5em}
    \item High contrast against surrounding topography
    \item Well defined/sharp feature like a corner
    \item Features that protrude from the surface
\end{enumerate}
Choosing a feature which has at least two of the above properties is essential so that the calibration feature is not obscured by some combination of masking, shadows and changes in diffractive contrast as the sample is rotated, hence the importance of taking multiple overview images. Sub-optimal point selection, which introduces any of the aforementioned difficulties in re-acquisition of selected points, is the dominant source of error in the statistical fitting routine. 
\begin{figure}[ht]
    \centering
    \begin{subfigure}{0.2\textwidth}
    \includegraphics[width=\textwidth]{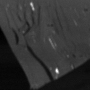}
    \subcaption[a]{$5\mu m$}
    \end{subfigure}
    \begin{subfigure}{0.2\textwidth}
    \includegraphics[width=\textwidth]{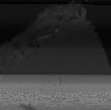}
    \subcaption[b]{$5\mu m$}
    \end{subfigure}
    \begin{subfigure}{0.2\textwidth}
    \includegraphics[width=\textwidth]{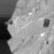}
    \subcaption[c]{$5\mu m$}
    \end{subfigure}
    \begin{subfigure}{0.2\textwidth}
    \includegraphics[width=\textwidth]{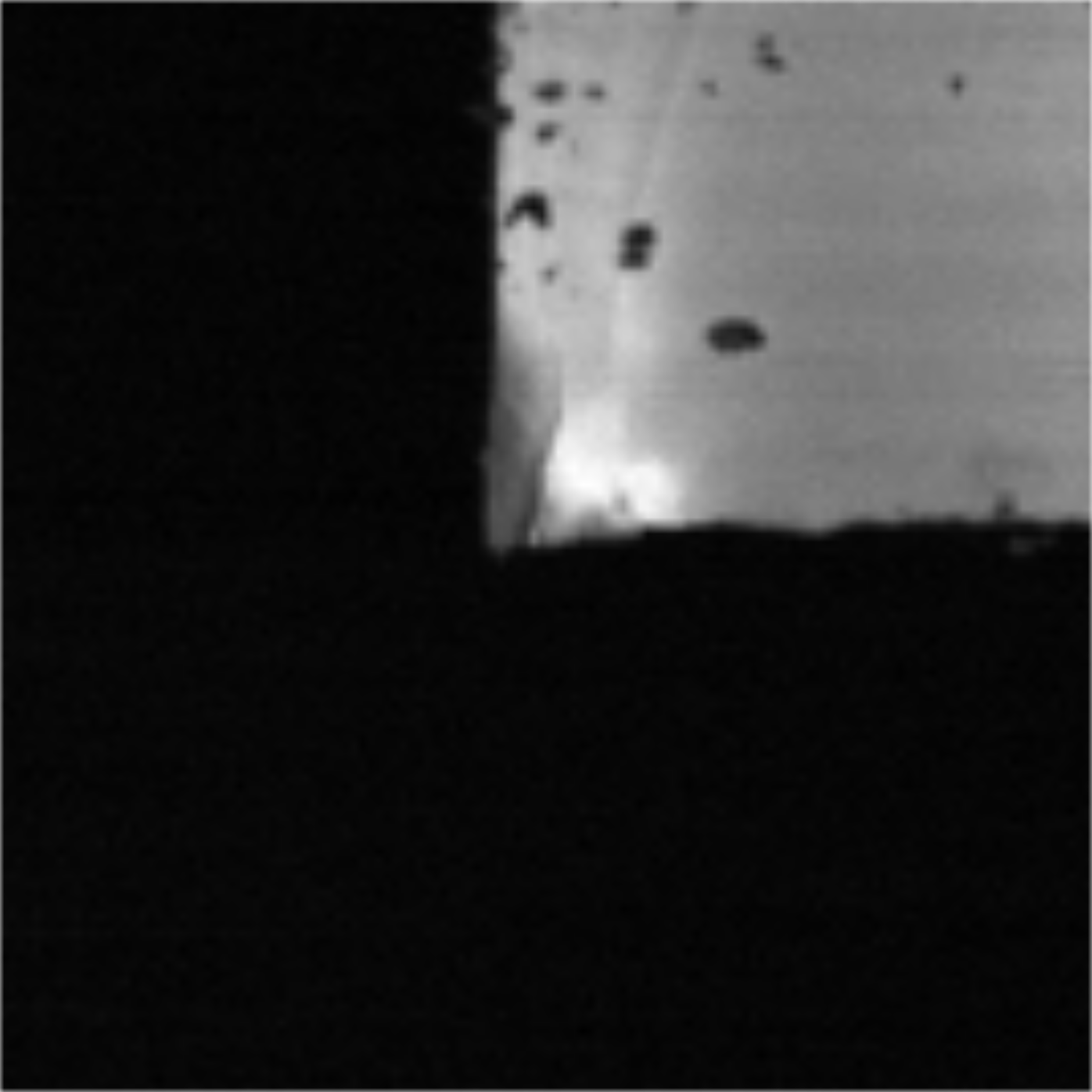}
    \subcaption[d]{$7.5\mu m$}
    \label{fig:lif_corners_d}
    \end{subfigure}
    \caption{Close up images of tracking candidates (corners of sample) with pixel sizes listed in sub-captions. Panel \ref{fig:lif_corners_d} was selected for the point tracking calibration routine because it best matches the selection criteria with high contrast to the surrounding area, a sharp corner and protrudes from the sample.}
    \label{fig:lif_corners}
\end{figure}
Having identified the optimal feature for tracking in Figure \ref{fig:lif_corners_d}, a high-resolution image was taken to identify the exact pixel at the corner which will be tracked.
\begin{figure}
    \centering
    \includegraphics[width=0.36\textwidth]{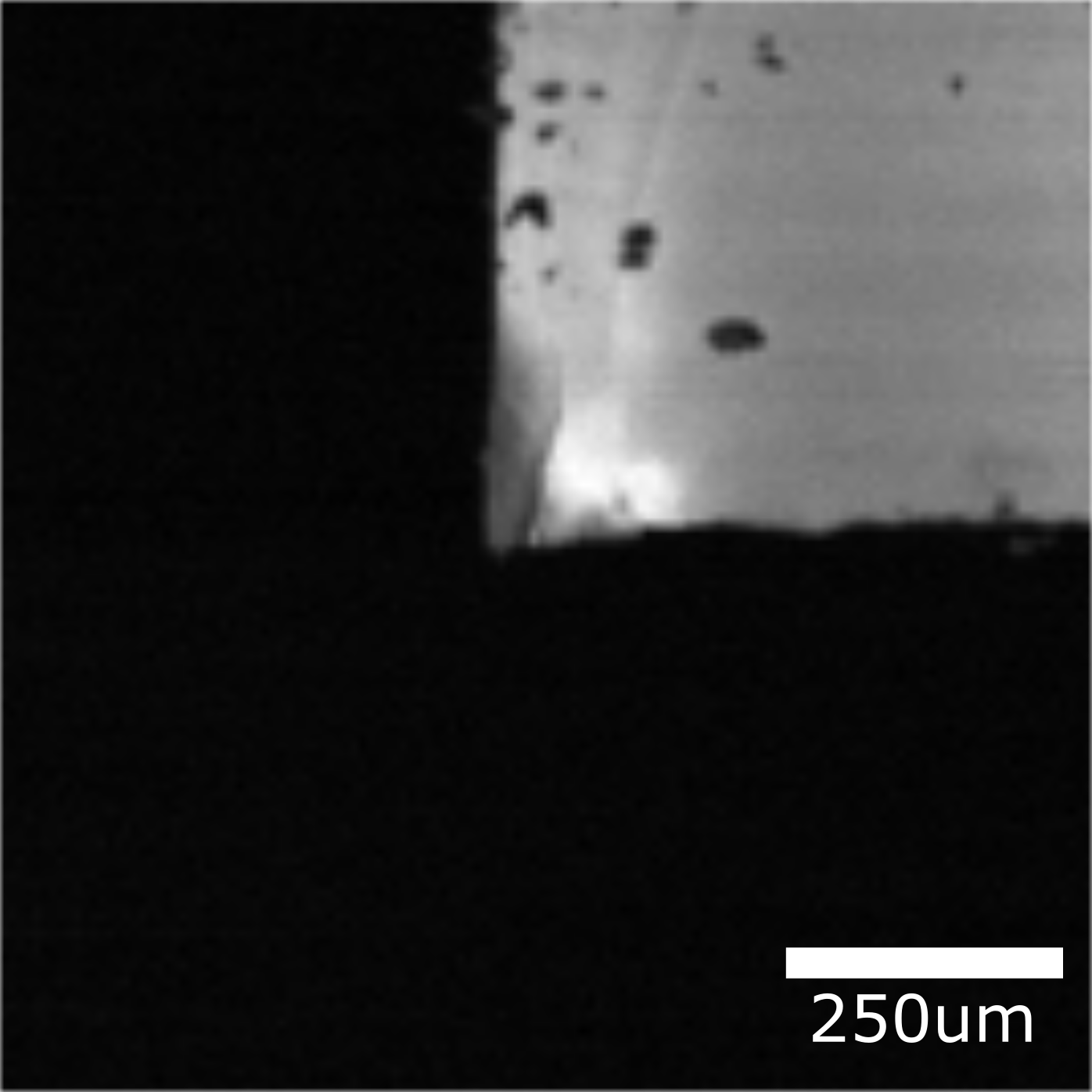}
    \caption{$7.5\mu m$ pixel size image used to identify the exact pixel for tracking. First image in centre of rotation calibration routine data set.}
    \label{fig:lif_tracked_corner}
\end{figure}
A minimum of 5 images are required for the least-squares fitting algorithm to effectively find the centre of rotation. Figure \ref{fig:fit_images} shows 6 images were taken at \ang{60} increments to optimise between total imaging time and improved statistical accuracy.
\\[12pt]
In the current work, all calibration images must be acquired manually by taking overview images after each rotation and then finding the selected feature to take a high resolution image. Calibration of the alignment of the sample stages only needs to be carried out once for a given configuration of pinhole plate, l-bracket and sample stages, not for every new sample, because the stages used maintain absolute position by self-correcting. 
\\[12pt]
Converting the pixel coordinates of the points across the calibration images to real-space, the coordinates can be input into the least-squares elliptical fitting algorithm shown in Appendix \ref{appendix:least_squares_ellipse} \cite{Gal2021}. Generally, if the calibration points deviate from the resulting fit by more than 2 standard deviations a different calibration point should be selected. 
\begin{figure}
    \centering
    \includegraphics[width=0.7\textwidth]{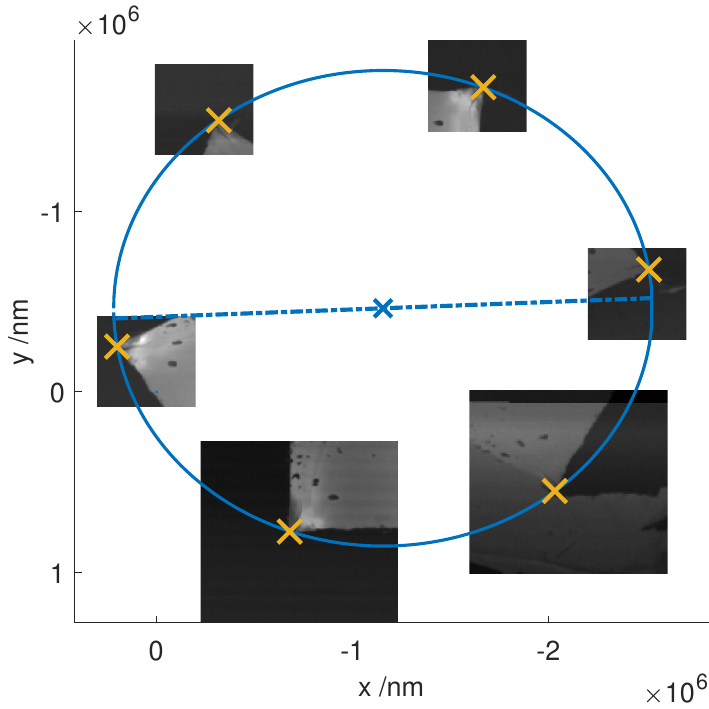}
    \caption{$10\mu m$ pixel size calibration images of the \ch{LiF} corner from Figure \ref{fig:lif_corners_d}. An ellipse has been fitted to the calibration feature, with the centre of rotation and semi-major axis shown. Calibration images have been rotated to have accurate position in real-space.}
    \label{fig:fit_images}
\end{figure}
The data shown in Figure \ref{fig:fit_images} fits an ellipse to sub-pixel precision, providing further supporting evidence that the analytical model has fully captured the motion of the feature and suggesting that pixel size is the dominant factor in point tracking error.
\begin{equation}
\begin{bmatrix}
    x_{rot} \\ y_{rot}\\ 1\\
    \end{bmatrix}
    =
    \begin{bmatrix}
    1 & 0 & x_{current}\\
    0 & 1 & y_{current}\\
    0 & 0 & 1 \\
    \end{bmatrix}
    \begin{bmatrix}
    cos(\theta) & sin(\theta)\cdot k & 0\\
    -sin(\theta)/k & cos(\theta) & 0 \\
    0 & 0 & 1
    \end{bmatrix}
    \begin{bmatrix}
    1 & 0 & -x_{centre}\\
    0 & 1 & -y_{centre}\\
    0 & 0 & 1 \\
    \end{bmatrix}
    \label{ellipse_transform}
\end{equation}
\\[12pt]
Using the centre of rotation and the eccentricity of the ellipse, $e = 1/k$, the rotational matrix shown in Equation \ref{ellipse_transform} can be constructed to predict the location of any point on the sample in real-space, regardless of its compatibility with the calibration point selection criteria. The angle $\theta$ is the angle of the azimuthal sample rotation.
\newpage
Taking the eccentricity of the ellipse, $e$,  the tilt angle, $\chi$, which projects a circular motion onto the observed ellipse can be calculated using

\begin{equation}
 \chi = cos^{-1}(1/k)
 \label{equation_chi}
\end{equation}
\\[6pt]
to give $\chi = 16 \pm \ang{3}$ for the data shown in Figure \ref{fig:fit_images}. The main use of finding $\chi$ and the direction it lies in, $\phi$, is in debugging the experimental setup. Knowledge of the magnitude and direction of the stage tilt helps to identify where in the sample manipulator the tilt originates, and therefore aides in correcting it. Applying point tracking to fixing stage axis alignment is crucial because non-zero $\chi$ represents misalignment of the sample with the scattering plane, affecting all SHeM measurements. The physical origins of the discussed angles can be seen in Figure \ref{fig:stages_tilts}.
\\[12pt]
The excellent precision of the statistically fitted ellipse, Figure \ref{fig:fit_images}, provides strong evidence that an analytical solution whose parameters return a single ellipse can be derived for the \textalpha-SHeM sample motion. 
\\[12pt]
\section{Analytical Tilt Model}
\begin{figure}[h]
    \centering
    \includegraphics[width=0.7\textwidth]{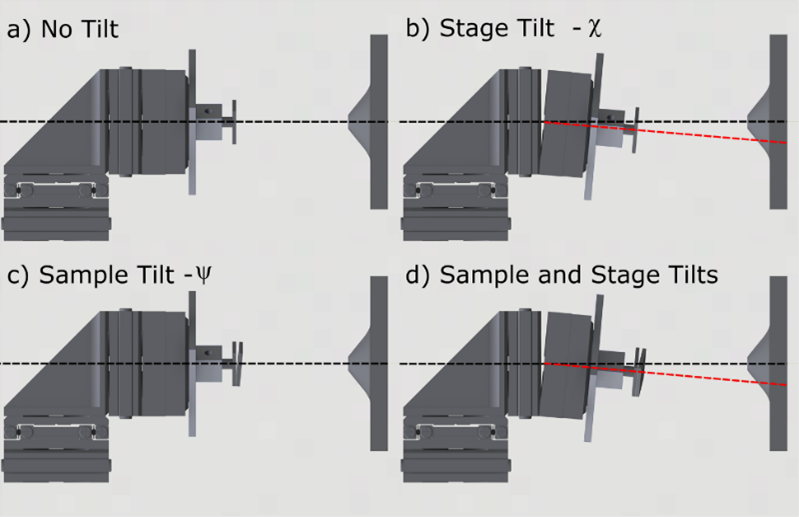}
    \caption{CAD models of the \textalpha-SHeM sample manipulator stages with dashed lines representing the rotational axis of the sample. The black dashed line represents the rotational axis when $\chi = 0$ such that it is normal to the scattering plane. The red dashed line illustrates $\chi \neq 0$, a misalignment of the rotational axis with respect to the normal of the scattering plane.}
    \label{fig:stages_tilts}
\end{figure}
\begin{figure}
    \centering
    \includegraphics[width=0.5\textwidth]{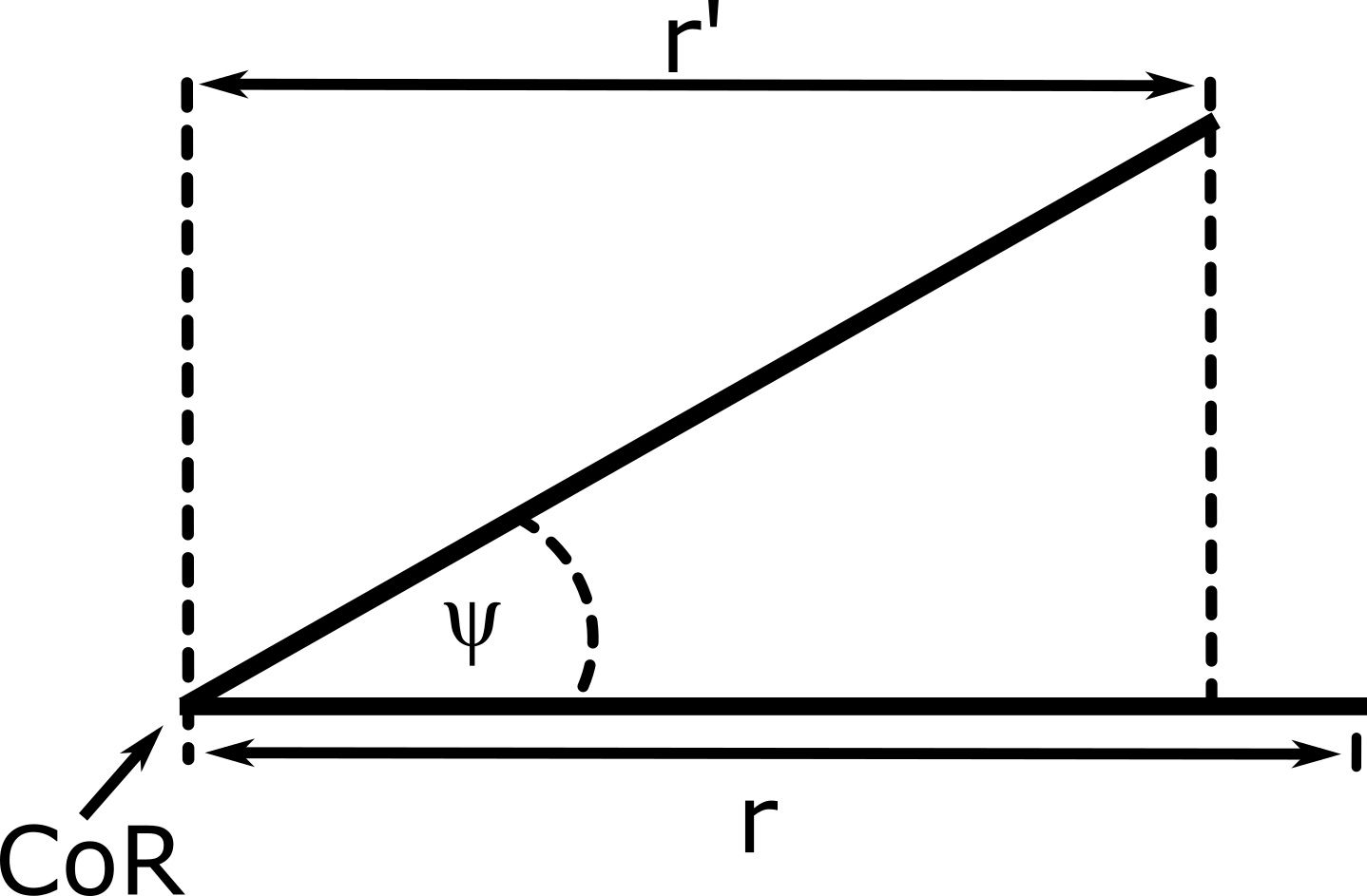}
    \caption{Diagram of how sample tilt, $\psi$, reduces the apparent size of a sample by a factor $cos(\psi)$. When tilted about the centre of rotation (CoR) by $\psi$, the distance of a feature from the CoR becomes $r'=r\cdot cos(\psi)$. Therefore, when the sample is rotated azimuthally the diameter of the circular path traced goes from $2r\, \text{to}\,2r'$.}
    \label{fig:cos_psi}
\end{figure}
\begin{figure}
    \centering
    \includegraphics[width=0.7\textwidth]{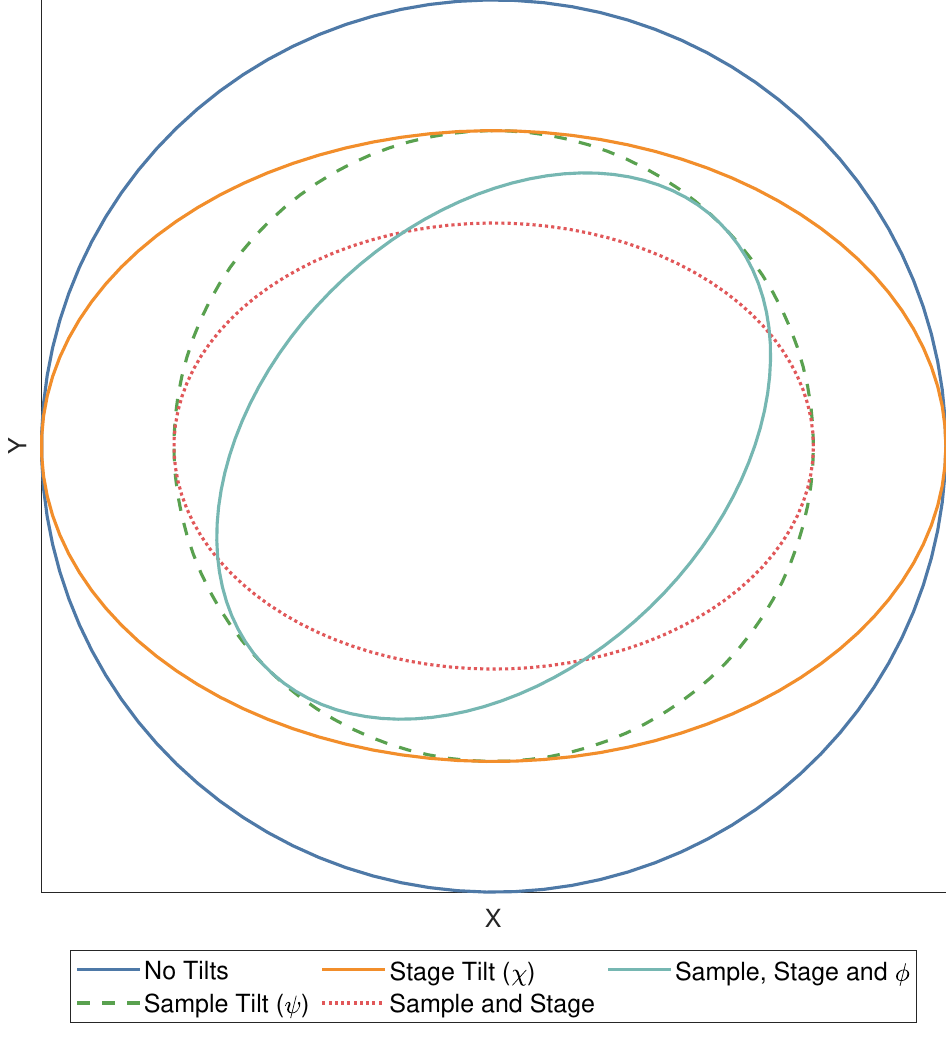}
    \caption{Plot showing the effect of the three types of sample tilt, $\psi$, $\chi$ and $\phi$ have on the observed path of a sample feature using the analytical model developed in the presented work.}
    \label{fig:tilt_paths}
\end{figure}

\begin{flushleft}
Starting from the geometry of the sample manipulator stages in \textalpha-SHeM, shown in Figure \ref{fig:stages_tilts}, one finds that there are three degrees of freedom which fully capture the motion of a point on the sample. Angles $\chi$, $\psi$, and $\phi$, representing stage, sample, and direction of stage tilt, respectively. $\chi$ and $\psi$ differ fundamentally in that $\chi$ is a function of any misalignments originating behind the rotational stage with respect to the master geometry, the scattering plane. $\psi$ on the other hand is the sum of angles after the rotational stage, whether that be from facet angles on the sample’s surface, or sample mounting misalignments. $\phi$ is simply the direction in which the $\chi$ misalignment lies with respect to the normal rotational axis. Using Figure \ref{fig:tilt_paths}, $\phi \neq 0$ results in a non-horizontal major axis for the elliptical path traced.  Within the three quantities, it is impossible to determine with exactness where misalignment originates, although estimates can be made using $\phi$.
\end{flushleft}
Panel a) of Figure \ref{fig:stages_tilts} shows the trivial case, the tracked point’s path will clearly be circular and true-to-size in the image relative to the real-space 3-dimensional path because the rotational axis of the sample is normal to the pinhole plate.
\\[12pt]
Panel b) shows pure stage tilt, $\chi$, the case when the rotational axis lies outside the scattering plane. Given that the point on a sample still moves in a circle, but the entire rotating geometry is inclined with respect to the pinhole plate, the point’s path is projected an angle, $\chi$, onto the image taken. Hence, the path appears elliptical and is defined by equation \ref{eqn_ellipse}.
\\[12pt]
\begin{equation}
    x^2/a^2 + y^2/b^2 = 1
    \label{eqn_ellipse}
\end{equation}
Where, $a$ and $b$are the major and minor axes respectively. Relating to the eccentricity by 
\\[12pt]
\begin{equation}
    e = \sqrt{1-b^2/a^2}.
    \label{eqn_eccentricity}
\end{equation}
Panel c) shows pure sample tilt which can be due to any combination of sample mounting inaccuracy and facet angles of the sample’s surface. As the rotational axis remains normal to the pinhole plate, because the angle originates after the rotation stage, a point’s path will remain circular, but the radius will be scaled by a factor $cos(\psi)$, as shown in Figure \ref{fig:cos_psi}. $\psi$ can only be found with the aid of an absolute measurement of the surface size normal to the imaging apparatus, which can be achieved with the addition of a goniometer sample stage.
\\[12pt]
\\[12pt]
Panel d) shows a combination of the $\chi$ and $\psi$ tilts. As they are linearly independent operations acting on the tracked point’s path, the result is simply an ellipse which has been scaled by a factor $cos(\psi)$.
\\[12pt]
The final degree of freedom, $\phi$, corresponding to the direction of the $\chi$ tilt manifests in images as the angle off the horizontal by which the major axis of the ellipse is displaced.

\section{Measuring facet angles using the point tracking method}
The standard technique that is used to find facet angles is x-ray diffraction (XRD) rocking which can additionally reveal the exact crystalline structure of the facets. However, XRD uses high energy x-rays that can be destructive to the sample, unlike SHeM. 
\\[12pt]
One can use the knowledge that there are two fundamentally distinct types of tilt possible in the \textalpha-SHeM sample environment, $\chi$ and $\psi$, in conjunction with an absolute measurement of surface features to find the magnitude of sample tilt, $\psi$. Such an absolute measurement can also be performed in \textalpha-SHeM with the aid of a goniometer stage to guarantee images taken normal to the sample surface. The Newcastle (Australia) SHeM \cite{Fahy2018} has a sample manipulator with a goniometer stage for use in 3D surface reconstructions using the stereogrammetry method \cite{Myles2019} which tilts the sample instead of azimuthally rotating it as in the Cambridge SHeM. By then performing point tracking on individual crystal domains, one can find the facet angles of a multi-faceted crystalline surface. The point tracking procedure has already been applied to cleaved lithium fluoride $\langle 111 \rangle$ sample shown in Figure \ref{fig:lif_overviews} with the elliptical fit parameters in Figure \ref{fig:fit_images} returning the sample (surface) tilt as $16\pm \ang{3}$. Repeating the procedure for a multi-faceted sample would be time intensive, but SHeM provides a range of unique imaging properties which have particular relevance in technological and biological samples like polycrystalline doped perovskites, biological crystals and synthetic substrates, all of which are difficult to image using tradition microscopy techniques.

\section{Statistical Analysis}
The most straightforward quantification of the accuracy of point prediction is to find the mean difference between a given experimental point and its predicted counterpart. Since the point prediction works by taking the \ang{0} experimental point and matching it to the nearest point on the fitted curve itself and then rotating to a chosen angle, the first (\ang{0}) point pair will be ignored in statistical analysis as it is not a true prediction. The mean difference comparison of experimental points to their nearest matching point on the curve provides insight into the goodness-of-fit in the form of residuals rather than providing information on accuracy of point prediction. Therefore, only five out of six point pairs are used in the following analysis.
\\[12pt]
It is important to note that although the data conforms to an elliptical fit, it is still non-directional because the investigation is not sampling some probability density function which is a function of the angle the sample is rotated through. Hence, standard non-directional statistical tools can be used without need for adaptation to directional statistics.
\\[12pt]
Calculating the mean distance between an experiment point and its corresponding predicted point, from now on referred to as the “mean point delta” (MPD), yields $\pm 17\mu m$ for the elliptical fit.
\\[12pt]
There is a moderate correlation $(0.47)$ between the total angle the sample has been rotated and the corresponding point deltas for the elliptical fit, demonstrating that the overall error in point prediction compounds with each successive iteration. One way to reduce the compounding error is to employ a “semi-automatic” prediction system which does a nearest point search to match every new predicted point back to the fitted function before applying another rotation. The “automatic” method used so far only matches the first experimentally acquired point to the fit and finds all predicted points based on the first, leading to a significant compounding error. Utilising the semi-automatic method yields an 11\% decrease in mean point delta, from approximately $17\,\mu m$ to $15\,\mu m$.
\\[12pt]
The method works to reduce the magnitude of the pixel selection error, by far the dominant component in the total error. Pixel selection error has two degrees of freedom, angular and radial. By re-matching the predicted point to the fit using a nearest point search before applying the next rotation, the radial error component of the pixel selection is greatly reduced. The angular component remains unaffected because the nearest point on the fit must lie on a straight line between the centre of rotation and the predicted point.
\\[12pt]
Given that the elliptical fit from Figure \ref{fig:fit_images} matches the experimental points to sub-pixel accuracy, the improvement in MPD from the semi-automatic method effectively quantifies the exact proportion of the original MPD which is due to radial error. Any further optimisations to the method should therefore focus on reducing the angular component of the MPD. The semi-automatic method is ideal for usage in data sets where the images cannot be retaken, or improvements to the experimental procedure, e.g. resolution enhancements, are not possible.
\\[12pt]
The more globally effective way to reduce the mean point delta is to identify the main sources of the error and make improvements in the experimental procedure. The dominant component of the error comes from the manual point selection during calibration. Some of the error can be attributed to human error, however the majority is from the resolution of the images and masking on the sample which can obscure features. Assuming the calibration point is optimal under the previously detailed criteria, an improvement to resolution remains by far the dominant component of the MPD.
\\[12pt]
As of June 2021, the highest resolution achieved, taken as the full-width half-maximum, is approximately $350\,nm$ under highly optimised conditions. A valuable continuation of the work would be to repeat the point tracking method at a range of pixel sizes from the $10\,\mu m$ images from Figure \ref{fig:fit_images} used here, down to $\approx 350\,nm$ to inform the scaling of the MPD with pixel size.

\section{Further Statistical Analysis}
\label{subsection:further_stats}
If a larger data set were to be taken, the mean point deltas could be used to extract further statistical information on the nature of the error through the standard deviation/variance, and the more advanced skewness and excess kurtosis. The standard deviation/variance quantify the spread of the mean point delta, effectively revealing the precision of point prediction whereas the MPD corresponds to the accuracy.
\\[12pt]
Skewness and excess kurtosis are used together to quantify the “shape” of a normal distribution and can also be used to identify what sub-category of normal distribution the data falls under, or in fact if it is one at all, using a skewness-kurtosis plot. Their graphical interpretations are that skewness describes the asymmetry of a normal distributions, where 0 is perfectly symmetric, negative skews left and positive skews right. Excess kurtosis describes the weight of the distribution present in the tails, 0 indicates a perfect normal distribution, positive values give a narrower, pointed distributions and negative ones tend to a step function.
\\[12pt]
Note that the method for further statistical analysis has been implemented in MATLAB, ready for future measurements.
\\[12pt]
Implementing the further analysis detailed would allow for deep insight into the source and nature of errors in point tracking, potentially allowing for improvements to the method which do not require significant changes to the SHeM hardware, such as the semi-automatic point prediction.

\section{Normal Incidence Imaging}
The pinhole plate is arguably the defining feature of SHeM implementation and defines much of its performance. The beam’s angle of incidence on the sample and the spatial and angular resolutions are all a direct product of the design of pinhole plate used. 3D imaging on \textalpha-SHeM using the heliometric stereo method outlined in \cite{Lambrick2021} can be implemented using a rotational stage to image the sample at different azimuthal angles, the procedure is more simple using a beam normally incident on the sample. Which can be achieved by extending the source cone in the usual \ang{45} incidence pinhole plates.
\newpage
The key difference between \ang{45} and normal incidence plates is the lengths of their source cones. To achieve a normally incident geometry, the source cone of the pinhole plate must be extended towards the sample which introduces a potential problem in the pumping speeds down the length of the cone. If the pumping through the source cone were insufficient, the beam incident on the back of the aperture would stagnate and interfere with the beam by forming a relatively high-pressure region in the source cone. The effect of differential pumping through the pinhole has always been assumed to be negligible and the stagnation of the beam behind the pinhole plate has never been investigated prior to the current work.
\\[12pt]
There are three possible regimes of beam-background interaction in the source cone, the first is effectively no interaction because the background pressure in the source cone is sufficiently low throughout that the mean free path does not lead to significant beam-background interaction. The second is when the background pressure is high enough to “significantly” attenuate the beam whilst still letting most of the beam propagate through and form an image. The third is complete destruction of the beam from a high background pressure that does not allow any image to form with only an effusive beam entering the sample chamber.
\\[12pt]
A candidate normal incidence pinhole plate was designed and a computational method was developed to evaluate the pumping speed.

\subsection{Normal Incidence Differential Pumping Simulations}
\begin{figure}
    
    \centering
    \begin{subfigure}{0.8\textwidth}
    \includegraphics[width=1\textwidth]{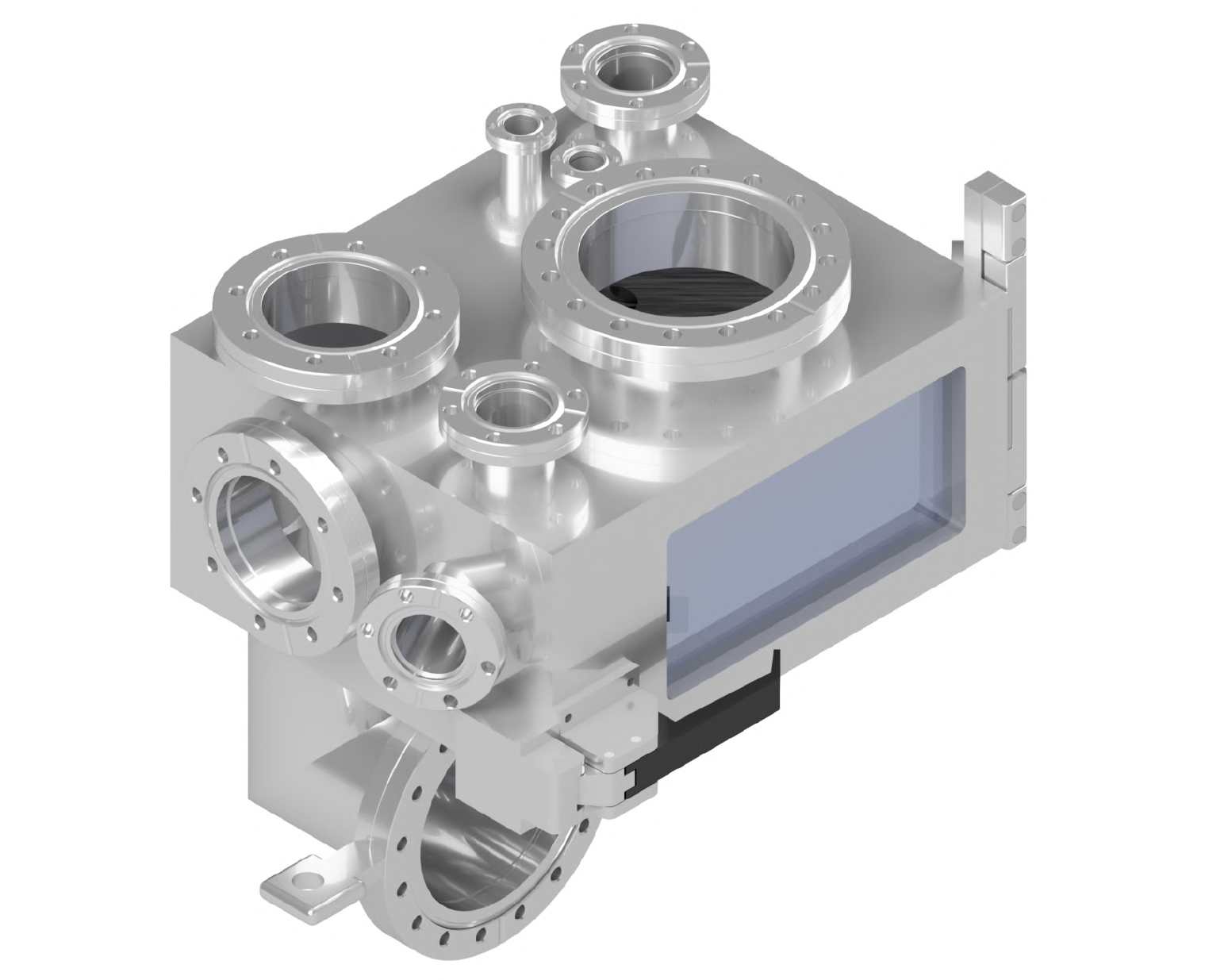}
    \end{subfigure}
    \begin{subfigure}{0.8\textwidth}
    \includegraphics[width=1\textwidth]{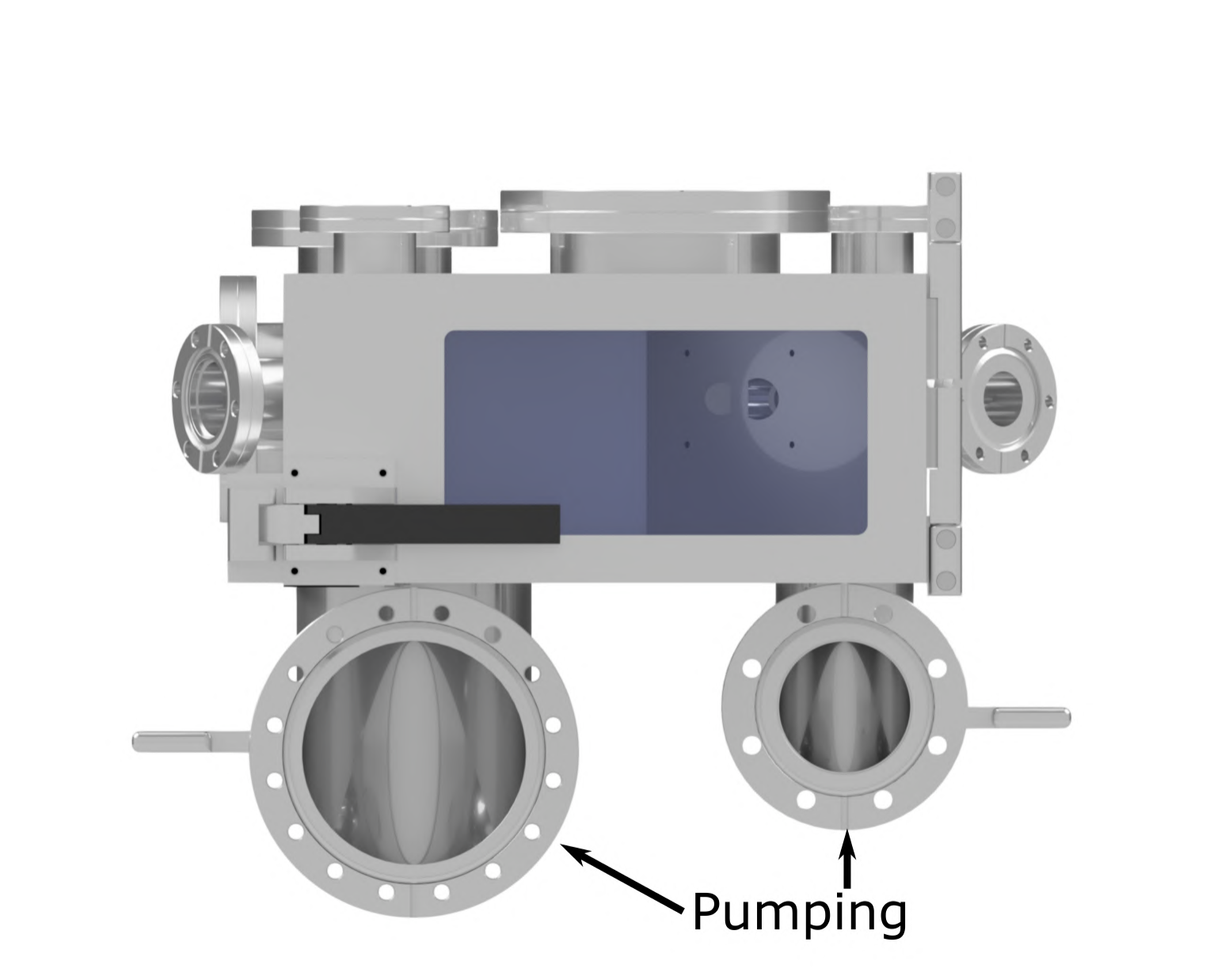}
    \end{subfigure}
    \caption{Rendered CAD model of real life \textalpha-SHeM sample chamber with no pinhole plate mounted}
    \label{fig:chamber_actual}
\end{figure}

\begin{figure}
    \begin{subfigure}{0.9\textwidth}
    \includegraphics[width=\textwidth]{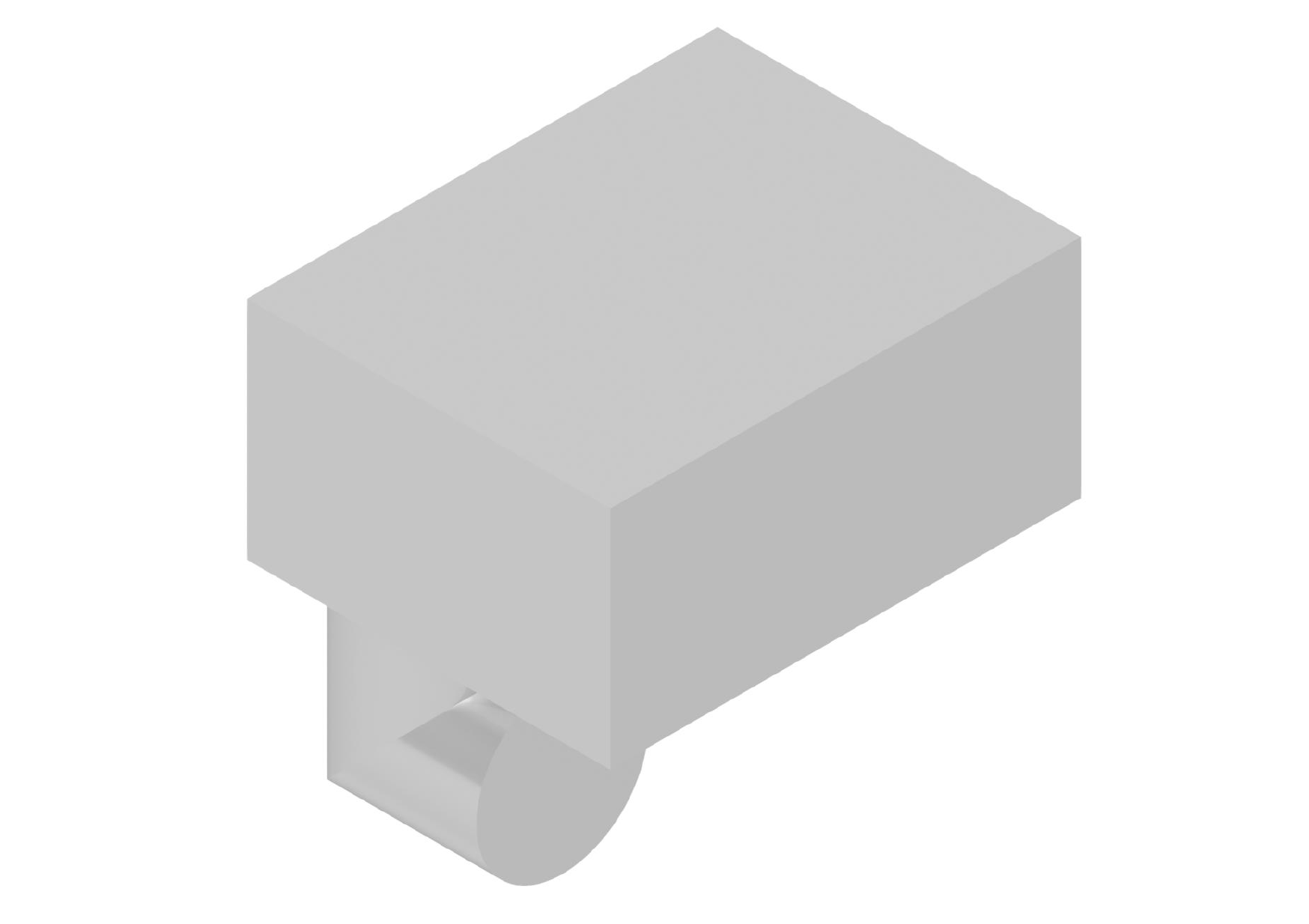}
    \end{subfigure}
    \begin{subfigure}{0.9\textwidth}
    \includegraphics[width=\textwidth]{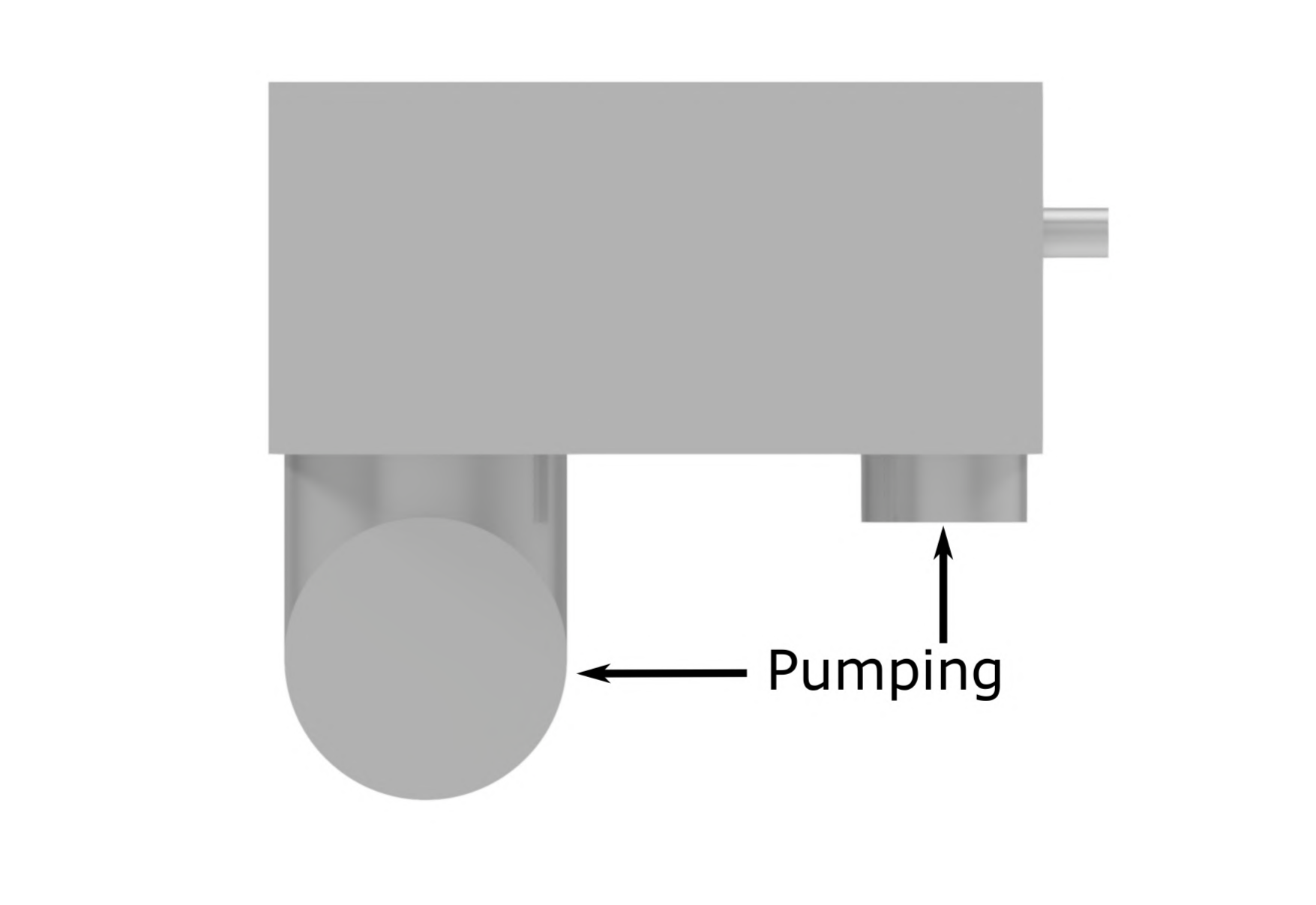}
    \end{subfigure}
    \caption{CAD model of \textalpha-SHeM sample chamber with simplified geometry to decrease computation time in Molflow. All non-essential exterior geometry has been removed, leaving only the pumping sources.}
    \label{fig:chamber_simplified}
\end{figure}

One could approach the problem of pumping speed through a cone analytically using an extension of the widely used Oatley method \cite{Oatley1957} for calculating conductance of low pressure gases through composite systems, as described by Mercier \cite{Mercier2006}. However, to employ Mercier's extended Oatley method, the pumping speeds at both ends of the source cone would need to be known accurately. Due to the complex structure of the vacuum system in \textalpha-SHeM, knowledge of precise pumping speeds within the pinhole plate is unrealistic. Most significantly, Mercier only deals with cones whose centre-line is perpendicular to their base whereas all \textalpha-SHeM pinhole plates have non-perpendicular centre-lines. Furthermore, the \textbeta-SHeM pinhole plate geometries are even more complex with non-perpendicular, curved centre-lines. The combination of these factors makes the Monte-Carlo based Molflow ideal for investigating SHeM pinhole plate pumping speeds.
\\[12pt]
The complexity of the pinhole plate geometry involved required a numerical solution to find pumping speeds. ``Molflow'' \cite{Ady}\cite{Kersevan2019}, a test-particle Monte Carlo (TPMC) algorithm to simulate vacuum systems developed by CERN, was chosen to investigate the effect of the length of the source cone on pumping speeds. Molflow allows great flexibility in setting the parameters of the simulation such that a real system can be modelled as accurately as possible, namely with species-specific pumping speeds and outgassing. A key feature of Molflow is that it simulates one gas species at a time meaning all outgassing, pumping speed and pressure values relate solely to helium-4. Effectively making these values partial relative to the real-world system which contains other vacuum contaminants. The only significant limitation is that Molflow does not have the ability to form a directed beam that one would need to fully simulate the SHeM set-up during measurements. The limitation can be avoided by setting an absolute outgassing to the back of the pinhole as an analogue to a stagnant beam.
\\[12pt]
It was determined that three scenarios would give sufficient information on the relative pumping in the source cone to conclude whether the additional length of the source cone would affect beam propagation. The baseline to determine the maximum possible pressure a stagnant beam could achieve is when the pinhole is completely blocked, allowing no effusive or direct beam to pass into the sample chamber. Next is a $5\,\mu m$ pinhole which represents the actual planned experimental set-up. Finally, a completely blown out membrane with the surrounding scaffold left intact provides a baseline at the opposite end of the spectrum where pumping would not be a limiting factor because the size of the missing membrane is large compared to the 5um pinhole and not insignificant in comparison to the total area of the source cone.
\\[12pt]
Although Molflow cannot form a directed beam, one can freely set outgassing coefficients, both per unit area and as absolute values, to each surface of the imported geometry. By setting an absolute outgassing to the back of the aperture which is the same as the flux of the incident beam, one can simulate the relevant portion of the beam which remains in the source cone. The value of the outgassing can then be adjusted as a ratio of the open to closed areas at the back of the aperture. Note that the outgassing was not adjusted for the $5\,\mu m$ diameter pinhole because its area is negligible compared to the $2\,mm$ diameter total area of the end of the source cone. The incident beam flux was approximated using equations \ref{eqn:flow_const} - \ref{eqn:phi_pinhole}. First the nozzle flow constant \cite{1988Aambeqn} \cite{jellis} , $C_{he}$, is found by

\begin{equation}
    C_{he} = 45\cdot \sqrt{4/M_{he}}
    \label{eqn:flow_const}
\end{equation}
Where $M_{he} = 4\,\text{amu}$, is the molecular mass of helium-4. Theoretical flow rate through the nozzle is then calculated using

\begin{equation}
  \phi_{nozzle} = C_{he}\sqrt{T_c/T_0}P_0d^2,
  \label{eqn:phi_nozzle}
\end{equation}
with the pressure at the nozzle, $P_0 = 100\,bar$, nozzle diameter, $d = 10\,\mu m$, nozzle temperature $T_0 = 295\,K$ and room temperature $T_c = 295\,K$. Next, we find the estimated forward beam intensity with
\begin{equation}
    I_0 = 2\cdot \phi_{nozzle}/\pi.
    \label{forward_int}
\end{equation}
Due to the presence of background gas in the source chamber there is also an empirical attenuation factor of 2.5 to account for the beam-background interactions in the source chamber, which gives the realistic forward intensity as

\begin{equation}
    I_{0_{real}} = I_0/2.5.
    \label{forward_int_real}
\end{equation}
Using the solid angle of the pinhole we can calculate the beam flux through the pinhole with
\begin{equation}
    \phi_{pinhole} = \Phi\cdot I_{0_{real}}
    \label{eqn:phi_pinhole}
\end{equation}
giving a pinhole beam flux $\phi_{pinhole} = \num{5.6e-10}\,mbar$.
\\[12pt]
One can now follow the method outlined by Bergin \cite{Bergin2019} which gives the spot size of the beam incident on the back of the pinhole as $\approx 2\,mm$ in diameter, approximately the same as the $2 \,mm$ of the mounted aperture which is the narrowest point of the source cone, on which the beam is incident. Assuming that the helium intensity is uniform across the entire beam, one can use a ratio of areas between the pinhole and entire membrane to estimate the total beam flux incident on the membrane using the beam flux through the pinhole. The exact distribution of the incident helium flux on the back of the pinhole is not important because the resulting stagnant helium in the source depends solely on the magnitude of the flux, hence can be assumed to be uniform here.
\\[12pt]
Taking the total beam flux, $\phi_t$, and therefore absolute outgassing on the membrane, as $\num{5.77e-5}\,mbar\, l\,  s^{-1}$ for the blocked and $5\,\mu m$ pinholes. The transmitted beam flux through the open membrane is $\phi_m = \num{8.47e-6}\,mbar\, l\,  s^{-1}$ from the following ratio where the area of the open membrane is $A_m = \num{4.61e-7}\,m^2$,
\begin{equation}
    \phi_m = (\phi_t \cdot A_m)/A_t
    \label{eqn:phi_membrane}
\end{equation}
Which gives the stagnated beam flux for no membrane as $\phi_s = \num{4.93e-5}\,mbar\, l\,  s^{-1}$.
\\[12pt]
Figure \ref{fig:chamber_actual} shows a rendered CAD model of the \textalpha-SHeM sample chamber, a simplified version of which, shown in Figure \ref{fig:chamber_simplified}, was imported into Molflow with a normal incidence plate mounted visible on the right. Cosmetic modifications to the CAD model were made to simplify the overall geometry of the chamber to speed up the computation time by reducing the total polygon count of the imported model.
The final geometry modification was made within Molflow, a $2\,mm\times50\,mm$ facet which is transparent to gas was added which stretches down the length of the source cone, extending through the aperture into the sample chamber. Monitoring the pressure profile down the length of the source cone by ``texturing'' the facet gives the desired information on relative pumping speeds and the stagnated beam, whereas the section protruding into the sample chamber gives data on the effusive beam produced by the differential pumping of the stagnant gas through the aperture. The texture size of $5\,\mu m \times5\,\mu m$ was used, making each cell approximately the size of the aperture being investigated. Preliminary testing showed that such cells are sufficiently dense for the current investigation and provide the optimal resolution in the data with respect to computation time and the overall (millimetre) scale of the investigation. One limitation in sampling the effusive beam in the sample chamber directly in Molflow is that one cannot set different outgassing coefficients to opposing sides of the same facet, so the sample chamber background pressure will be slightly lower at the aperture because that facet has the beam stagnation outgassing set to the other side. It is possible however to use the pressure at the end of the source cone to analytically calculate the magnitude of the effusive beam.
\newpage

\begin{figure}
    \begin{subfigure}{0.85\textwidth}
    \includegraphics[width=\textwidth]{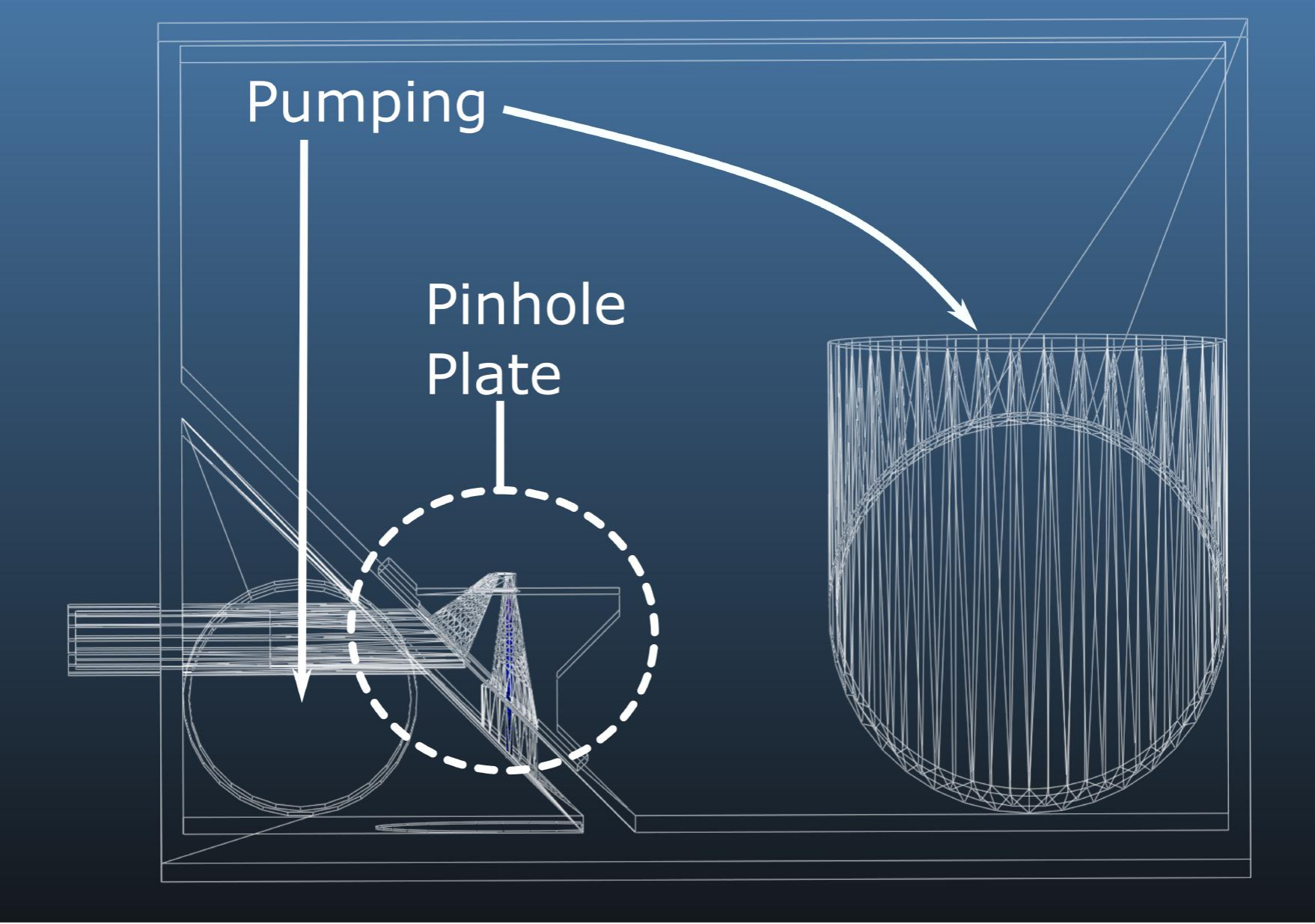}
    \end{subfigure}
    \begin{subfigure}{0.85\textwidth}
    \includegraphics[width=\textwidth]{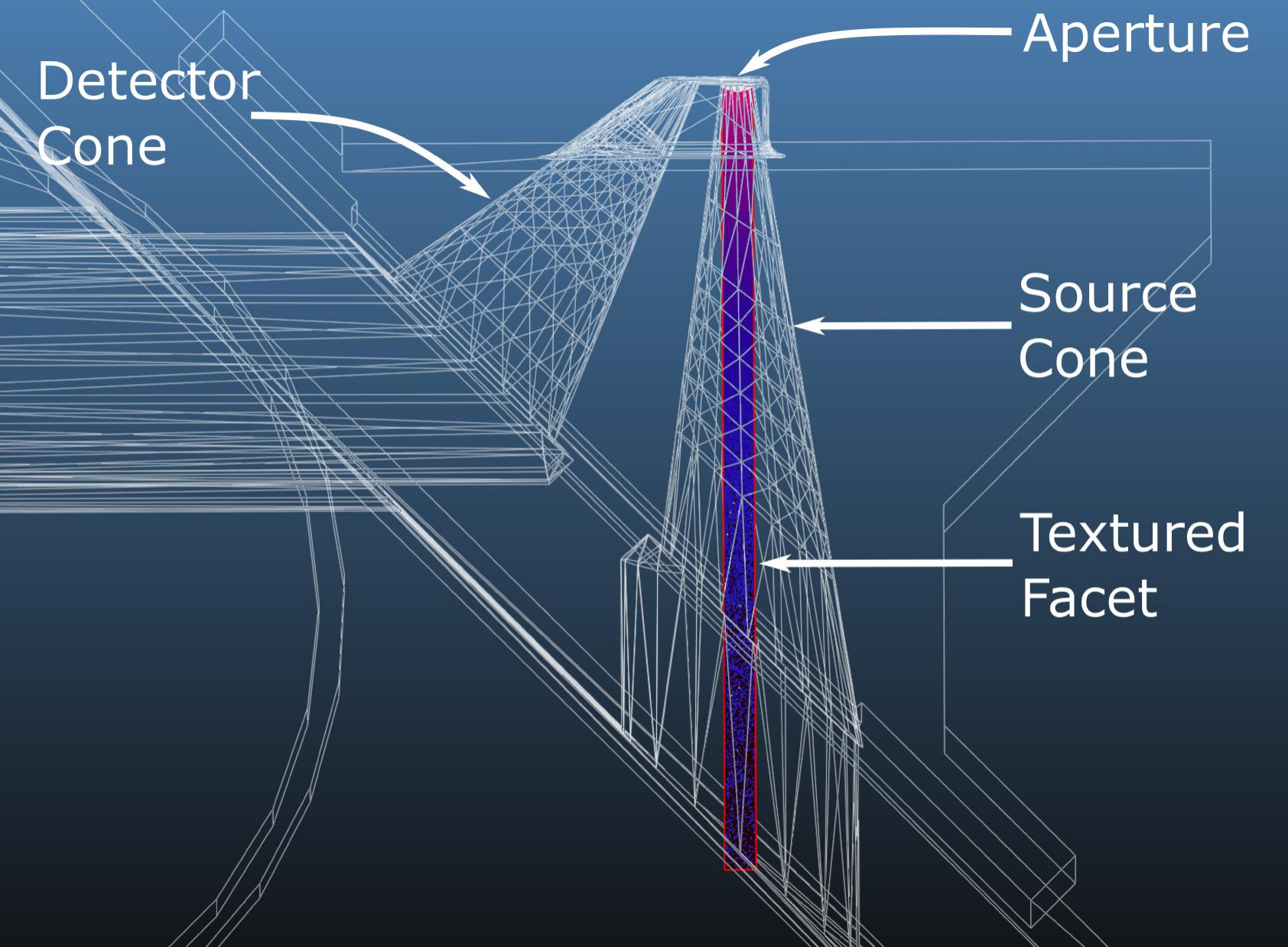}
    \end{subfigure}
    \caption{Screenshots of the simulation set-up within Molflow showing the \textalpha-SHeM sample chamber wire-frame geometry with a normal incidence pinhole plate installed. The coloured region down the length of the source cone, labelled ``Textured Facet'', is the only area in which pressure data is collected during the simulation.}
    \label{fig:molflow_setup}
\end{figure}

\begin{sidewaysfigure}
    \includegraphics[width=\textwidth]{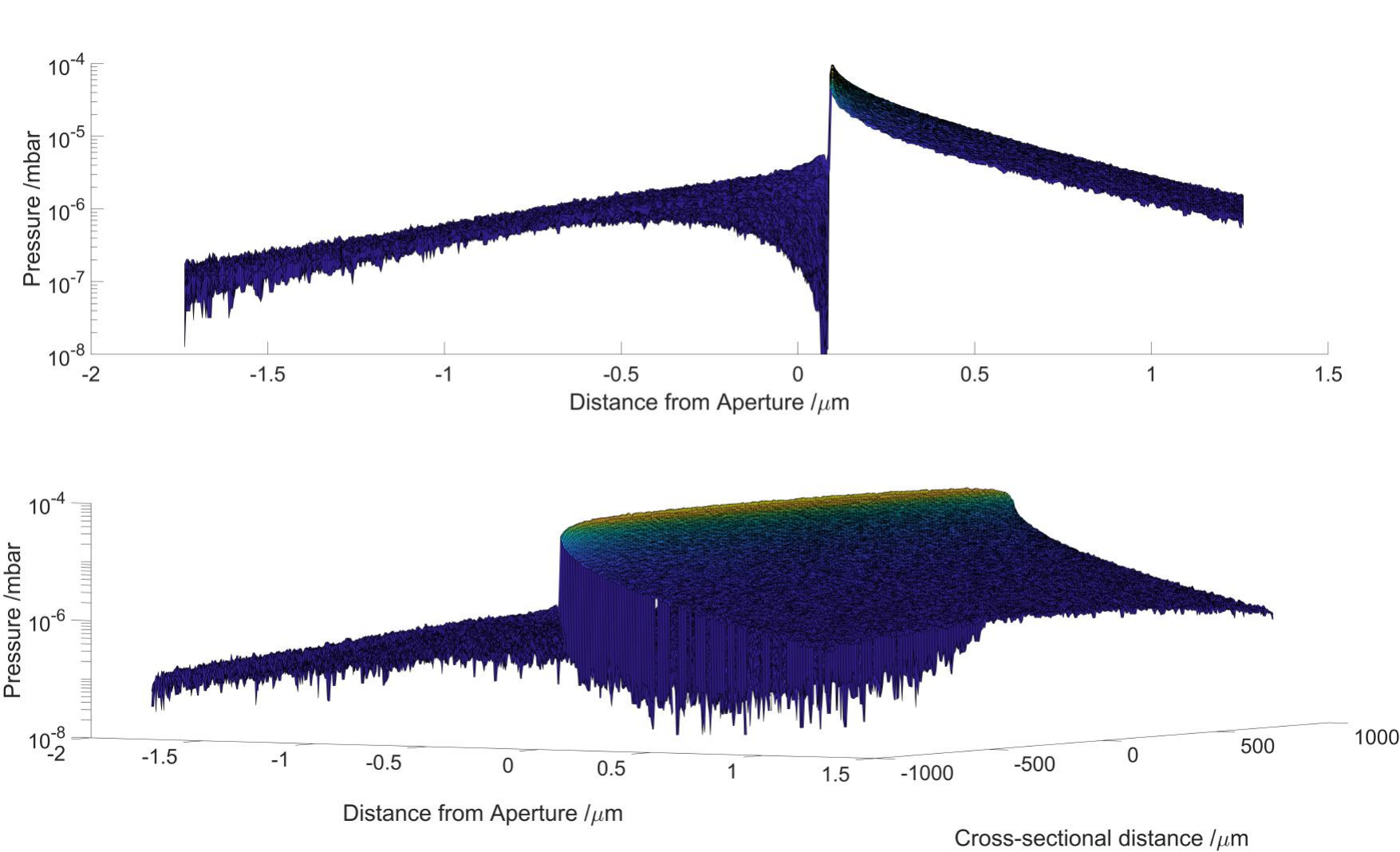}
    \caption{Surface plots showing the pressure profile down the centre of the normal incidence plate source cone, starting from within the sample chamber and ending in the pinhole plate, for the blocked aperture/no hole scenario where none of the beam is transmitted to the sample chamber. The large pressure differential between the sample chamber and the source cone, visible at the membrane itself as noisy data dropping into the $\num{e-8}\,mbar$ range, illustrates how the outgassing values for the two sides of a facet are not individually addressable, a limitation of Molflow.}
    \label{fig:molflow_no_hole}
\end{sidewaysfigure}

\begin{sidewaysfigure}
    \includegraphics[width=\textwidth]{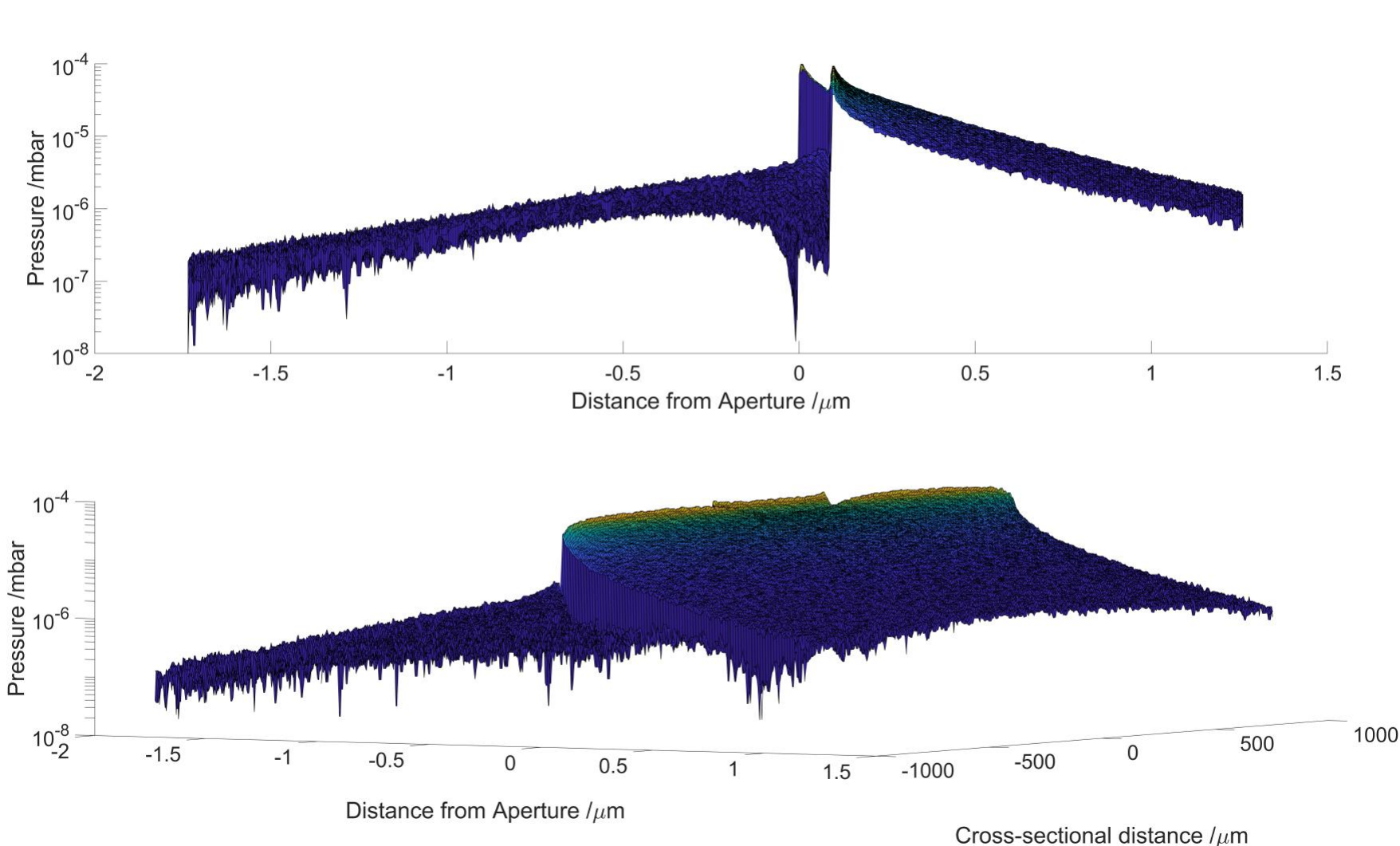}
    \caption{Surface plots showing the pressure profile for the $5\,\mu m$ aperture scenario. The recession in the aperture distance axis, visible in the side profile, is present because the wider membrane scaffold which is the size of the end of the source cone has a recess into which the membrane containing the aperture is mounted. The total outgassing, as calculated using equation \ref{eqn:phi_membrane}, was therefore assigned one part to the membrane scaffold like in Figures \ref{fig:molflow_no_hole}, \ref{fig:molflow_membrane}, the other part to the membrane itself.}
    \label{fig:molflow_5um}
\end{sidewaysfigure}

\begin{sidewaysfigure}
    \includegraphics[width=\textwidth]{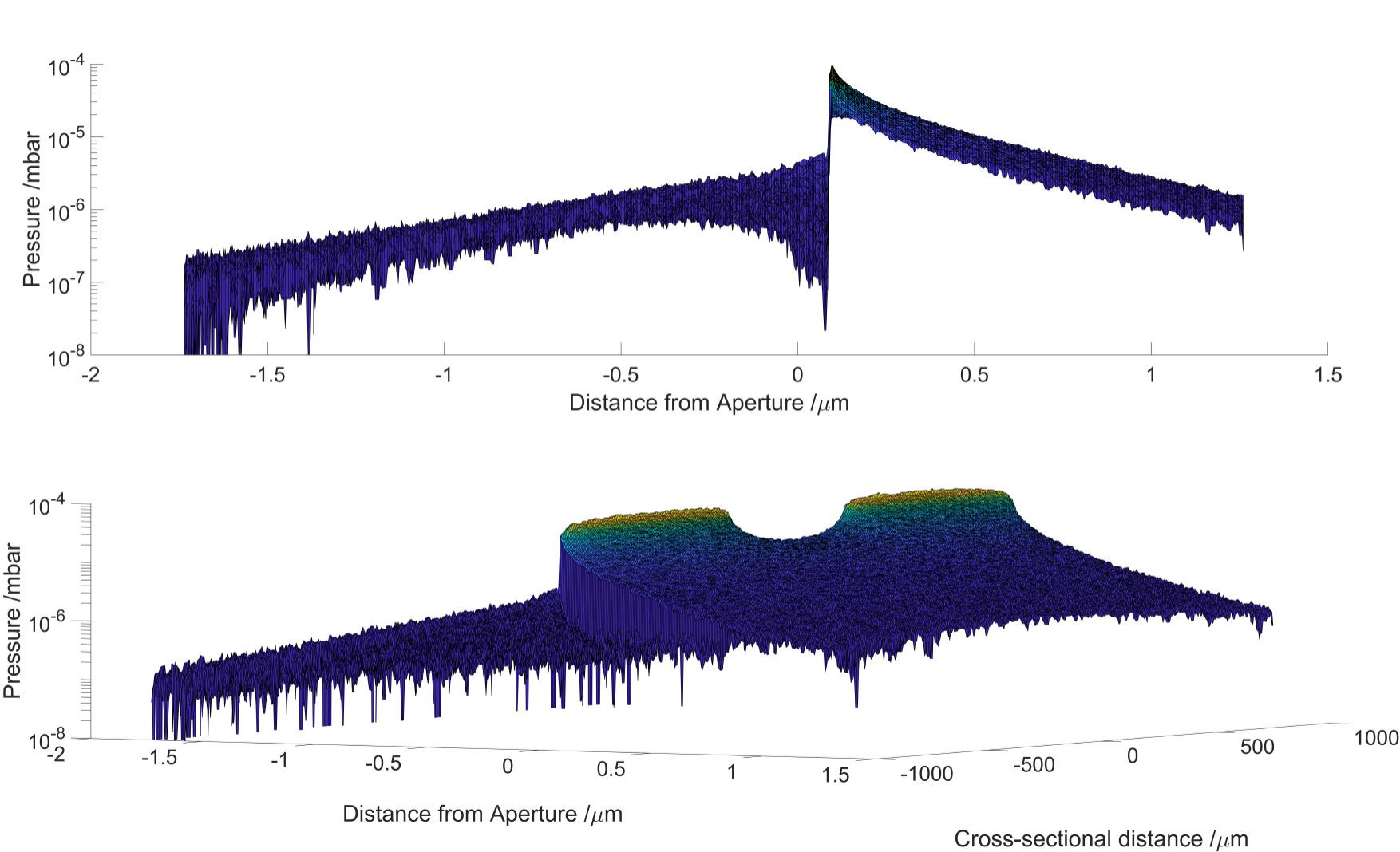}
    \caption{Surface plots showing the pressure profile for the blown out membrane scenario where the maximum differential pumping through the ``aperture'' into the sample chamber is possible. Note that imaging in this scenario was achieved.}
    \label{fig:molflow_membrane}
\end{sidewaysfigure}

\begin{flushleft}
The simulations were all run with the parameters shown in Appendix \ref{appendix:molflow}, in Figures  \ref{fig:molflow_a} and \ref{fig:molflow_b}, with only the geometry at the end of the source cone (blocked, $5\,\mu m$ pinhole and no membrane) differing. Extracting the pressure data collected on the transparent facet, location shown in Figure \ref{fig:molflow_setup}, and generating a surface plot for each data set gives Figures \ref{fig:molflow_no_hole}-\ref{fig:molflow_membrane}.
\end{flushleft}

\newpage
\begin{flushleft}
Comparing the maximum pressures across the width of the source cone, they are constant across both the width of the cone and across each scenario. In the blocked and $5\,\mu m$ aperture cases, there is a 5\% decrease in the pressure in front of the $5\,\mu m$ aperture compared to the completely blocked case, greatly exceeding the percentage difference in beam flux assigned. Hence, the difference must be due to differential pumping through the aperture.
\end{flushleft}

\begin{figure}[h]
    \centering
    \includegraphics[width=\textwidth]{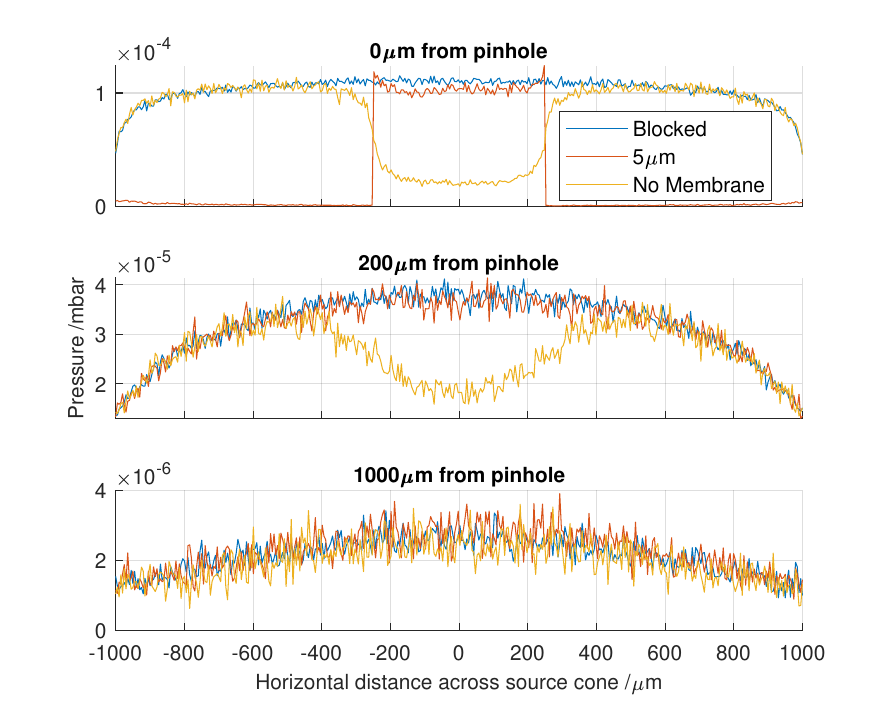}
    \caption{Cross-sections of the pressure profiles from Figures \ref{fig:molflow_no_hole}-\ref{fig:molflow_membrane}. The already small 5\% difference in pressure at the aperture/membrane itself quickly disappears only $200\,\mu m$ into the source cone. Pressure difference due to completely blown out membrane also becomes negligible by $1\,mm$. Note that signal-to-noise ratio increases proportional to distance from the outgassing facet at the aperture because sampling statistics become poorer.}
    \label{fig:molflow_cross_sections}
\end{figure}
\begin{flushleft}
However, the effect of differential pumping through the aperture is quickly dominated by the outgassing from the rest of the membrane, with indistinguishable cross-sections at $200\,\mu m$ from the aperture. Demonstrating that although differential pumping through the pinhole is present and not insignificant at the pinhole itself, it is negligible on the length scale of the entire source cone. Even considering the extreme example of a completely blown out membrane, a 15\% difference in area from the other two cases, the effect of differential pumping becomes negligible after 1mm as shown in Figure \ref{fig:molflow_cross_sections}.
\end{flushleft}
Although the pressure is high relative to the background pressure, both the simulated and real microscopes, at $\num{1e-4}\,mbar$, the incident flux on the end of the source cone is constant without major changes to the source configuration. Recalling that the maximum typical beam pressure of $100\,bar$ was used in the estimation of beam flux, the pressures at the end of the source cone are also maximised to give a worst-case scenario.
\\[12pt]
Having determined that the differential pumping through the aperture has only a 5\% effect on helium partial pressure at the pinhole and no effect after $200\mu m$, and that the absolute peak pressure at $\num{1.2e-4}\,mbar$ decreases to near-background pressures within $2\,mm$ of the aperture, one can conclude that extending the source cone to enable normal incidence will not introduce significant beam attenuation and impact beam propagation.

\section{Conclusion \& Further Work}
Least-squares fitting has been used successfully to calibrate the centre of rotation for an arbitrary sample to enable high-accuracy point tracking in real-space. Currently, a mean point delta of $17\,\mu m$ has been achieved at image resolution of $10\,\mu m$. Furthermore, a novel imaging technique for SHeM has been developed in extracting the tilt of a surface from point tracking calibration, with the aid of an absolute measurement of feature size on the sample.
\\[12pt]
Additionally, the method has proven useful in experiments measuring spatially resolved spot diffraction of lithium fluoride where the resulting pseudo-diffraction pattern appears to be affected by the tilt of the surface, hiding the specular condition.
\\[12pt]
Finally, possibly the most ubiquitous usage of point tracking is in troubleshooting the experimental setup. Using the parameters of the elliptical fit, one can determine the alignment of the rotational axis relative to the scattering plane and therefore locate the source of any misalignment.
\\[12pt]
The current work has been applied to novel point diffraction measurements on lithium fluoride $\langle 111 \rangle$ \cite{vonJeinsen2021} with plans to replicate the measurements on 2D materials with industrial/technological applications. Candidate samples include water ice, graphene and transition metal dichalcogenides (TMDs). As we move to better spatial and angular resolutions to enable such measurements, pinhole plate geometry becomes increasingly complex. The current “simple” pinhole plates have already taken traditional machining techniques to their limit with regards to feature size and complexity, so novel manufacturing techniques must be explored.
\\[12pt]
There is further experimental work planned to develop the point tracking method by fabricating a multi-faceted sample with surfaces with well-defined tilts which will be used to determine the accuracy of tilt calculations and the effect of image resolution on mean point deltas. Additionally, known tilts allow us to work out the component of the tilt which is due to factors outside of the sample itself like mounting.
\\[12pt]

\begin{figure}[h]
    \centering
    \begin{subfigure}{0.4\textwidth}
    \includegraphics[width=\textwidth]{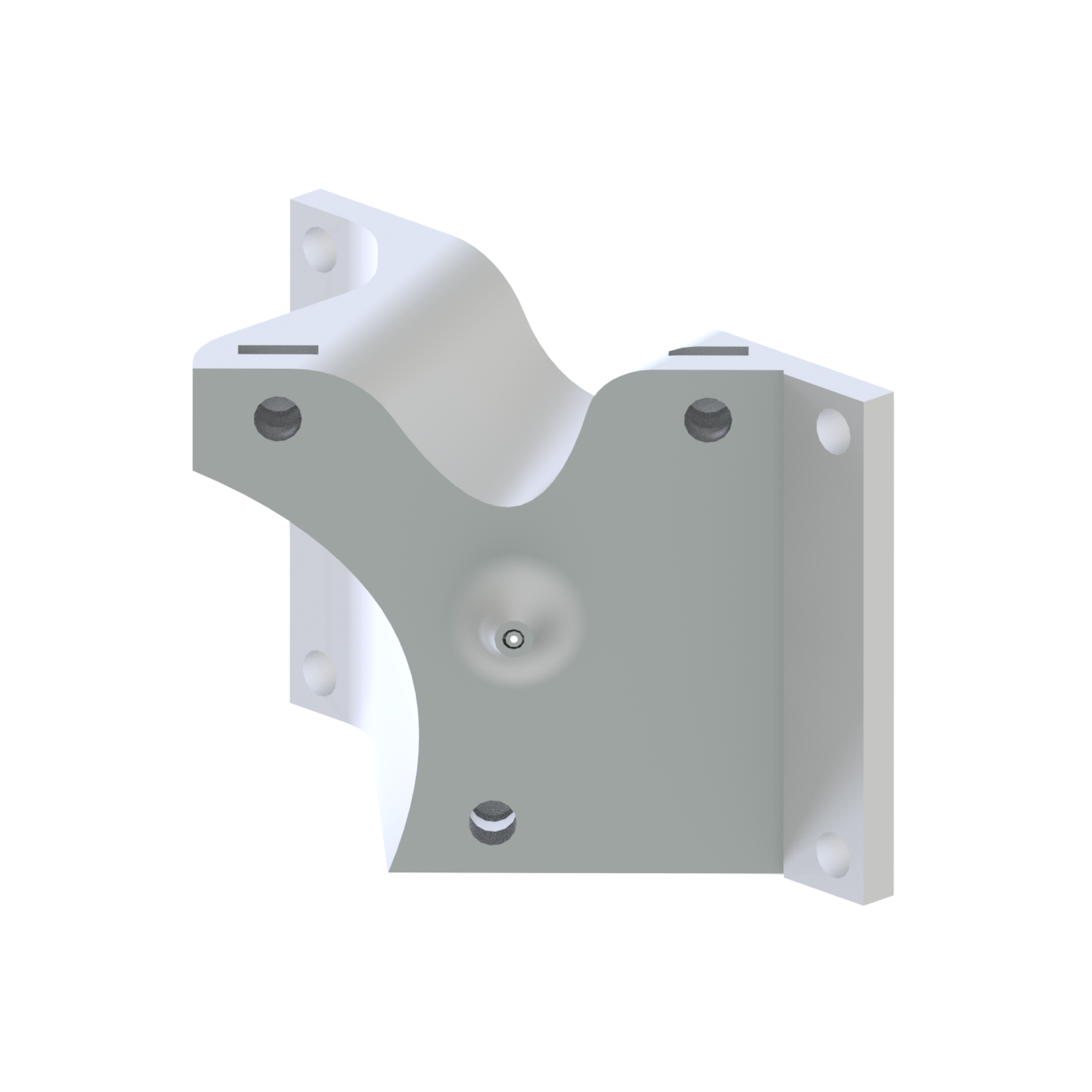}
    \subcaption[a]{}
    \label{fig:annular_plate_a}
    \end{subfigure}
    \begin{subfigure}{0.4\textwidth}
    \includegraphics[width=\textwidth]{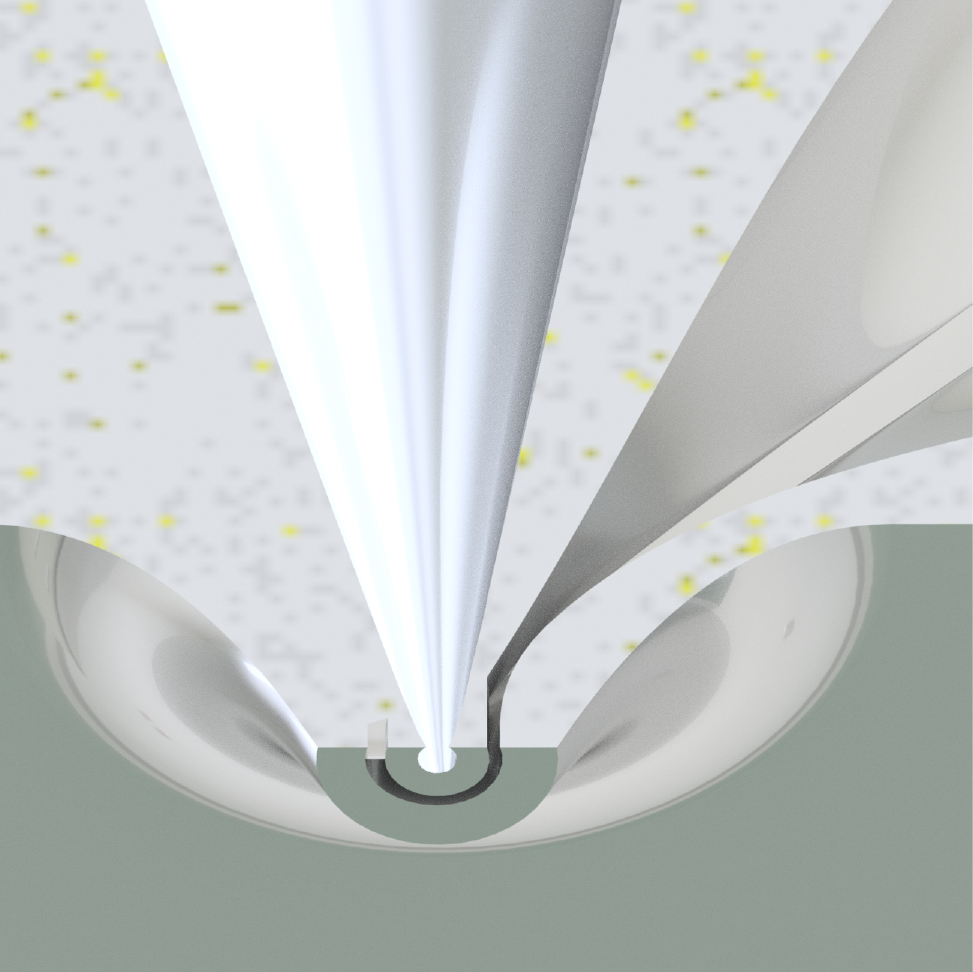}
    \subcaption[b]{}
    \label{fig:annular_plate_b}
    \end{subfigure}
    
    \caption{Panel \ref{fig:annular_plate_a} shows a normal incidence pinhole plate design with annular detector aperture completely surrounding the source aperture. Panel \ref{fig:annular_plate_b} shows a close-up half section view of the complex internal structure where the source cone is enveloped by the detector cone.}
    \label{fig:annular_plate}
\end{figure}

Additionally, measurement of facet angles on polycrystalline technological, biological and synthetic samples with micron-scale domain sizes is now possible using SHeM. As further improvements are made to SHeM resolution, having currently achieved a FWHM of $350\,nm$, the minimum size of domain will only decrease. Such resolution improvements would be an important step in establishing SHeM as a delicate alternative facet angle measurement tool to traditional XRD rocking curves which can be damaging to samples or alter their chemical composition \cite{Stuckelberger2020}\cite{Zheng2003}.
\\[12pt]
The previously mentioned interference fit sample mount has been manufactured and will be used in any further tilt-sensitive measurements.
\\[12pt]
Upon completion of the further work, the much larger data set from both a fabricated and “natural” sample will allow for the further statistical analysis to yield meaningful insight into the nature of the error, allowing for further optimisation of the technique.
\\[12pt]
A continuation of the present work using Molflow would be to repeat the same three scenarios with a \ang{45} incidence pinhole plate for completeness for a direct comparison. Additionally, Molflow can be used to probe the pumping speeds in pinhole plates with extremely complex internal structures on both the source and detector cone sides. An immediate candidate is  detector pinhole plates, which are yet to be fabricated, because the detector cone must split and surround the source aperture with a detector aperture, introducing a difference in pumping speeds, and therefore signal (helium flux into detector), if the pumping speeds are not equal all around the detector aperture. Figure \ref{fig:annular_plate} shows a design for a normal incidence annular pinhole plate.


\chapter{3D Printed Polymer Pinhole Plates}
\label{section:polymer_plates}
\section{3D Printed Plastics in Vacuum}
Traditional machining techniques are already a limitation in \ang{45} incidence plates because relatively long and narrow cones need to be cut in aluminium, tapering down from approximately $30\,mm$ to $2\, mm$ diameter, at $50\,mm$ deep. Usually with thin walls between the source and detector cones and accurately positioned apertures and recesses on the front of the pinhole plate. Furthermore, the back side of the plate has o-ring grooves which must have a smooth surface to form a seal with the wall of the sample chamber, preventing leakage of the beam into the detector. The additional complication of a longer source cone required for \ang{90} incidence plates for 3D imaging makes the process more complex using traditional tooling as smaller tools operated at more challenging attack angles are needed. 
\\[12pt]
3D printing has been considered as a possible fabrication method which can bypass limitations of traditional machining techniques and allows for the presence of internal structure in the part, if they can be made with the same precision and finesse as machined parts. Complex internal structures are a particular problem in pinhole plates for multiple detector 3D imaging in \textbeta-SHeM which has twisting source and detector cones. Sintered metal printing methods can be used and have the potential for vacuum compatibility, only depending on ultimate surface finish and porosity \cite{Zwicker2015}\cite{Chaneliere2017}\cite{Povilus2014}. The option explored here is SLA printing in plastic because it is ubiquitous, cheap and has very short lead times whilst maintaining excellent feature size with minimum layer thickness of $25\,\mu m$ using a FormLabs Form 3 printer and FormLabs “Clear Resin” \cite{clear_data}\cite{plastic1}.
\begin{figure}
    \centering
    
    \begin{subfigure}{0.7\textwidth}
    \includegraphics[width=\textwidth]{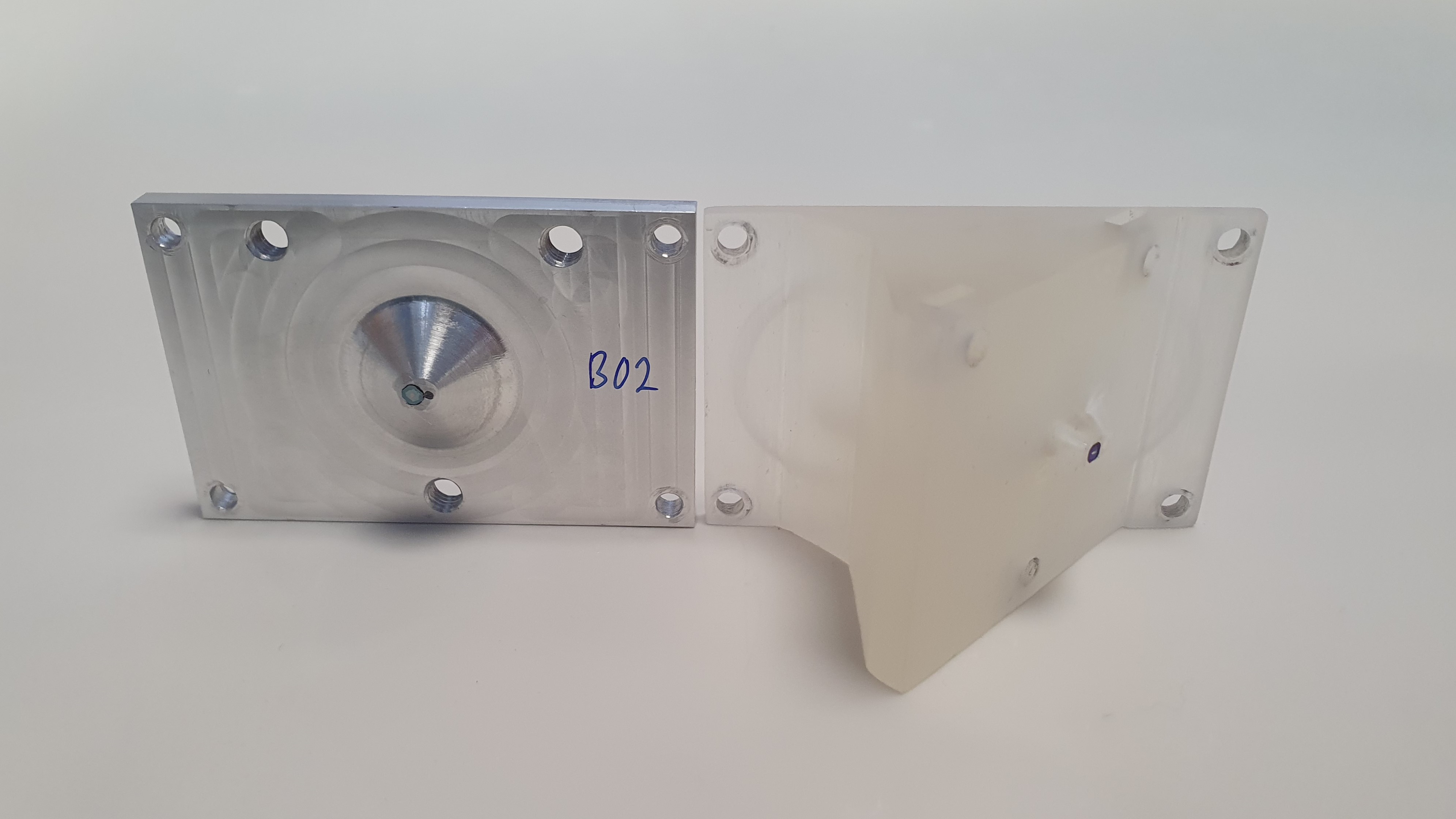}
    \end{subfigure}
    
    \begin{subfigure}{0.7\textwidth}
    \includegraphics[width=\textwidth]{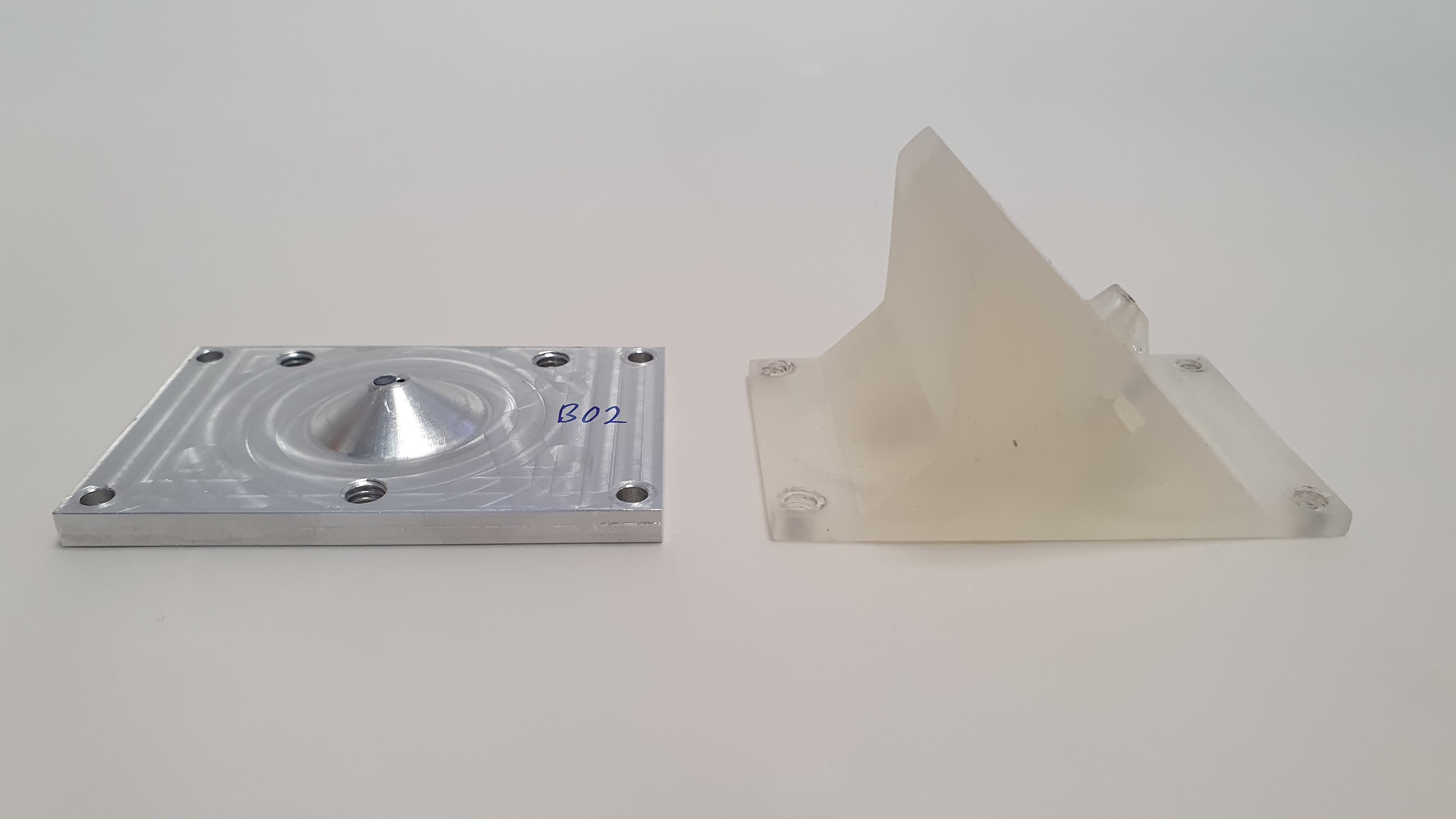}
    \end{subfigure}
    
    \begin{subfigure}{0.7\textwidth}
    \includegraphics[width=\textwidth]{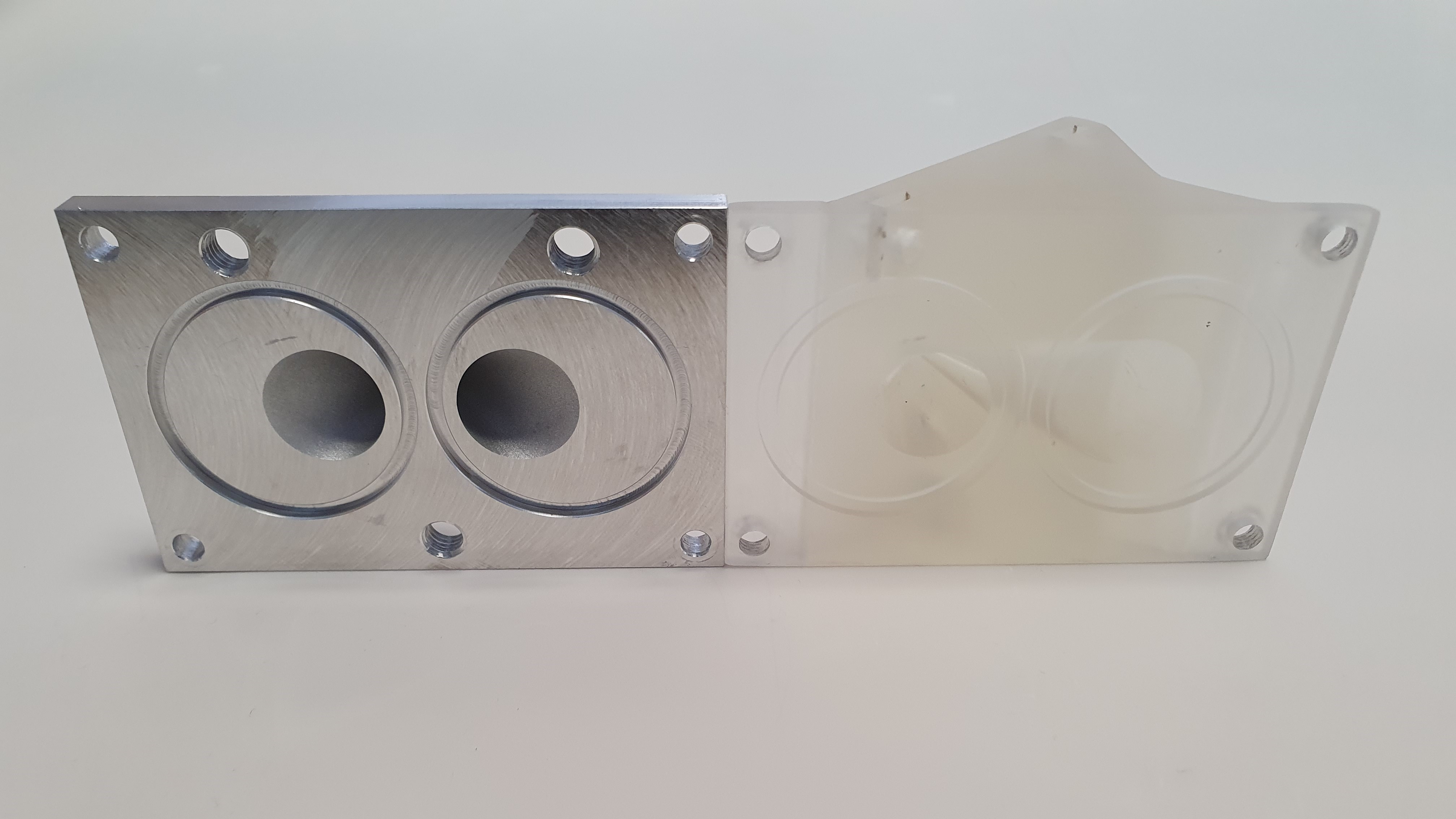}
    \end{subfigure}
    
    \caption{Comparative images between a \ang{45} incidence pinhole plate fabricated using traditional machining techniques and a \ang{90} incidence pinhole plate printed in Clear Resin using SLA. The source cone length differs by approximately a factor of 5 between the two plates, with the normal incidence plate also having captive nuts.}
    \label{fig:4590_comparison}
\end{figure}
Upon ordering the normal incidence plate pictured in Figure \ref{fig:4590_comparison}, whose pumping properties were previously tested in Molflow in Chapter \ref{section:enhancements_heliometric}, the dimensions of the component were tested to determine the fabrication accuracy of Formlabs SLA printing. A visual inspection of the sealing surfaces found some plastic dust, which was easily cleaned with compressed air, presumed to be residue from the manufacturing process. The small and complex features appeared to be as specified along with a smooth and flat back side in which the o-rings lie. The actual dimensions of the part were all found to be within the quoted $\pm0.15\,mm$ tolerance, significantly less than $\pm1\%$. Note that the pinhole itself is not printed or machined, but milled into a silicon nitride membrane using an ion beam, and the entire membrane is then glued to the pinhole plate.
\\[12pt]
A concern prior to printing was structural integrity because in the Cambridge \textalpha-SHeM the entire sample manipulator mounts onto the pinhole plate as shown in Figure \ref{fig:scattering_plane_stages}. However, the material properties datasheet \cite{plastic1} states that the clear resin has a tensile strength of $65\,\text{MPa}$ and Youngs' Modulus of $2.8\,\text{MPa}$, and proved to be completely rigid once installed in the microscope. For reference, the clear resin's tensile strength and Youngs' Modulus values are of a similar magnitude to the extremely strong PEEK which has approximately $98\,\text{MPa}$ and $4.0\,\text{MPa}$, respectively \cite{VICTREX2019}.
\\[12pt]
The most significant issue with using plastics in vacuum is that their vacuum properties are generally accepted as being very poor due to degassing. Installing the plastic normal incidence plate as delivered raised the base pressure of \textbeta-SHeM from $\num{2e-8} \,mbar$ to $\num{4e-7} \,mbar$ after comparable pump down times. Typical operating pressures in the \textalpha-SHeM sample chamber are around $\num{5e-8}\,mbar$, with the detector, which is separated by a differential pumping stage, operating optimally at around $\num{3e-11}\,mbar$. Installing the plastic normal incidence plate into \textalpha-SHeM raised the pressures of the sample and detector chambers to $\num{5e-7}\,mbar$ and $\num{3e-10}\,mbar$, respectively. The increased pressure in the detector resulted in huge instability and an inability to form images.  Mass spectra ranging from mass $1-300$ using a Hiden “HAL/3F RC301 PIC300” quadrupole mass spectrometer revealed large peaks associated with water using the integrated mass peak identification in Hiden’s “MASsoft10 Professional”. The spectra showed only water as a major vacuum contaminant, with no evidence of hydrocarbons evolving from the plastic itself, suggesting that vacuum baking the plastic components could be a viable solution to the poor vacuum properties if the structural integrity and fidelity of fine features is preserved.
\\[12pt]
The normal incidence plate was heated to \ang{120}C over 3 hours in a vacuum oven and left to bake for 48 hours after which the heating filament was turned off and the plate left to cool in vacuum over 2 hours. The pressure in the oven was not logged, however an approximate increase of 3 decades was observed during heating to maximum temperature, with the final pressure being a decade lower than the initial.
\\[12pt]
Placing the baked plate immediately back into \textbeta-SHeM yielded the normal base pressure of $\num{2e-8}\, mbar$ within 24 hours pumping down. Further mass spectra were taken which only showed the presence of atmospheric contaminants using Hiden’s MASsoft10 peak identification software, validating the idea that the unbaked plastics predominantly degas water. 
\\[12pt]
Upon re-installing the plastic normal incidence plate into the microscope after a few days exposure to atmospheric conditions in a pinhole plate storage box, the pump down speeds and chamber base pressure were significantly poorer than when the plate had been freshly baked.  Such a decline in vacuum performance suggests a re-wetting process occurs when exposed to the atmosphere for prolonged periods.

\section{Vacuum Compatibility of SLA 3D Printed Plastics for Fast Prototyping}
\begin{figure}
    \centering
    \begin{subfigure}{0.45\textwidth}
    \includegraphics[width=\textwidth]{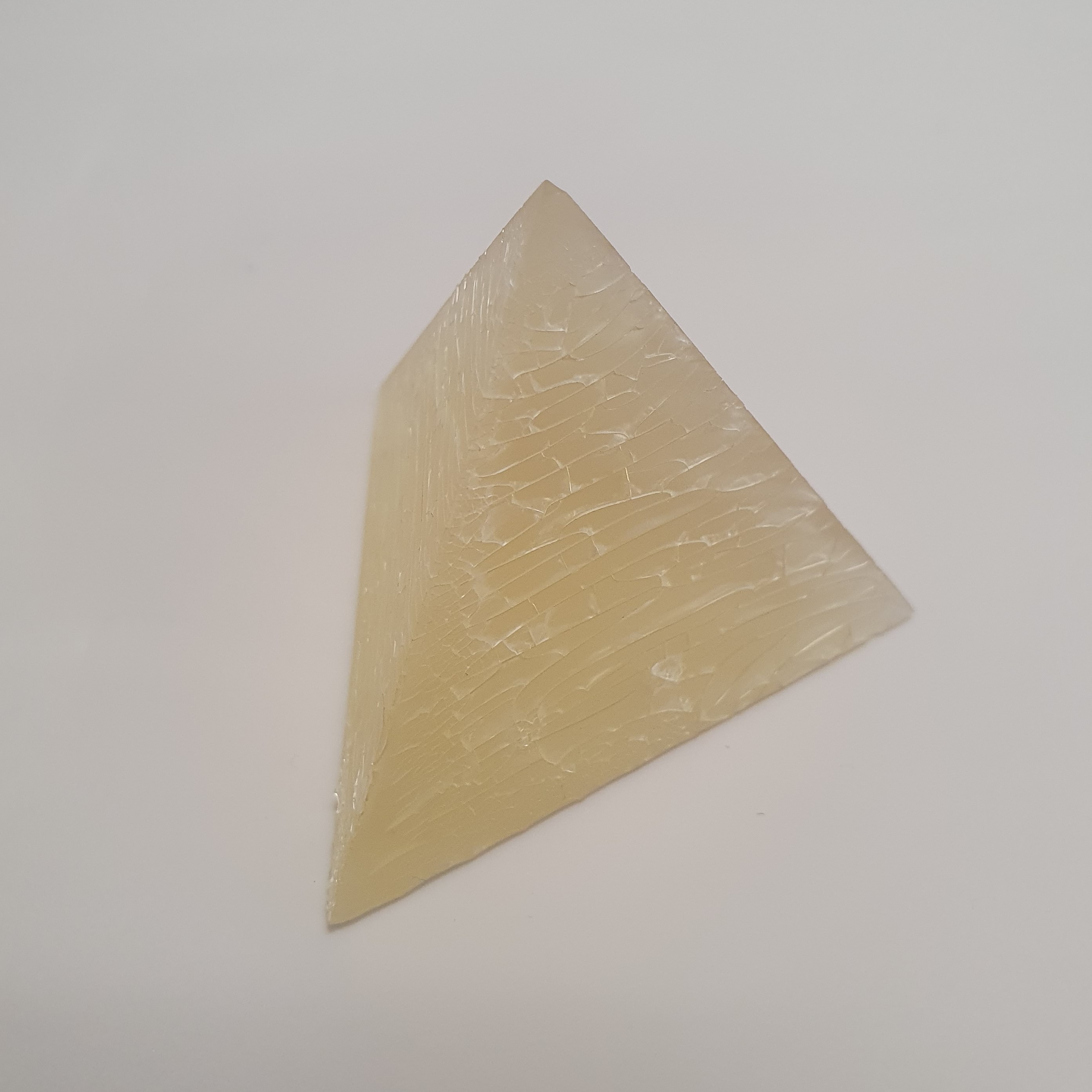}
    \subcaption[a]{}
    \label{fig:delamination_a}
    \end{subfigure}
    \begin{subfigure}{0.45\textwidth}
    \includegraphics[width=\textwidth]{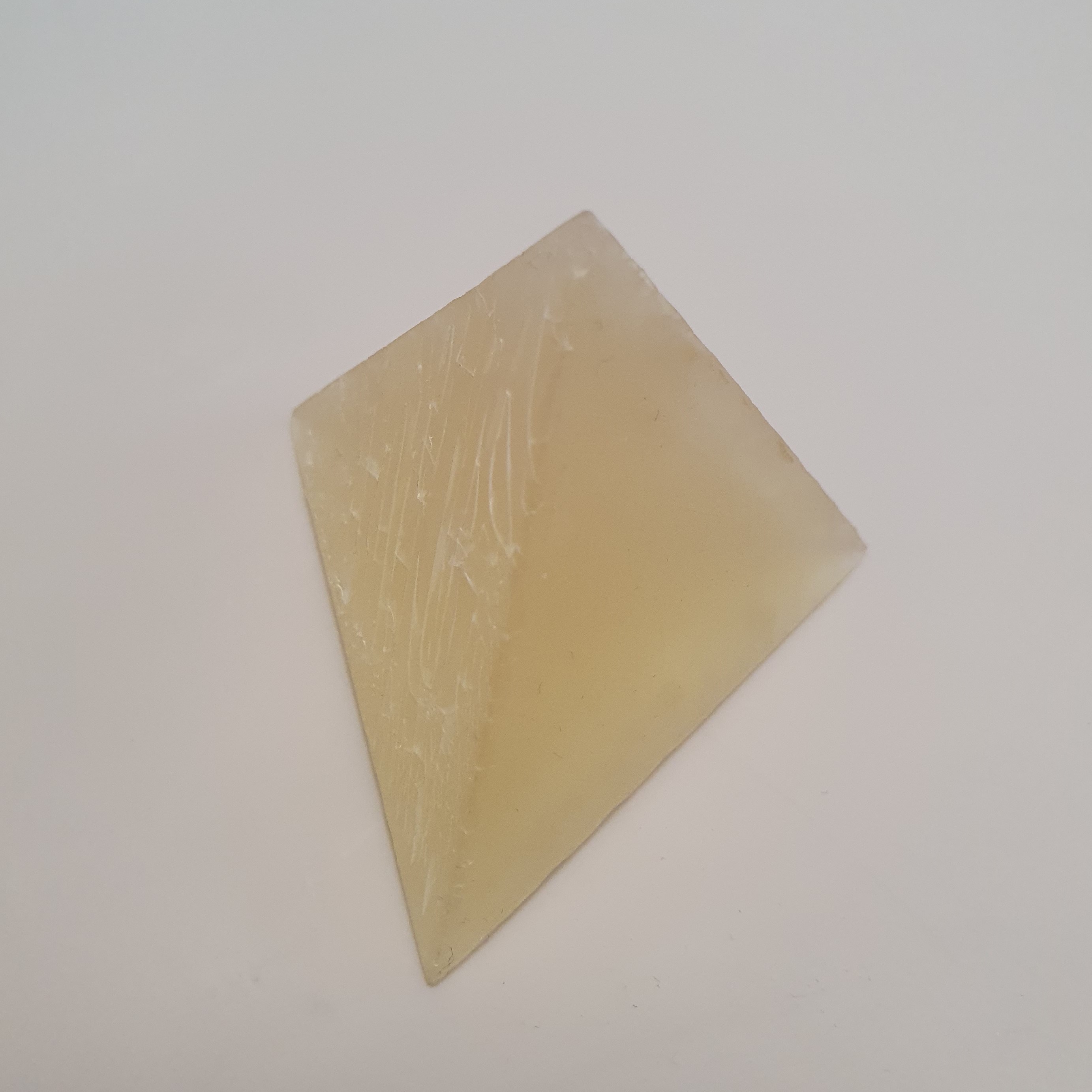}
    \subcaption[b]{}
    \label{fig:delamination_b}
    \end{subfigure}
    \caption{Images show the surface delamination of the sample that was baked at \ang{170}C to achieve UHV pressures. The side of the tetrahedron that was unaffected, shown in Panel \ref{fig:delamination_b}, was in full contact with a metal shelf in the chamber.}
    \label{fig:delamination}
\end{figure}

An investigation into the re-wetting characteristics of the SLA printed “Clear Resin” was deemed critical if 3D printed plastics were to take over as the primary material and manufacturing process for pinhole plates and miscellaneous small vacuum components. Pertaining to pinhole plates especially, the typical procedure when using a new pinhole plate involves extended periods of exposure to atmosphere during aperture mounting and prior to installation into the microscope. Additionally, pinhole plates are currently also exposed to atmosphere between installations in the microscope.  Knowledge of the rate of re-wetting is essential in the storage of pinhole plates, and more widely in characterising the vacuum compatibility of SLA “Clear Resin” for general vacuum usage.
\\[12pt]
The experiment designed to investigate the rate and nature of re-wetting in SLA Clear Resin was as follows:
\begin{enumerate}
\setlength\itemsep{-0.5em}
    \item Design and fabricate identical standard, non-SHeM specific samples
    \item Clean with IPA upon arrival
    \item Record masses and dimensions of samples
    \item Bake samples together for 48 hours at \ang{120}C, increment by \ang{20}C every half hour until maximum temperature is reached. Allow to cool slowly by turning off heating filament and allowing for natural cooling.
    \item Immediately visually inspect and measure the first tetrahedron, placing into \textbeta-SHeM sample chamber and evacuating as quickly as possible. Visual inspection and measurement of remaining samples, store in an opaque storage container with temperature and humidity monitoring.
    \item Planned atmospheric exposure times are 0, 1, 2, 4, 7, 14 days, nomenclature for samples is therefore S0, S1, S2 etc.
    \item Log \textbeta-SHeM sample chamber pressure
    \item Collect profiling scan mass spectra using aforementioned Hiden quadrupole as late as possible before chamber is vented and next sample is inserted. Mass spectra then used to find the partial pressure of water degassing from each sample by integrating the spectrum to the total pressure.
    \item Repeat for all samples.
\end{enumerate}
Since the investigation presented was conducted, the above preparation and baking procedure has been used tens of times and implemented as the standard operating procedure for using SLA printed components in SHeM. 
\\[12pt]
A tetrahedron with $(4.80\pm0.05)\,cm$ edges was chosen as the standard sample to maximise surface area to volume ratio with a regular polyhedron. The exact shape of the sample is unimportant, only that the samples are identical. The samples were SLA printed using Clear Resin with no post-printing curing/surface treatment procedures, as with the normal incidence plate.
\\[12pt]
The samples were washed in IPA upon arrival to clean them of contaminants they picked up during manufacturing and shipping so that both the microscope vacuum chamber and mass spectra are not affected unexpectedly by contamination.
The mass of each sample was measured 5 times and the mean recorded as $(17.27\pm0.05)\,g$. The samples were weighed again post-bake with a mean mass of $(17.22\pm0.05)\,g$, changing by $-(0.29\pm0.005)\%$. Dimensions of all samples were within stated manufacturing tolerances of $\pm0.15$mm before and after baking, with no change outside of error bounds of calipers. No colour change or distortion was noted either.
\\[12pt]
The exposure times of the samples was designed such that they are more tightly grouped at small exposure times, as one would expect that the rate of re-wetting is fastest initially and slows with time as the material saturates. Additionally, \textless 1 week is the most common length of time pinhole plates are left in atmosphere during aperture mounting or between measurements, therefore that period is most valuable to pinhole plates specifically. Ideally, one would have even more dense grouping between 0 and 2 days to further inform aperture mounting procedures and pinhole plate transfer time-windows between vacuum environments, but \textbeta-SHeM takes approximately 24 hours to reach base pressure with an empty chamber so inserting samples for shorter times would not yield valuable results. One could bake the samples separately so that <1 day exposures can be investigated, but the decision was made to bake them all at the same time for consistency in sample preparation.
\\[12pt]
The container the samples were stored in maintained an average temperature of \ang{21.5}C and a relative humidity of 32.7\% with standard deviations of \ang{0.6}C and 1.1\%, respectively, over the two week storage period. Both values measured using a Pico Technology Humidiprobe \cite{JB2005}.
\\[12pt]
All analysis of vacuum properties was done in the \textbeta-SHeM sample chamber ($20\,l$ volume with a nominal $450\,ls^{-1}$ turbo pump) with the aforementioned Hiden quadrupole mass spectrometer used for residual gas analysis. It should be noted that the spectrometer's filament was left turned on from a few days prior to the first sample being placed in the chamber to allow the filament to degas. The gate valve between the detector and sample chamber was closed to allow the filament to stay on during sample transfers to maintain consistent detector sensitivity. It should be noted that samples were left to pump down for as long as possible before taking mass spectra because that allows us to probe the composition of the vacuum at comparable total pressures which are approximately the test chamber's base pressure.
\\[12pt]
Pump down curves for each sample were recorded and used to demonstrate the re-wetting effect of exposing the samples to air. One can plot the pumping curves directly for comparison, but a more valuable quantitative metric is the degassing values as a function of time for each sample.
\begin{figure}
    \centering
    \includegraphics[width=0.9\textwidth]{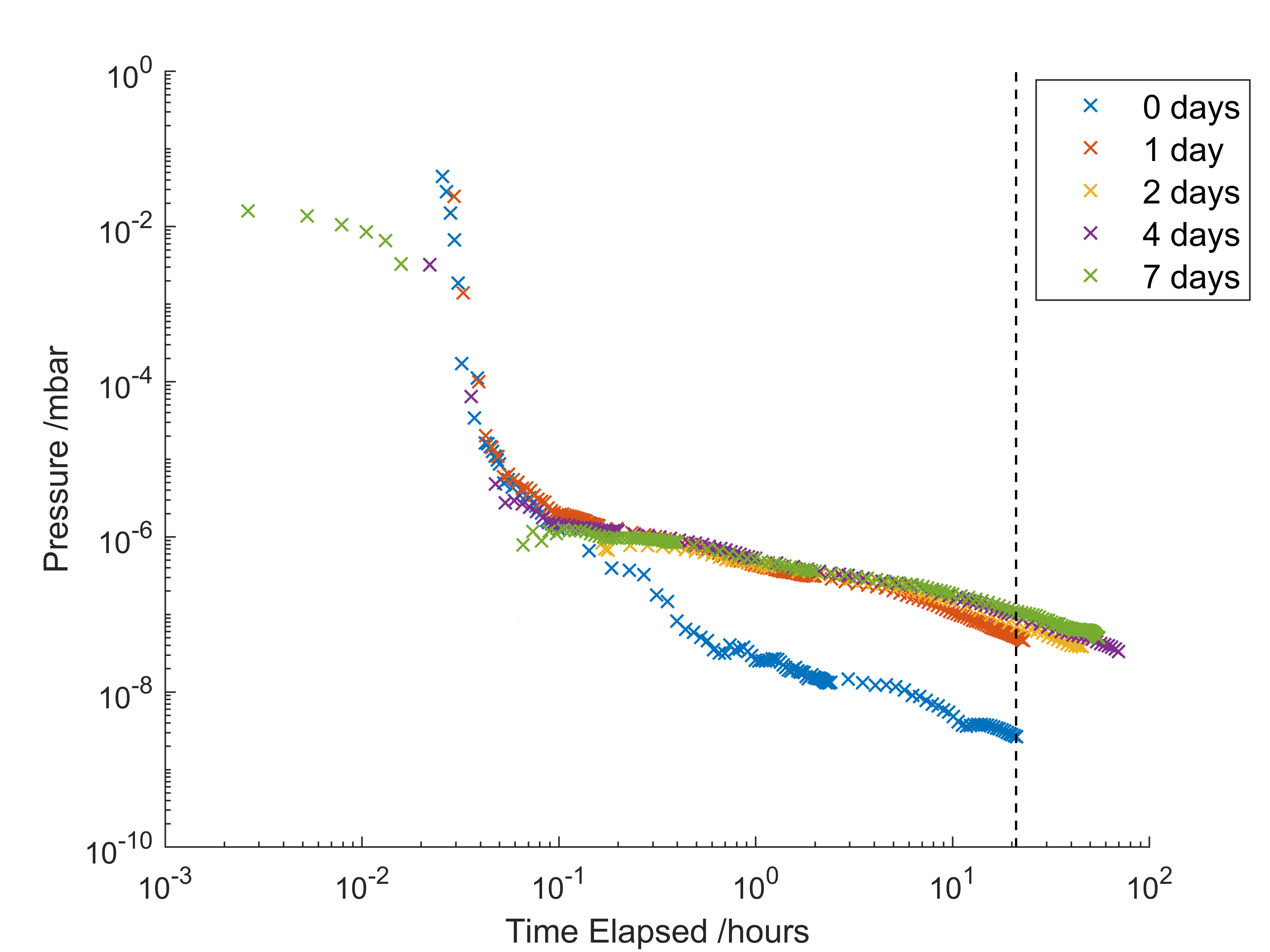}
    \caption{Degassing curves for samples exposed to air for lengths of time between 0-7 days. An empty chamber pumping curve was subtracted from the samples' pumping curves to estimate the total degassing per sample. Vertical slice (dotted line) through data is plotted in Figure \ref{fig:degassing_slice}. The 14 day data has been omitted due to significant database errors whilst logging pump down curves.}
    \label{fig:degassing_plot}
\end{figure}
Figure \ref{fig:degassing_plot} shows a strong correlation between exposure time and re-wetting of the samples, with most of the re-wetting occurring in the first 24 hours of exposure. Taking a vertical slice through the figure at $t\approx 21$ hours gives Figure \ref{fig:degassing_slice}.
\begin{figure}
    \centering
    \includegraphics[width=0.9\textwidth]{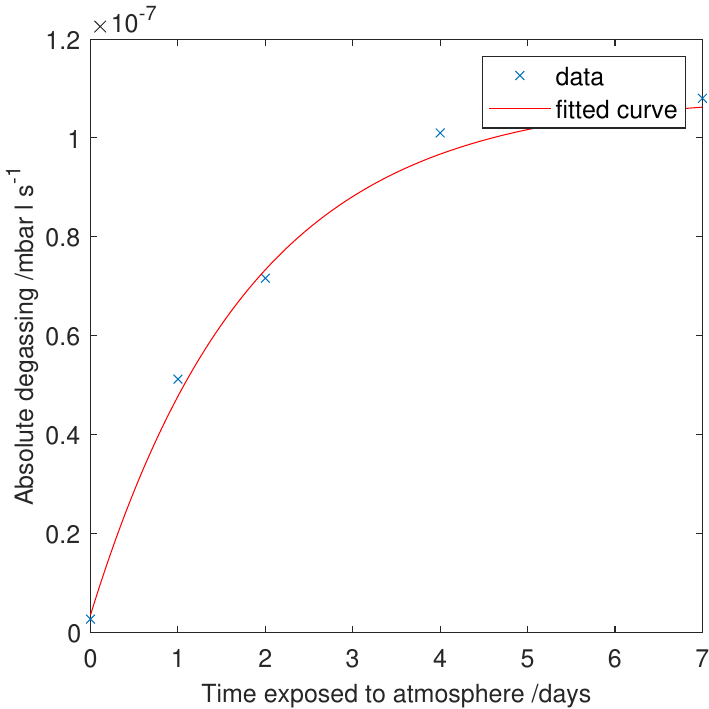}
    \caption{Plot showing a vertical slice through Figure \ref{fig:degassing_plot} at $t = 21\,hours$. A single exponential models the re-wetting process with high agreement ($R^2 = 0.9934$).}
    \label{fig:degassing_slice}
\end{figure}
Figure \ref{fig:degassing_slice} shows that the re-wetting process is accurately described by a single exponential, and that almost half the total re-wetting occurs within the first 24 hours of exposure, making a matter of hours the critical time for exposure to achieve high-vacuum in a time comparable to the empty chamber, or the usual metal pinhole plates.
\newpage
Mass spectra were also taken after as long a pump down time as possible to probe the “base” pressure for each sample, within the limitations of \textbeta-SHeM’s pumping speed. Taking spectra at the lowest possible pressure for each sample allows for the comparison of the chemical composition of the vacuum contaminants at the base pressure for each sample by normalising the spectra to the respective total absolute pressure by integrating the counts over the entire spectrum and thereby converting from counts per species to partial pressure per species. Achieving a stable ``base'' pressure is also important so that the total chamber pressure does not change significantly during the course of a set of mass scans.
\begin{figure}
    \centering
    \includegraphics[width=\textwidth]{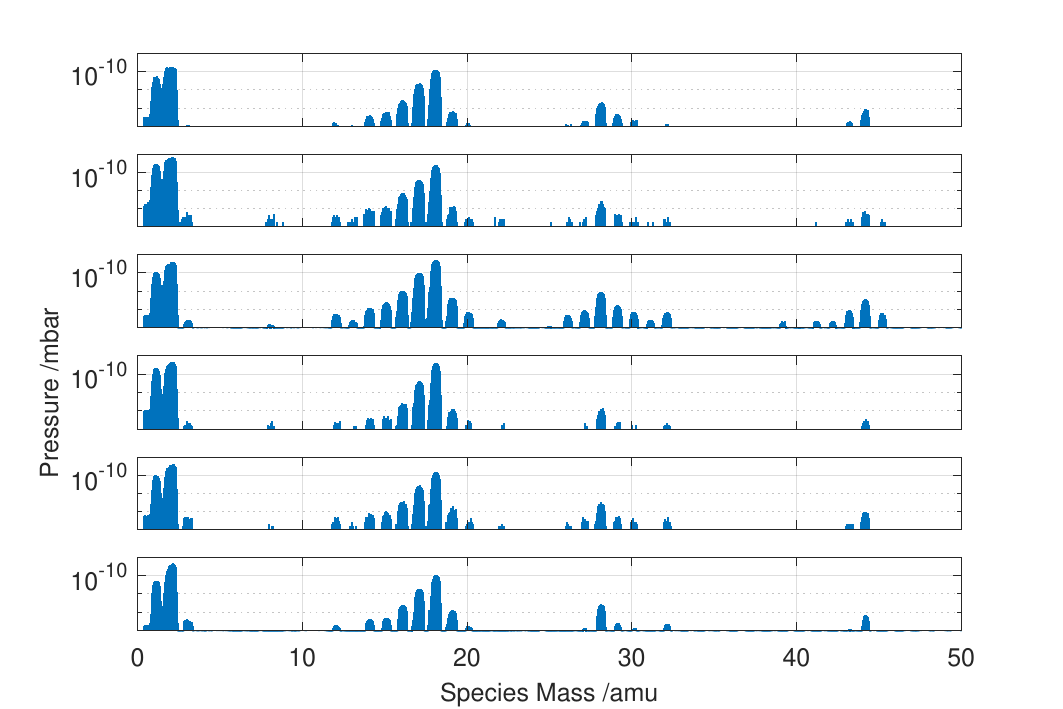}
    \caption{Normalised mass spectra of samples exposed to atmosphere for 0, 1, 2, 4, 7 and 14 days (listed top to bottom) showing that the vacuum properties of all baked samples are consistent provided pump down times approximately equal to the time the samples were exposed to the atmosphere.}
    \label{fig:stacked_rga_tetra}
\end{figure}
\newpage
\begin{flushleft}
The similarity between all normalised mass spectra in Figure \ref{fig:stacked_rga_tetra} indicates that, although the pump down speed slows as exposure time increases, re-wetting is exclusively a surface process on the time scales investigated. Baking effectively removes water from the bulk of the material and prevents water from reabsorbing into it. Therefore, although the majority of re-wetting occurs in the first week of exposure, water can only saturate the surface which functions as a finite volume of water. In contrast to the unbaked plastic whose bulk water content acts as an infinite volume of water which cannot be pumped out in a practical length of time, hence the base pressure of only $\num{4e-7}\,mbar$ achieved in the first instance after almost 2 weeks pumping down.
\end{flushleft}
\begin{figure}[h]
    \centering
    \includegraphics[width=0.9\textwidth]{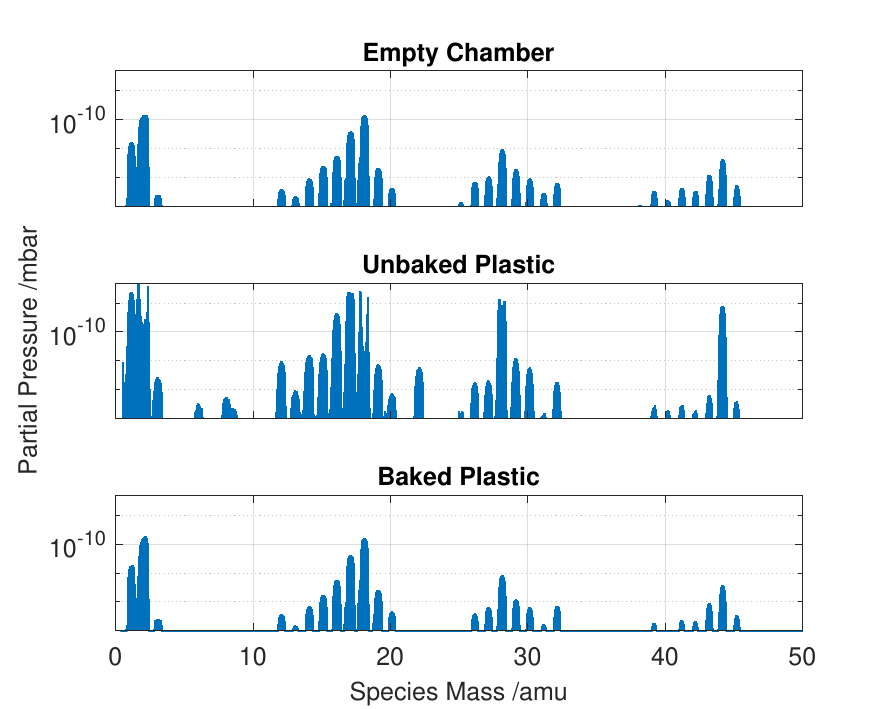}
    \caption{Normalised mass spectra for an empty chamber (top), unbaked sample (middle) and baked sample with no exposure time (bottom). The spectra for empty chamber and baked sample are near identical, both at the base pressure of $\num{1.9e-8}\,mbar$. Note: the electron multiplier likely saturated during the unbaked measurement, taken at $\num{4e-7}\,mbar$, resulting in artifacts in the 2, 18 and 28 amu peaks.}
    \label{fig:stacked_rga_plates}
\end{figure}
\newpage
Figure \ref{fig:stacked_rga_plates} shows again that water is the primary contaminant degassing from the plastic samples, and validates that the vacuum properties of a baked plate are indistinguishable from the empty chamber down to high-vacuum at $\num{1.9e-8}\,mbar$.
\\[12pt]
A further experiment was conducted to evaluate the ultrahigh-vacuum compatibility of SLA Clear Resin wherein the entire vacuum oven was baked during the sample baking process to give a true UHV platform to find the true base pressure of the baked plastic. The same baking procedure was followed except the maximum temperature was increased to \ang{170}C because the metal of the oven must be baked too. An ultimate base pressure of $\num{9.9e-10}\, mbar$ was achieved, validating the UHV compatibility of SLA Clear Resin. It should be noted however that upon removal from the chamber, the surface of the sample appeared to have delaminated with thin surface cracks appearing, shown in Figure \ref{fig:delamination}.
\newpage
\section{Conclusion \& Further Work}
FormLabs ``Clear Resin” printed using SLA technology has been shown to be an excellent material and printing method combination for cheap, rapid prototyping of small vacuum components. Out of the available FormLabs materials, clear resin has the ideal material properties based on published data sheets. It has been determined that the primary vacuum contaminant from the plastic to be water degassing, using a simple baking protocol one can achieve ultimate pressures in the low-HV to UHV range.
\\[12pt]
Additionally, there is strong evidence to support that re-wetting under atmospheric conditions, for exposures up to two weeks, only allows re-absorption of water into the surface layers of the plastic. Thus, the benefits to the vacuum properties of the plastic provided by baking are determined to be semi-permanent, with further investigation needed to probe longer timescales and quantify the re-absorption rate of water into the plastic’s bulk.
\\[12pt]
Further investigation is needed to determine whether the delamination during the \ang{170}C bake was caused by the increased maximum temperature or heating/cooling rate, and whether any curing process offered by the manufacturer could prevent it.
\\[12pt]
The base pressure of the unbaked \textbeta-SHeM sample chamber proved to be a limiting factor in determining the absolute degassing of water from the short exposure samples. If the experiment were to be repeated, it would be valuable to bake the chamber in parallel with the samples. A sample transfer unit can then be used so that the baked testing chamber isn't vented to atmosphere post-bake.


\newpage

\chapter{Experimental Realisation of Heliometric Stereo}
\label{section:exp_heliometric}
Combining the instrumentation work on point tracking and normal incidence pinhole plates presented in Chapter \ref{section:enhancements_heliometric} with the validation of the vacuum compatibility of SLA printed plastics in Chapter \ref{section:polymer_plates}, the current work moves on to acquire the first normal incidence SHeM images, and following that, the first experimental verification of the heliometric stereo method as detailed by Lambrick et al.

\section{Normal Incidence First Images}

\begin{figure}[h]
    \centering
    \includegraphics[width=0.65\textwidth]{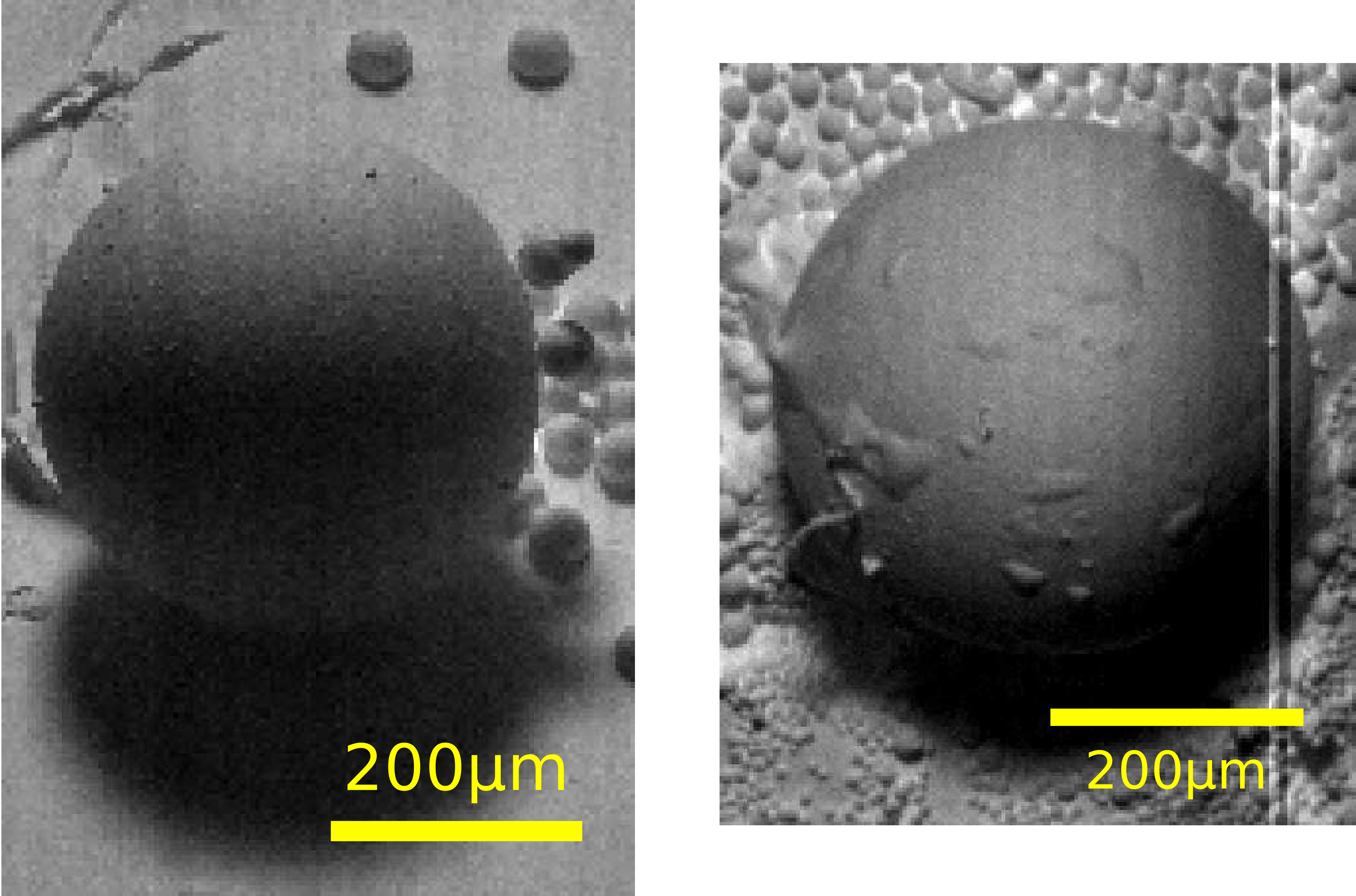}
    \caption{\ang{45} incidence image (left) compared to \ang{90} incidence image (right) of $400\,\mu m$ glass spheres taken in \textalpha-SHeM. Figure adapted from Lambrick et al. \cite{LambrickDiffuse}.}
    \label{fig:45_90_spheres}
\end{figure}
\newpage
Figure \ref{fig:45_90_spheres} shows the first normal incidence image acquired using the previously discussed 3D printed pinhole plate. Notice the difference in illumination at the top of the spheres, where the normally incident geometry allows for a more complete view of the scattering distribution.

\section{Heliometric Stereo Reconstruction}
\begin{figure}[p]
    \centering
    \begin{subfigure}{0.55\textwidth}
    \includegraphics[width=\textwidth]{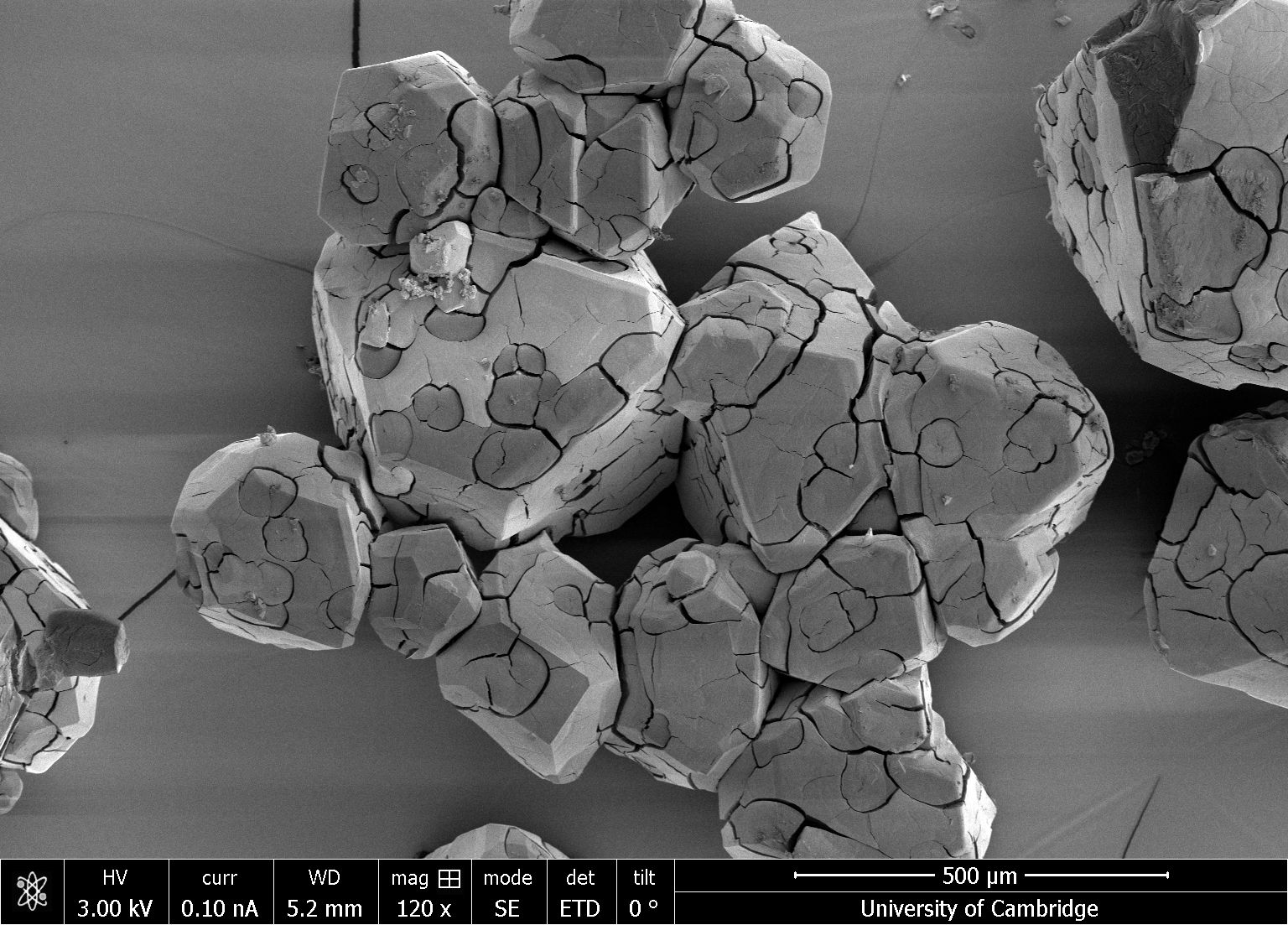}
    \subcaption[a]{}
    \label{fig:crystals_sem_a}
    \end{subfigure}
    \begin{subfigure}{0.55\textwidth}
    \includegraphics[width=\textwidth]{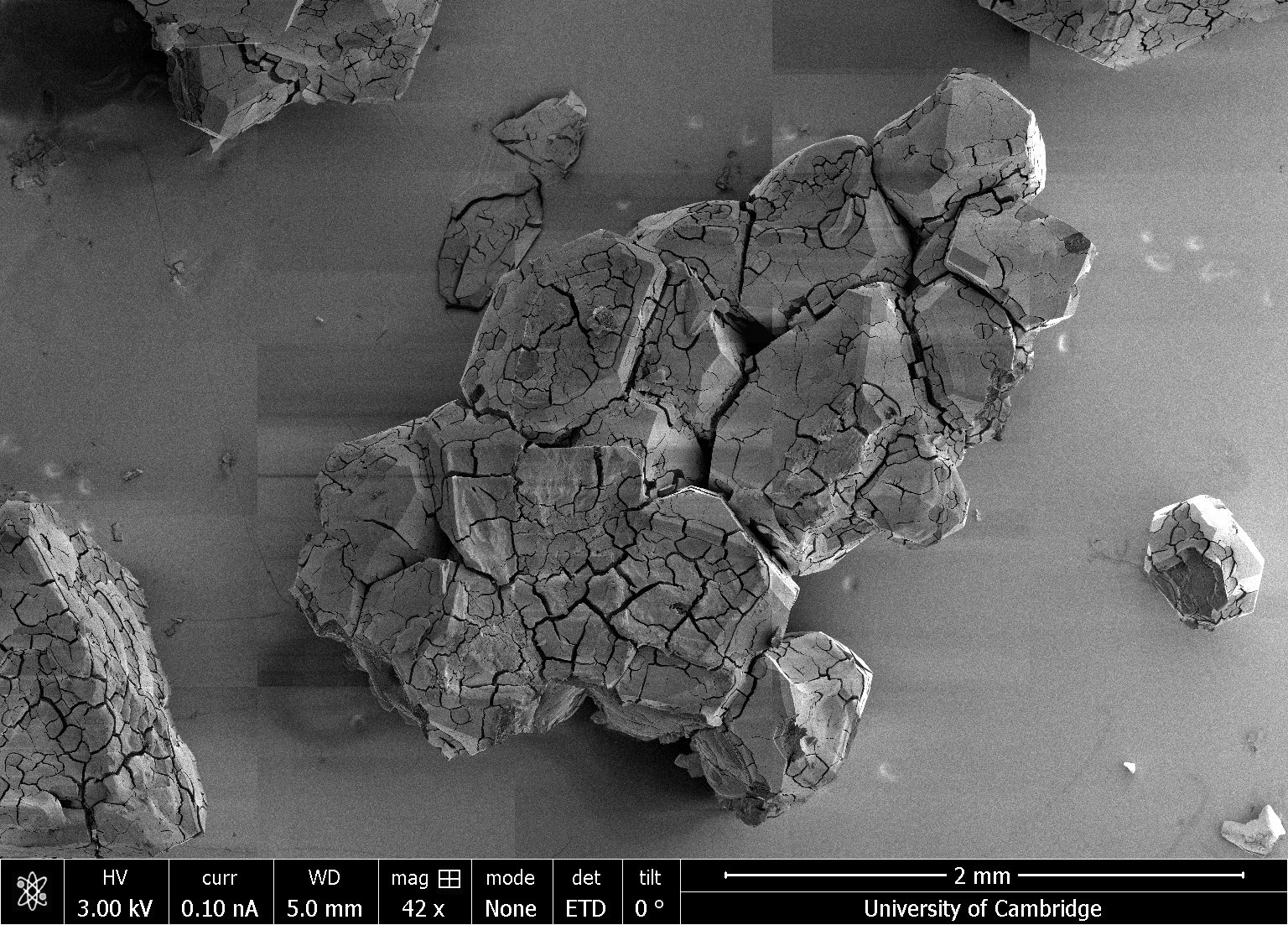}
    \subcaption[b]{}
    \label{fig:crystals_sem_b}
    \end{subfigure}
    \centering
    \begin{subfigure}{0.55\textwidth}
    \includegraphics[width=\textwidth]{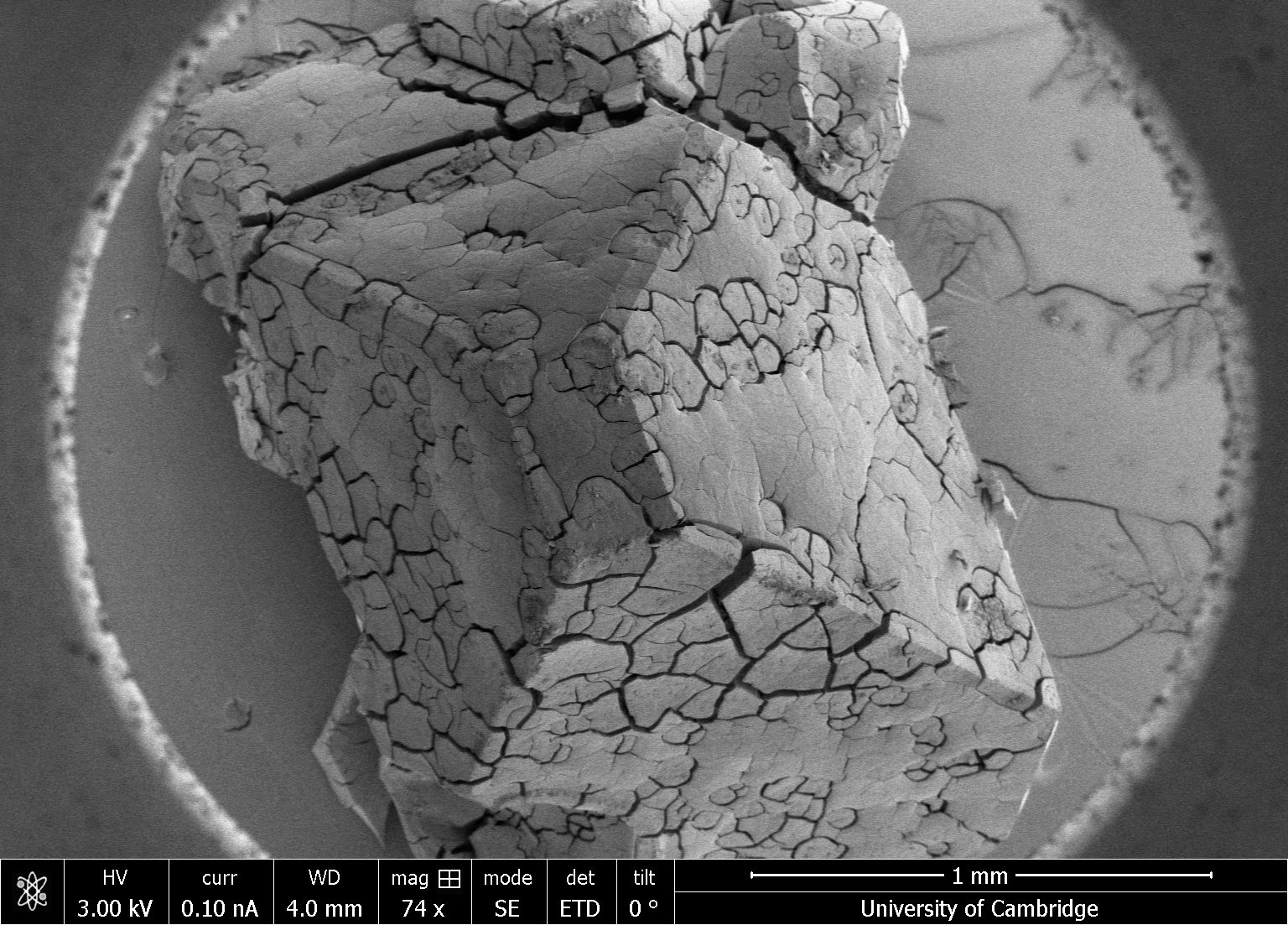}
    \subcaption[c]{}
    \label{fig:crystals_sem_c}
    \end{subfigure}
    \caption{SEM images of the grown aluminium potassium sulphate crystals coated in gold. Panel a) small cluster, b) large cluster and c) pyramid.}
    \label{fig:crystals_sem}
\end{figure}

\begin{figure}[h]
\centering
\includegraphics[width=.3\textwidth]{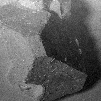}\quad
\includegraphics[width=.3\textwidth]{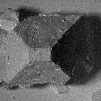}\quad
\includegraphics[width=.3\textwidth]{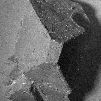}

\medskip

\includegraphics[width=.3\textwidth]{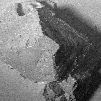}\quad
\includegraphics[width=.3\textwidth]{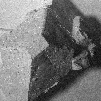}

\caption{$20\,\mu m$ pixel size SHeM images of the large pyramid, from Figure \ref{fig:crystals_sem_c}, taken at the optimal working distance for the normal incidence plate used, $\approx 4\,mm$ at the top of the pyramid, rotated by \ang{72} between images. }
\label{fig:pyramid_rotated}
\end{figure}

\begin{figure}
    \centering
    \includegraphics[width=0.5\textwidth]{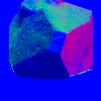}
    \caption{Overlaid correlated images of the crystals from Figure \ref{fig:pyramid_rotated}. The high contrast colour maps used in the overlaid images helps to highlight misalignment from the image correlation. The dark blue region surrounding the pyramid is where there is not enough data to attempt a reconstruction because the images don't overlap.}
    \label{fig:normal_pyramid}
\end{figure}

\begin{figure}
    \centering
    \includegraphics[width=0.5\textwidth]{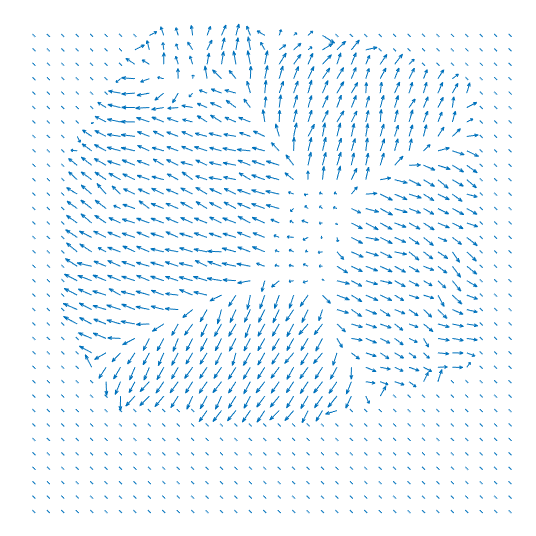}
    \caption{Normal vectors for each pixel of the correlated image. At this stage the structure of the pyramid becomes clearly visible. The field of vertical normal vectors surrounding the pyramid are inserted by the algorithm to cover areas where there is insufficient data for a reconstruction to be attempted.}
    \label{fig:vectors_pyramid}
\end{figure}

\begin{figure}
    \centering
    \includegraphics[width=0.5\textwidth]{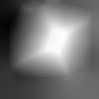}
    \caption{Height map of the pyramid generated from normal vectors where brighter regions are taller.}
    \label{fig:height_pyramid}
\end{figure}

 The unit test sample for 3D surface reconstruction, shown in Figure \ref{fig:heliometricstereo_algorithm}, proved difficult to fabricate using focused ion beam (FIB) milling so different candidate samples were explored. Standard metric samples used for microscope calibration were explored because of their precisely known dimensions and existing CAD models, making a comparison between the nominal geometry of the sample and the reconstructed sample. The metrics, however, generally had feature sizes below SHeM's current resolution capabilities. They were additionally unsuitable because most contained features which are highly symmetric, which makes correlating points across the images in the reconstruction difficult, or had vertical edges for the purposes of accurate microscope calibration, which introduces delta functions into the reconstruction algorithm. 
\\[12pt]
An ideal sample for heliometric stereo reconstruction therefore has:
\begin{itemize}
\setlength\itemsep{-0.5em}
    \item Low aspect ratio.
    \item No vertical edges.
    \item Minimal masking.
    \item Identifiable features.
\end{itemize}
It was determined that a ``natural''  sample was necessary, so aluminium potassium sulphate, \ch{KAl(SO_{4})_{2}. 12 H_{2}O}, crystals were chosen because of their regular octahedral structure and ability to form crystals on the length-scale of hundreds of microns to millimetres.
\\[12pt]
$25\,g$ of Anhydrous potassium aluminium sulphate,  \ch{KAl(SO_{4})_{2}}, was stirred into $100\,ml$ heated de-ionised water to form a super-saturated solution. The solution was removed from the heating source and left to precipitate out crystals over 24 hours. The solution was then drained and a selection of crystals ranging from approximately $200\,\mu m$ to $2\,mm$ were removed and placed on an SEM sample stub. Figure \ref{fig:crystals_sem} shows SEM images of three spots on the sample that are good candidates for reconstruction because they don't have vertical edges, highly symmetric features or heavy masking. A TEM grid was also added to the sample to aid in centre of rotation and working distance calibrations once inserted into SHeM.
The cluster of crystals shown in Figure \ref{fig:crystals_sem_a} is a promising candidate for reconstruction because it has a low aspect ratio and doesn't have any tall features near it that would mask the cluster when the sample is rotated azimuthally. However, Panel \ref{fig:crystals_sem_a} does have a low region in the centre of the cluster which, depending on the height of the surrounding features, could be masked significantly and hinder the reconstruction. 
\\[12pt]
The larger cluster in Panel \ref{fig:crystals_sem_b} satisfies the majority of the listed ideal reconstruction criteria, making it the an ideal candidate. However, the size of the cluster poses a problem, due to the size of the feature each image would take 8-12 hours to complete. Long imaging times increase the probability that the reconstruction will be affected by non-topographical contrast because the algorithm assumes the instrument has constant contrast sensitivity in time. Since the nature of heliometric stereo is to reconstruct a structure by using contrast across images, such changes would seriously affect the result.
\\[12pt]
The large pyramid in Panel \ref{fig:crystals_sem_c} provides a near-ideal reconstruction candidate in that it has no masking from surrounding features and clearly defined edges and vertices. The main drawback of the large pyramid is that its large aspect ratio may prove problematic for the reconstruction as the detection angle will change by a non-trivial amount from the top to the base. 
\\[12pt]
The pyramid was chosen as the first candidate site for reconstruction, with the small cluster second. The large cluster was discounted due to long imaging times and the associated potential detector issues.
\begin{figure}
    \centering
    \begin{subfigure}{0.45\textwidth}
    \includegraphics[width=\textwidth, angle=90]{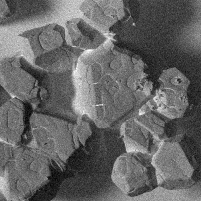}
    \subcaption[a]{Small cluster from Figure \ref{fig:crystals_sem_a}.}
    \label{fig:crystals_overview_a}
    \end{subfigure}
    \begin{subfigure}{0.45\textwidth}
    \includegraphics[width=\textwidth]{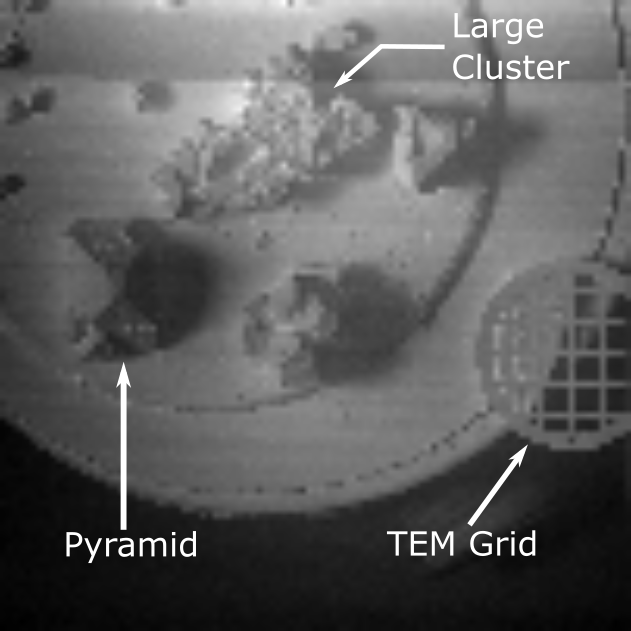}
    \subcaption[b]{Large cluster and pyramid from Figures \ref{fig:crystals_sem_b} and \ref{fig:crystals_sem_c}, respectively. Small cluster from Panel \ref{fig:crystals_overview_a} is out of frame at the top of the image.}
    \label{fig:crystals_overview_b}
    \end{subfigure}
    \caption{Overview \textalpha-SHeM images of candidate sites shown in Figure \ref{fig:crystals_sem}. The pyramid, large and small clusters are stuck to carbon tape in the centre of the SEM sample stub with a TEM grid added for point tracking calibration.}
    \label{fig:crystals_overview}
\end{figure}
\newpage
Starting with the pyramid, five images were taken at separations of \ang{72} azimuthal rotation to over-constrain the reconstruction algorithm, shown in Figure \ref{fig:pyramid_rotated}.
\newpage
The reconstruction process starts by manually identifying common points across the image to correlate them to each other. The pyramid's sharp vertices provide ideal correlation points that are visible across all images. Once the common points are matched, the images are overlaid to give a normal map with to give Figure \ref{fig:normal_pyramid}.
The normal map is then taken and normal vectors for each pixel are generated, as shown in Figure \ref{fig:vectors_pyramid}. The normal vectors represent the orientation of the normal of each pixel.
\newpage
Using the normal vectors each pixel is assigned a position and orientation in real-space. In Figure \ref{fig:height_pyramid} a height map is then generated from the normal vectors.
Finally, the reconstruction process produces a 3D surface reconstruction of the pyramid from the height map.
\begin{figure}
    \centering
    \begin{subfigure}{0.7\textwidth}
    \includegraphics[width=\textwidth]{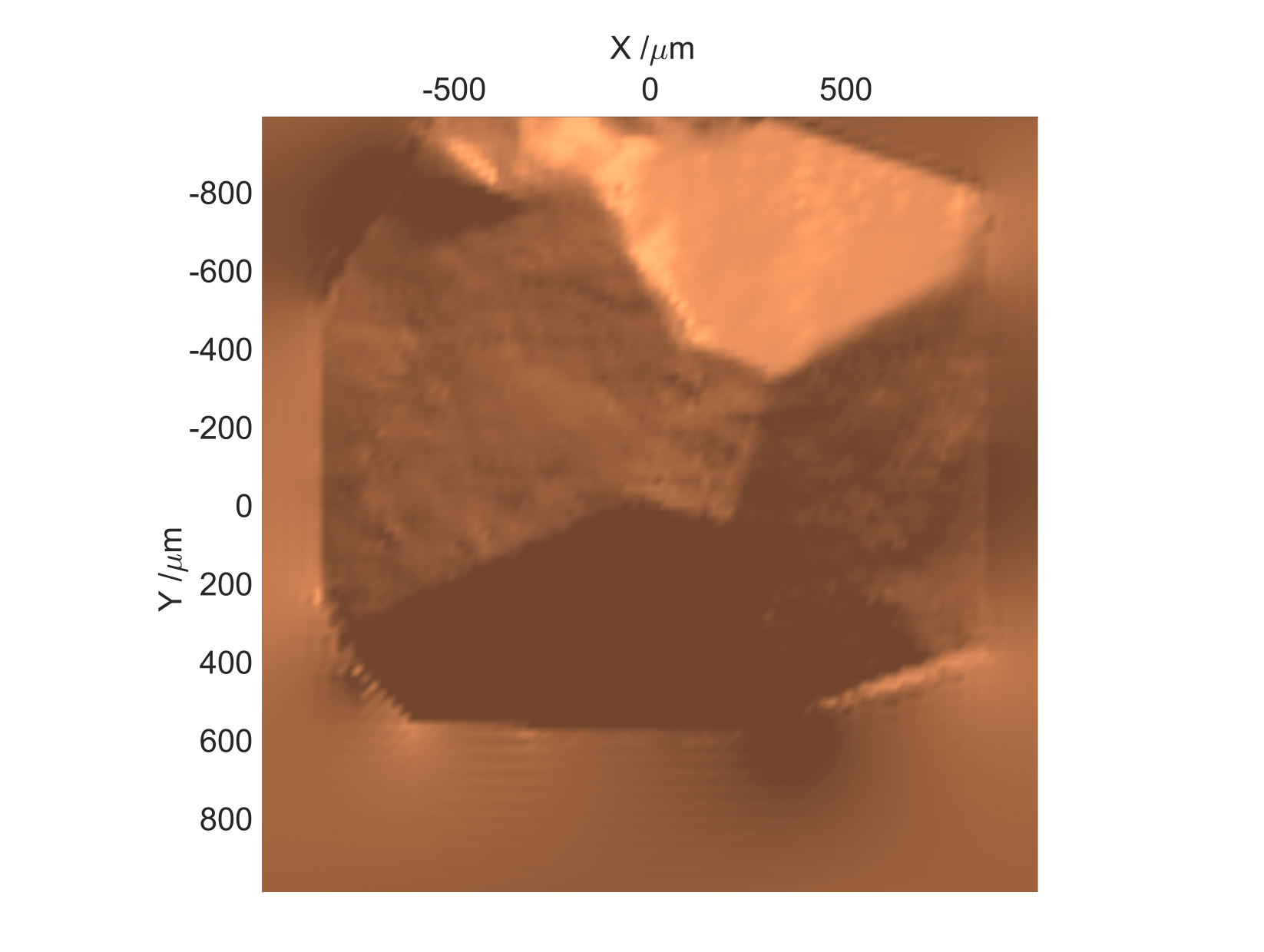}
    \subcaption[a]{Top-down view directly comparable to Figures \ref{fig:normal_pyramid} - \ref{fig:height_pyramid}.}
    \end{subfigure}
    \begin{subfigure}{0.8\textwidth}
    \includegraphics[width=\textwidth]{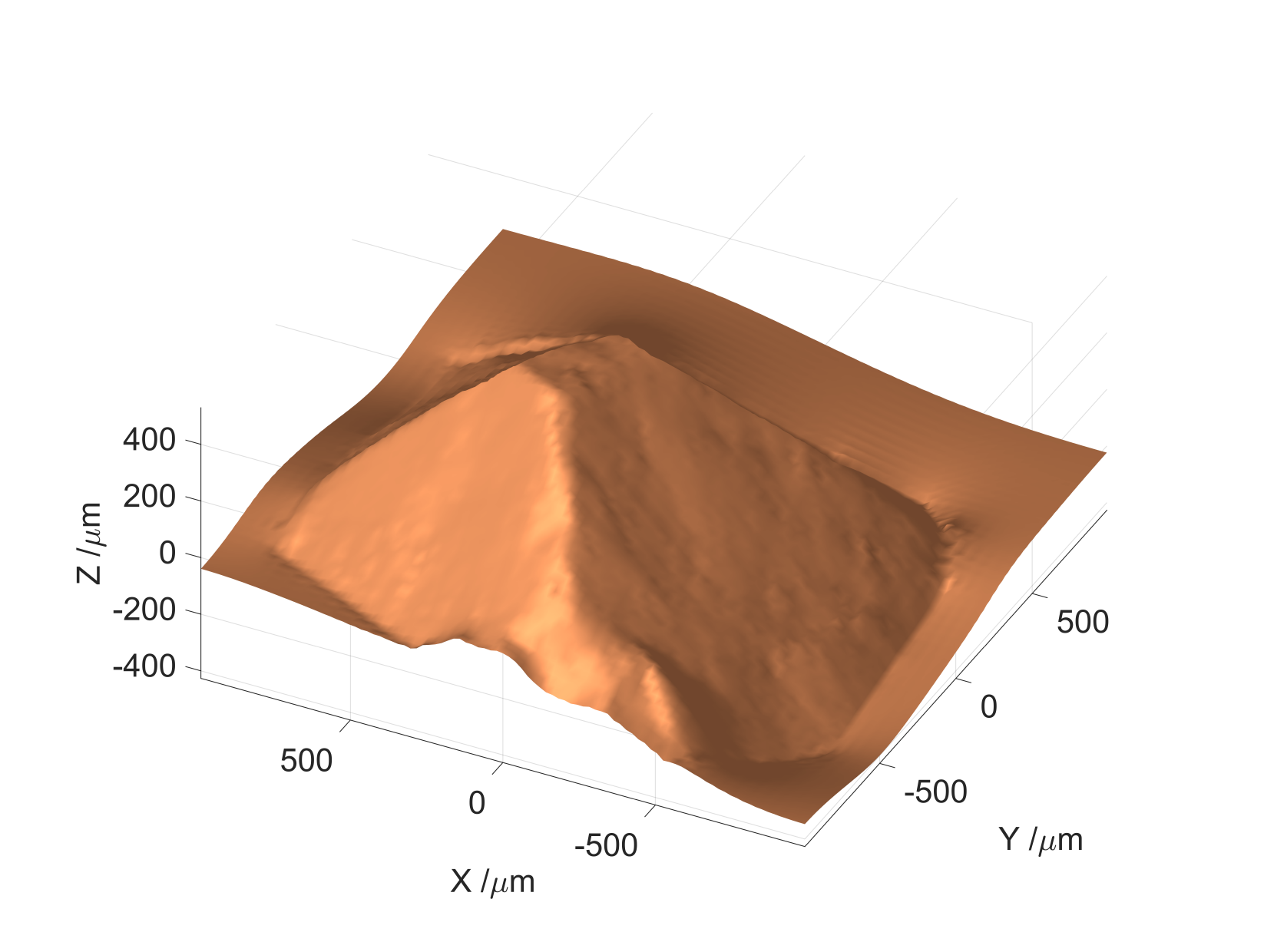}
    \subcaption[b]{Orthographic view of surface reconstruction.}
    \end{subfigure}
    \caption{Single detector heliometric stereo reconstruction of the large pyramid shown in Figure \ref{fig:crystals_sem_c}. The reconstruction returns that the pyramid is approximately $500\, \mu m$ tall with a base $1.3\,mm$ by $1.9\,mm$. The dimensions are in good agreement with measurements made using the SEM images in Figure \ref{fig:crystals_overview}. Note that the colour map does not correspond to a height map in this figure.}
    \label{fig:surface_pyramid}
\end{figure}
\newpage
An accurate measurement of the height of the pyramid in Figure \ref{fig:surface_pyramid} cannot be made because the false background added by the reconstruction algorithm obscures the true base of the pyramid where it attached to the substrate. Additionally, upon removal from the SEM, where the images in Figure \ref{fig:crystals_sem} were taken, the pyramid exploded, likely due to charging in the surface cracks. Therefore, complimentary techniques like optical profilometry or white-light interferometry could not be used to measure the pyramid either. The pyramid's top surface was selected for true-to-size quantification because it is approximately perpendicular to the beam in both SEM and SHeM, allowing for a reasonable comparison to be made.
\\[12pt]
The second candidate sample was also imaged to see how the reconstruction algorithm deals with sub-optimal conditions, namely heavy masking in the centre of the sample, shown in Figure \ref{fig:crystals_sem_a}. 
\begin{figure}[H]
    \centering
    
    \begin{subfigure}{0.65\textwidth}
    \includegraphics[width=\textwidth]{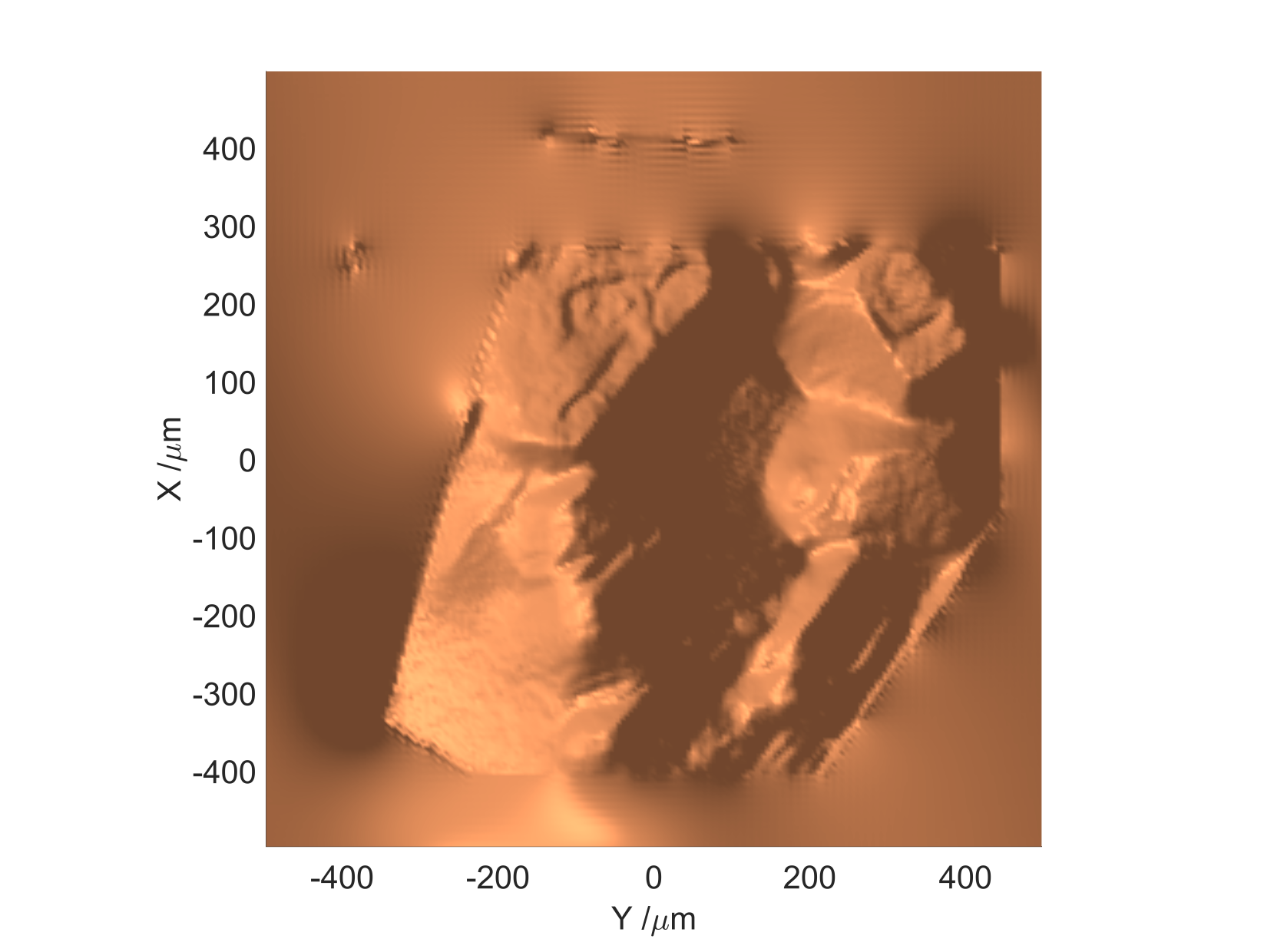}
    \subcaption[a]{Top-down view.}
    \end{subfigure}
    
    \begin{subfigure}{0.75\textwidth}
    \includegraphics[width=\textwidth]{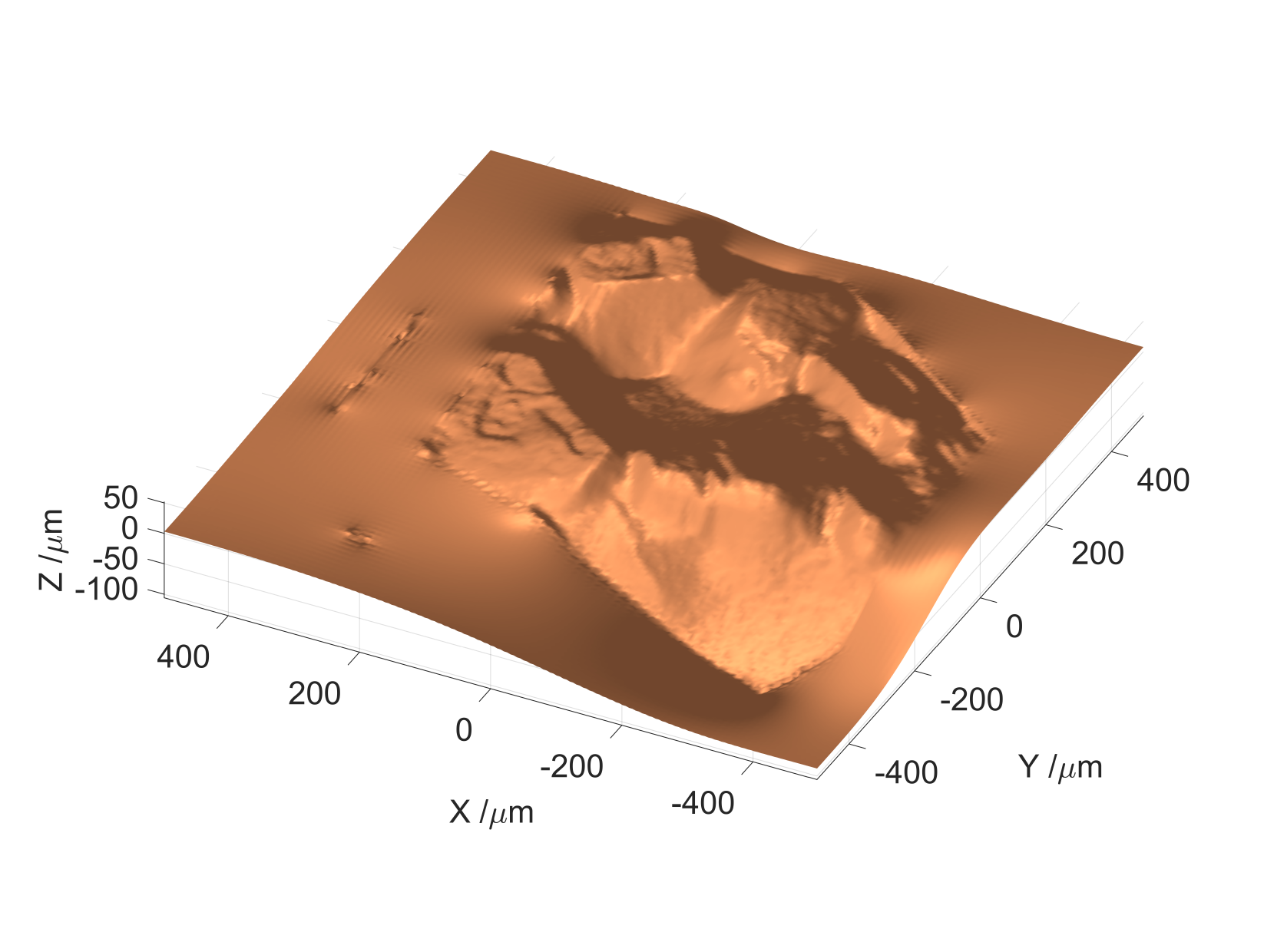}
    \subcaption[b]{Orthographic view of surface reconstruction.}
    \end{subfigure}
    
    \caption{Partially successful 3D reconstruction of the small cluster in Figures \ref{fig:crystals_overview_a} and \ref{fig:crystals_sem_a}. The rounded edges and faces in the bottom of the central region demonstrate the effect of heavy masking and vertical edges on the heliometric stereo algorithm.}
    \label{fig:surface_small_cluster}
    
\end{figure}
\newpage
The reconstruction of the small cluster was partially successful, some of the smaller features are visible and reasonably accurate, with the general topography of the region also clear. The heavy masking in the centre of the sample has softened the sharp edges at the bottom of the central area, giving sloped faces that ought to be vertical based on SEM images in Figure \ref{fig:crystals_sem_a}. The reconstruction in Figure \ref{fig:surface_small_cluster} clearly demonstrates the effects of strong masking, and to some degree vertical faces, on the heliometric stereo algorithm. 
\\[12pt]
By taking measurements from the SEM images in Figure \ref{fig:crystals_sem} one can quantify how true-to-size the reconstructions in Figures \ref{fig:surface_pyramid} and \ref{fig:surface_small_cluster} are. The model of Helios DualBeam FIB/SEM used has a beam size of approximately $1\,nm$ at $3\,kV$ and $5\,mm$ working distance \cite{helios5}, however, the spot size cannot be used as an error bound on measurement accuracy because the dominant error comes from pixel-based computer measurements made on the images themselves. The error in pixel-based measurement has therefore been taken as $\pm5\%$, or half the smallest division. The rectangular top face of the pyramid was measured as approximately $(0.10\pm0.005)\,mm^2$ in the SEM images and $(0.09\pm0.005)\,mm^2$ in the reconstruction, a 10\% difference. The rectangular shape at the top of the partially successful reconstruction, Figure \ref{fig:surface_small_cluster}, yielded a similar result of 8\%, with areas of $(0.017\pm0.001)\,mm^2$ and $(0.015\pm0.001)\,mm^2$, for SEM and reconstruction respectively. It should be noted that the false background, seen in Figure \ref{fig:normal_pyramid} as dark blue, obscures the position of the sample stub, making an accurate height measurement difficult.

\section{Conclusion \& Further Work}
In the current work a method for true-to-size 3D surface reconstruction, taking approximately 5 days of total imaging time from sample insertion to final reconstruction, as detailed by Lambrick et al. was successfully applied experimentally on a single detector SHeM and shown to have the ability to reconstruct a surface to approximately 10\% of the original size. The resulting reconstruction being approximately $10\%$ smaller than the actual sample is in good agreement with the $8\%$ figure quoted by Lambrick et al. using simulated data \cite{Lambrick2021}.
\\[12pt]
To develop the technique in the short term one could attempt a higher resolution reconstruction of a smaller sample to combine the current maximum resolution $\approx 350\,nm$ with the novel imaging mode. Equivalent, established techniques, like optical profilometry and white-light interferometry, could be used to create a CAD model of samples for comparison with both the experimental and simulated outputs of heliometric stereo. Additionally, side-on images can be taken to give more accurate height data. By using a CAD model from a method with known accuracy the effects of experimental factors on the heliometric stereo method, e.g. detector instability or masking, can be explored and quantified. At smaller length-scales, AFM could also be used as a method of generating a 3D reconstruction.
\newpage
Long term development of the technique should be approached from a hardware perspective. Improvements to detector sensitivity and stability would allow for higher resolution reconstruction and longer imaging times, and therefore larger images, respectively. Multiple physical detectors would greatly reduce imaging times and potentially remove the need for azimuthal sample rotation, depending on their number.
\\[12pt]
Simulating 3D reconstruction data using the in-house ray tracing framework typically takes 1 day on a modern quad-core desktop central processing unit (CPU). In order to more effectively investigate the effects of masking and sample topography on heliometric stereo, the ray tracing must be accelerated. 


\chapter{GPU Accelerated Ray Tracing}
\label{section:cuda}
Here the applicability of accelerating the ray tracing with a graphical processing unit (GPU) parallelisation is probed. The chapter continues by giving an overview of CPU and GPU architectures, outlining how each is optimised at a hardware level for serial and parallel tasks, respectively. Currently parallelised inefficiently on the CPU at a high-level through MATLAB directly, moving the parallelisation to a lower level of the code, on the GPU, and leveraging the ideally parallelisable nature of the simulation has been shown to yield a performance increase of up to 380 times in the current work. By using Nvidia's Compute Unified Device Architecture (CUDA) Application Programming Interface (API), preliminary performance testing on analogous problems has shown that GPU parallelisation can decrease the time complexity of ideal problem such as the in-house ray tracing. The analogous problems investigated promise a quantitative method to predict the computational load on High-Performance Computing (HPC) clusters once GPU parallelisation of the ray tracing is achieved.

\section{Overview of Ray Tracing}
The core in-house computational tool used in conjunction with SHeM uses a Monte-Carlo based ray tracing framework to fully simulate the experimental environment of \textalpha-SHeM \cite{Lambrick2018}. The ray tracing is instrumental in every stage of the SHeM imaging procedure, starting from simulating novel pinhole plate designs like those used in point diffraction \cite{vonJeinsen2021} and 3D image reconstruction, to investigating the microscope's unique contrast mechanisms and recreating/predicting real SHeM measurements. In the current work, the ray tracing is needed to simulate 3D image reconstruction so that the reconstruction algorithm may be tested and optimised on control data. The algorithm requires a minimum of 3 images to attempt a reconstruction, with 5 usually taken to over-constrain the problem and provide redundancy. A single high quality ray traced image typically takes 5 hours to complete, totalling to approximately 1 day of simulation for a complete reconstruction using a desktop-grade CPU. MATLAB itself currently accelerates the ray tracing by parallelising on the CPU by opening new instances of MATLAB for each CPU core utilised. Such parallelisation is coarse and inefficient, consuming resources on running multiple instances of MATLAB itself. Further acceleration of the ray tracing simulation would enable a variety of new simulations to be run such as more complex scattering distributions and sample topology, in addition to increasing the throughput of work that is currently possible.

\section{Overview of CUDA and CPU vs. GPU Architecture}
CUDA is an API developed by Nvidia for parallel computing using graphical processing units (GPUs) \cite{{cuda}}. CUDA gives developers access to the highly specialised architectures and instruction sets present in GPUs, typically used for rendering 3D graphics in computer games or professional art/design, such that they may be leveraged for general purpose computing. CUDA also allows seamless expansion to multi-GPU HPC systems to further accelerate computing. 
\\[12pt]
CUDA builds upon standard C/C++ which makes it easier to learn and use than alternative APIs like OpenGL or Direct3D. Although the alternatives are open-source and can be used on GPUs from any brand, unlike CUDA which can only be used with Nvidia GPUs, the ease of use provided by prior knowledge of C/C++ outweighs the proprietary nature of CUDA.
\\[12pt]
Additionally, CUDA provides vast libraries of highly optimised mathematical algorithms for commonly used functions, notably fast Fourier transforms (cuFFT), linear algebra (cuBLAS and CUTLASS) and pseudo-random number generators (cuRAND) to name a few. CUDA provides further performance improvements over simply porting functions written for a CPU to a GPU by leveraging low-level access to the GPU architecture with these optimised libraries.
\\[12pt]
CPU and GPU architectures possess the same core features but differ significantly in key aspects. CPUs are designed to handle a wide range of tasks within a computer, often referred to as the ``brain'' of a system, meaning they have many instruction sets available to them. CPUs are also optimised for fast serial operations, architecturally that means fewer Arithmetic Logic Units (ALUs) which execute instructions, with higher clock speeds so more instructions can be computed per clock cycle. In contrast, GPUs have highly specialised instruction sets, focusing on linear algebra and general arithmetic predominantly, so that a select few tasks can be performed with the utmost speed and efficiency. Their architectures also have many more ALUs than CPUs do, order of $10^2-10^3$ times more, with relatively slower clock speeds. Such design decisions make GPUs highly optimised for parallel computing tasks \cite{Owens}, where many highly specialised ALUs can execute mathematically oriented instruction sets in parallel. Furthermore, graphics cards are designed for high bandwidth between their memory and ALUs, rather than low latency between the system memory and central processor.
\begin{figure}[ht]
    \centering
    \begin{subfigure}{0.45\textwidth}
    \includegraphics[width=\textwidth]{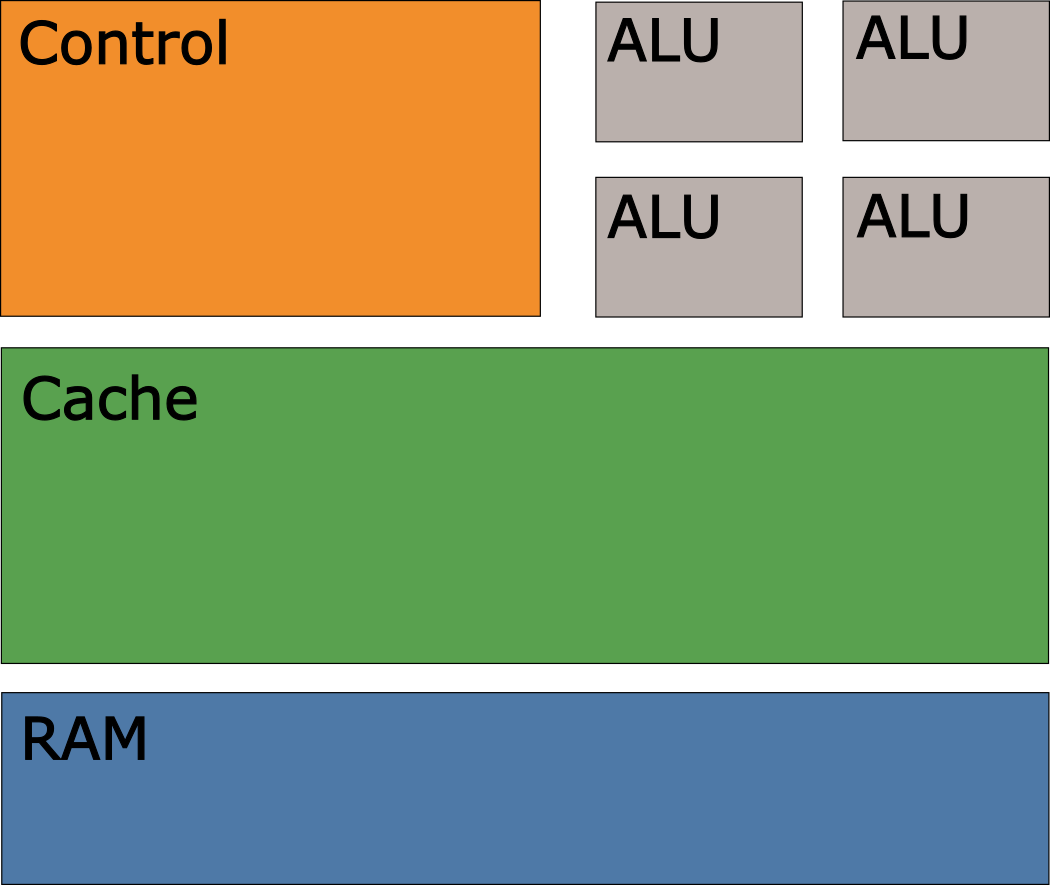}
    \subcaption[a]{CPU}
    \end{subfigure}
    \begin{subfigure}{0.45\textwidth}
    \includegraphics[width=\textwidth]{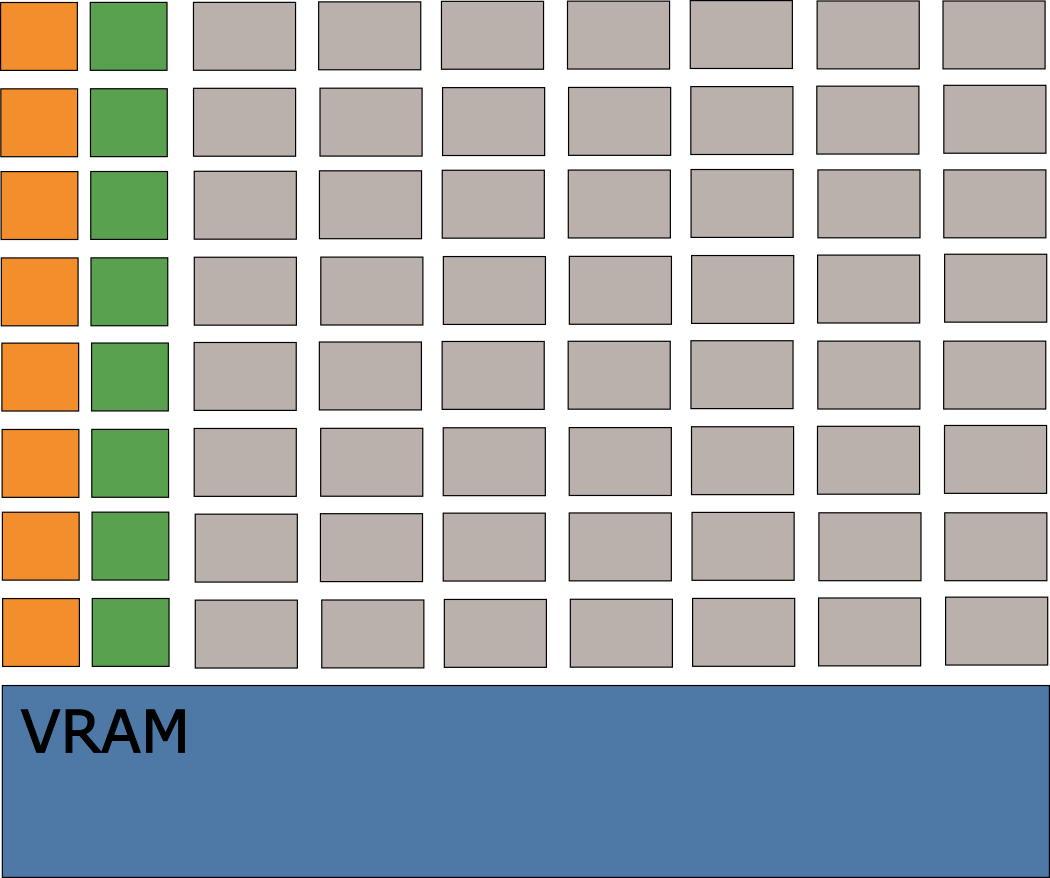}
    \subcaption[b]{GPU}
    \end{subfigure}
    \caption{Schematic diagram showing the key architectural differences between CPUs and GPUs. The GPU has many more ALUs, and additional controllers to issue commands to smaller groups of ALUs. Although GPUs have less cache per ALU than a CPU, they have access to a dedicated high-bandwidth pool of memory shared within the GPU called the Video RAM (VRAM) which is analogous to system memory (Random Access Memory - RAM) accessible to the CPU.}
    \label{fig:cpu_gpu_arch}
\end{figure}

\section{Applicability of CUDA to Ray Tracing}
 The in-house ray tracing is able to treat individual rays completely independently from one another because the SHeM helium-4 beam doesn't interact with itself in a significant way after the initial free jet expansion because the beam is not charged and does not cluster. The rays are created at a virtual source, moving to the sample surface and evaluating a scattered direction for each ray based on the sample's scattering distribution, finally being detected at a virtual detector, or being deleted after a threshold number of scattering events.
 \\[12pt]
 The completely independent propagation of each ray makes the SHeM ray tracing simulation ideally suited to GPU parallelisation. A problem is considered ideal when the overarching problem can be broken down into completely independent elements, also called subsets of the problem, which can be computed simultaneously without need for direct communication between ALUs working on different subsets. Such problems avoid the need for synchronisation or data transfer between ALUs, a common obstacle when tackling molecular dynamics using GPUs for example \cite{Liu2008}, which inevitably introduces significant overheads where parts of the system must wait for others to finish their operations.
 \\[12pt]
The immediately obvious approach to leveraging GPU parallelisation in the ray tracing is to adapt the existing code, designed for a CPU, to work on a graphics card by replacing the relevant functions with CUDA equivalents. Adapting existing code would certainly yield a performance increase, but would limit the opportunities for proper optimisation of the CUDA code because only chunks of the simulation have been converted. Additionally, previous attempts at directly converting CPU algorithms to GPU parallelisation have proven both difficult and sub-optimal \cite{Asadchev2012}\cite{Alam2007} with the best results coming from entirely new implementations on graphics cards.
\\[12pt]
A conversion to GPU ray tracing would start with the most computationally expensive step which solves a system of linear equations would be moved to the GPU, making use of the highly optimised linear algebra libraries CUDA offers. Such a partial implementation would result in some performance improvement, but realistically every function shown in the flowchart should be replaced by CUDA equivalents for maximal performance. If one considers replacing all the functions, and therefore surrounding code too, the entire core of the ray tracing has been re-written leaving only the top-level MATLAB scripts intact. Therefore, a full implementation that properly leverages the parallel computing power of graphics cards should be approached from the ground up by designing the ray tracing with GPU parallelisation in mind, avoiding the aforementioned overheads and synchronisation limitations whilst streamlining development process. Although a full implementation of the ray tracing is beyond the scope of the current work, preliminary performance testing on analogous problems is presented which demonstrates the potential performance improvements from GPU parallelisation.

 \section{Preliminary Performance Testing}
\begin{figure}[ht]
    \centering
    \includegraphics[width=0.75\textwidth]{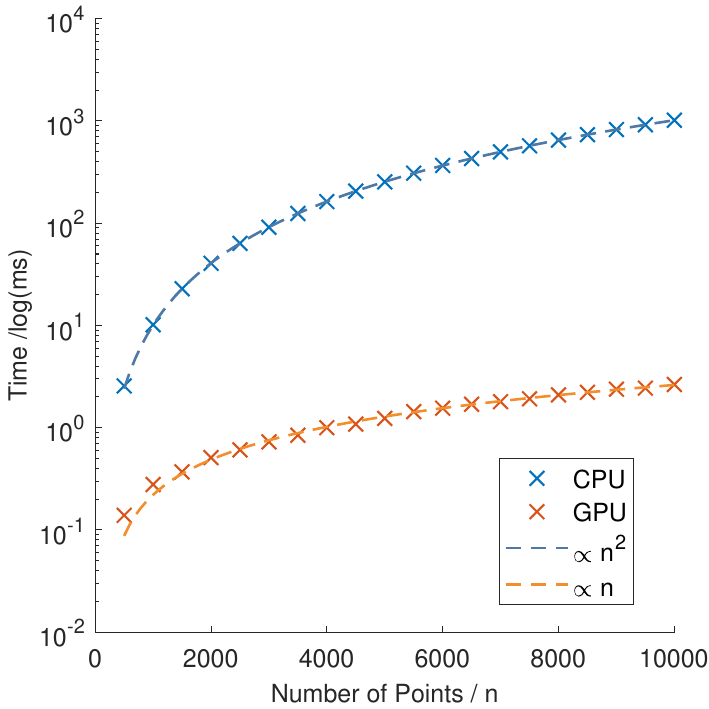}
    \caption{CPU and GPU compute times for the brute force nearest point search algorithm, code in Appendix \ref{appendix:pointsearch}, plotted with logarithmic y-axis. The serial version of the algorithm (CPU) has time complexity $\mathcal{O}(n^2)$, therefore fitting well quadratic polynomial. When the algorithm is executed in parallel by the GPU however, the time complexity becomes $\mathcal{O}(n)$, fitting a linear polynomial.}
    \label{fig:point_search_performance}
\end{figure}

\begin{figure}
    \centering
    \includegraphics[width=\textwidth]{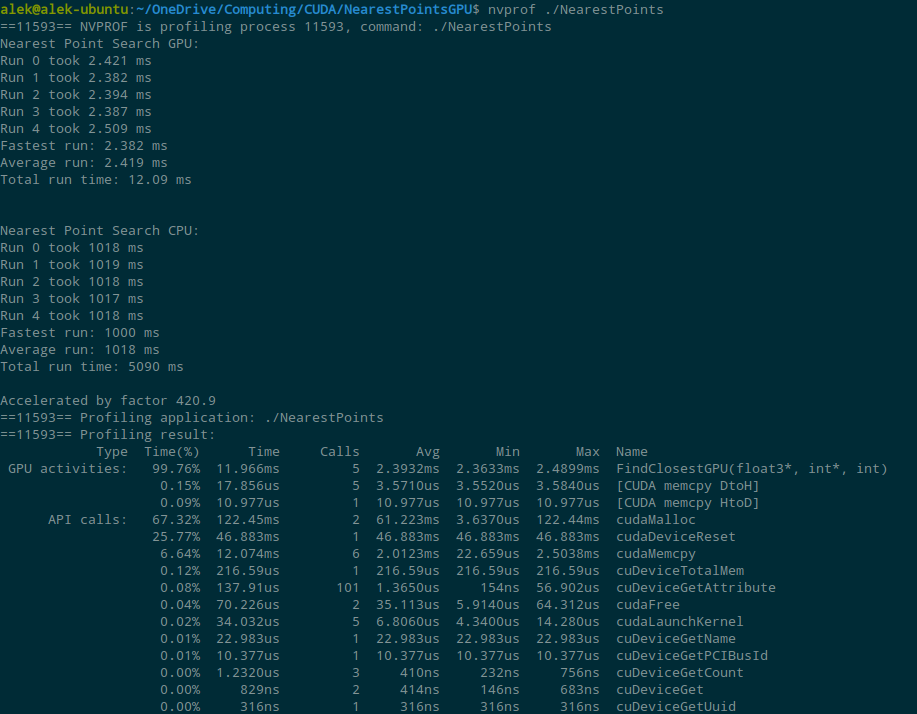}
    \caption{Screenshot of the console output when the nearest point algorithm is analysed by nvprof, run on a intel Core i5 6600k and Nvidia GTX 1070. The runtime profiler gives a detailed report of all function calls made, separating them into ``GPU activities'' and ``API calls'' which can be interpreted simply as simulation functions and non-simulation functions, respectively. The timing protocol used to time the runs and function calls is extremely accurate because everything is timed against the GPU's own clock cycles.}
    \label{fig:point_search_nvprof}
\end{figure}

 A unit test was designed as an analogue to the ray tracing to probe the potential upper-bound performance increase one could achieve by moving an ideal problem from CPU to GPU. A brute force nearest point search was chosen because the problem can be easy parallelised into independent subsets as there is no interaction between points, the algorithm naively searches through a list of points in 3-dimensional space from index $1\,\text{to}\, n$, where $n$ is the total number of points, and computes Pythagoras' theorem against all $n-1$ points sequentially to find the closest neighbouring point, meaning there are $(n-1)!$ total combinations of points. The algorithm has total time complexity $\mathcal{O}(n^2)$ because it involves one level of nested loop to execute. Time complexity quantifies the dominant factor in the scaling of a computational problem, hence $\mathcal{O}(n^2)$ means the problem is dominated by a quadratic factor.
 \\[12pt]
 Implementing the algorithm on both CPU and GPU, with code for both shown in Appendix \ref{appendix:pointsearch}, yielded an improvement of approximately 380 times for 10,000 points, averaged over 10 runs, completing $\approx \num{5.7e35660}$ point comparisons in $2.6\,ms$ compared to $1017.0\,ms$. 10,000 points was chosen because that is a typical number of rays propagated per pixel in the ray tracing. Figure \ref{fig:point_search_performance} shows the compute times for CPU against GPU for the point search algorithm. The CPU times fit a quadratic function as one would expect from the quadratic time complexity. The GPU results, however, show not only a large performance increase up 380 times, but a decrease in the order of the time complexity to linear, $\mathcal{O}(n)$. Reduction of the time complexity of the problem is vastly more significant than a linear performance improvement because it means the problem now scales more favourably, with the relative performance improvement only increasing as the simulation grows. 
 \\[12pt]
 Such an improvement is only possible because the CPU and GPU, and the code they execute, fundamentally differ. Parallelisation on the CPU, as is currently done through MATLAB for the ray tracing, does not change the time complexity and only decreases the compute time by a linear factor proportional to the number of CPU cores. Note that the timer was set so that only the actual computation time was measured rather than including library and point generation overheads which are equal in both cases, and negligible in comparison to total compute time.
\begin{flushleft}
One can then use Nvidia's runtime profiling tool, ``nvprof'' \cite{cuda}, to analyse function execution calls and times, latency and data throughput of the program, identifying the key optimisation areas. Figure \ref{fig:point_search_nvprof} shows the terminal output given by nvprof when applied to the nearest point search algorithm.
\end{flushleft}

\begin{flushleft}
The reduction in order of time complexity presented in Figure \ref{fig:point_search_performance} demonstrates the applicability of GPU parallelisation to an ideal problem with quadratic time complexity. The ray tracing, however, has linear time complexity, $\mathcal{O}(n)$. The time complexity is linear because each ray introduces a constant number of new calculations. Although it contains various algorithms at stages within the ray propagation that are higher order time complexity, such as Cramer's Rule ($\mathcal{O}((n+1)\,!)$ for solving systems of linear equations, the dimension of such problems remains constant as the number of rays propagated changes, giving those algorithms constant time complexity on the scale of the entire simulation, therefore overall $\mathcal{O}(n)$ is the highest order time complexity. Another unit test was designed to demonstrate the efficacy of GPU parallelisation on an ideal problem with linear time complexity.
\end{flushleft}
A program was written that designates two vectors, $a[n]$ and $b[n]$, and sums them together to give a third vector, $c[n]$, where $n$ is the length of the vector, executed on both the CPU and GPU. The algorithm, code available in Appendix \ref{appendix:vector_addition}, therefore scales linearly with $n$ giving linear time complexity, like the ray tracing itself.
\begin{figure}[ht]
    \centering
    \includegraphics[width=\textwidth]{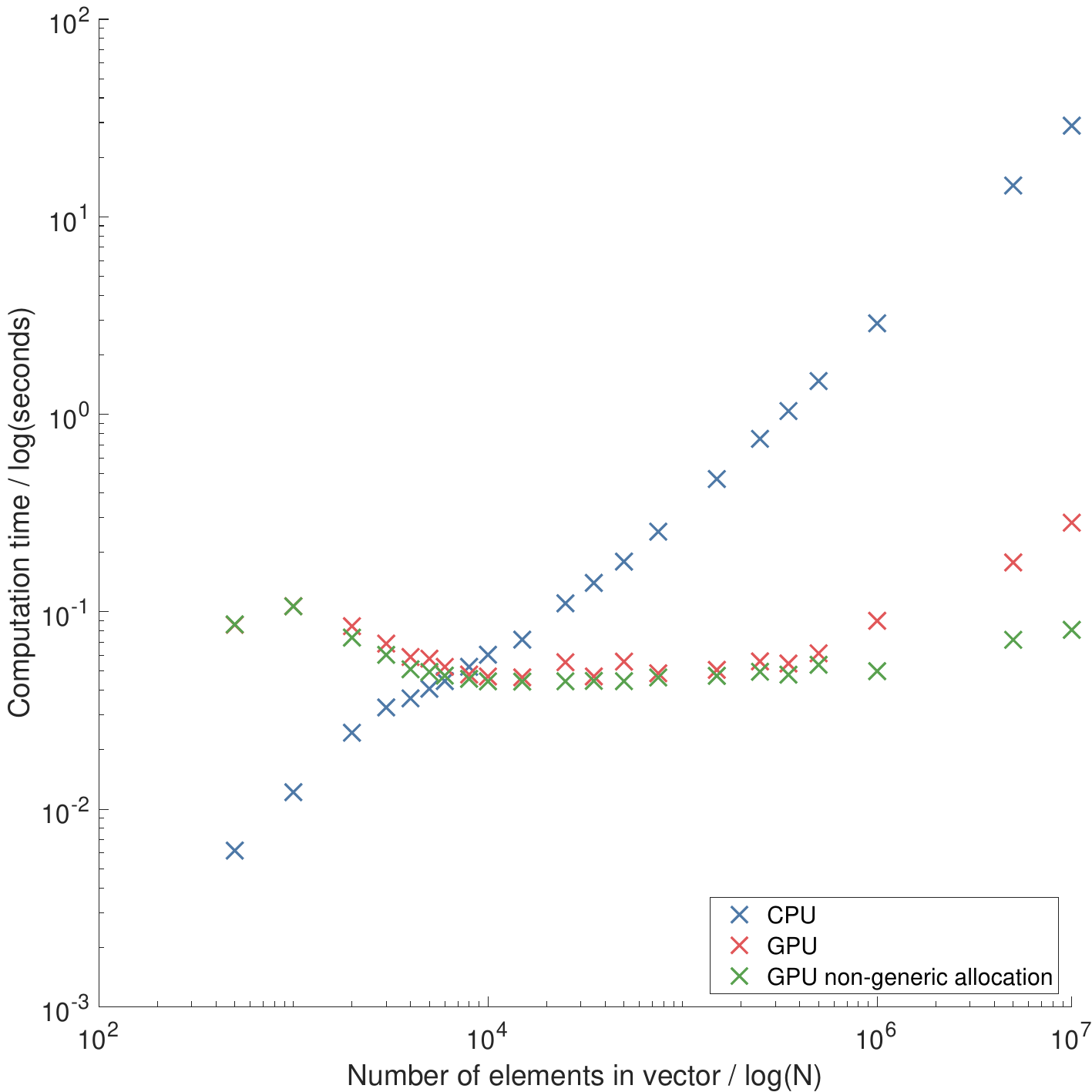}
    \caption{Double log plot showing the times taken for the CPU and GPU to execute addition of vectors of length $n$. The CPU execution times fit a linear polynomial as expected, but the GPU parallelisation, using dynamic resource allocation that scales with $n$, has once again reduced the time complexity by one order of magnitude, to constant time complexity $\mathcal{O}(1)$. The ``GPU non-generic allocation'' data shows the execution times when the allocation of resources is manually fixed at a value.}
    \label{fig:vector_addition_performance}
\end{figure}
\begin{flushleft}
The GPU is slower than the CPU initially when $n<5000$ because the graphics card's resources are under-utilised and overheads dominate the execution time since for small $n$ the computation time is of the same order of magnitude as the API calls, which take a constant amount of time. When the problem becomes large enough for parallelisation to be worthwhile, $5,000<n<500,000$, the time complexity of the problem becomes constant in time with no increase in compute time despite a factor 100 increase in operations. The orange data, in Figure \ref{fig:vector_addition_performance}, then regains linear time complexity at $n>500,000$ because the generic, dynamic allocation of threads per block, and total blocks, becomes inefficient in such a way that the problem is being calculated serially. The dynamic allocation used for the "GPU'' data is a function of the length of the input vector, increasing the threads per block and blocks proportionally to $n$. The critical point when the dynamic allocation becomes inefficient is dependent on the total threads a GPU can concurrently run, for the Nvidia GTX 1070 used here it is $n \approx 500,000$. The "GPU non-generic allocation'' data sets a fixed value of threads/blocks for the entire range of $n$, resulting in the problem remaining at constant time complexity for longer. By using multiple GPUs in an HPC, the region in which the time complexity is reduced will be extended significantly with each additional graphics card. The regression to the higher order time complexity was not visible in Figure \ref{fig:point_search_performance} because $n$ was not sufficiently large to expose the limitations of generic resource allocation.
\end{flushleft}
The improvement in time complexity from linear to constant, for the CPU and GPU respectively, shown in Figure \ref{fig:vector_addition_performance} demonstrates the efficacy of GPU parallelisation on an ideal problem that has linear time complexity, analogous to the ray tracing. The data further reinforces the performance increase shown by Figure \ref{fig:point_search_performance} and reveals that the gains made through parallelising an ideal problem are not a constant amount, but changes the time complexity of the problem which is a far higher impact improvement.

\begin{flushleft}
The preliminary performance results presented here demonstrate that an acceleration of up to 380 times when the length of vector, $n$, is approximately the number of rays per pixel in the ray tracing. More significantly, the results presented in Figure \ref{fig:vector_addition_performance} suggest that the ray tracing would gain constant time complexity. If constant time complexity holds true, up to values of approximately \num{1e7} rays per pixel, even the most complex current ray tracing simulations will take the same amount of time on a single desktop GPU regardless of the number of rays, complexity of scattering distribution or sample geometry.
\end{flushleft}

\section{Conclusion \& Further Work}
The combination of the ray tracing being identified as exceptionally suitable for parallelisation, the highly optimised mathematical CUDA libraries and moving the parallelisation from a high to low level programming language presents a strong case for the adaptation of the ray tracing to GPU parallelisation. In conjunction with the preliminary performance testing that yielded a reduction in the order of the total time complexity of two analogous algorithms, the work concludes that leveraging the parallel computing power of graphics cards is an important step in enabling previously unfeasible simulation using high resolution sample geometries, complex scattering distributions and high-throughput 3D image reconstruction.
\\[12pt]
Due to the fundamentally different approach in coding methodology required to extract maximal performance increase from GPU parallelisation, I would recommend for the ray tracing to be re-built entirely using CUDA instead of adapting the current version. A full rebuild would be both easier and quicker than making the required syntactic, technical and stylistic changes to the CPU version. Additionally, an entirely CUDA-based approach makes it trivial to tightly integrate new libraries or features that are released in regular updates to the CUDA framework. 
\\[12pt]
Overall, GPU parallelisation of the ray tracing would significantly advance the capability of the ray tracing to simulate high polygon density samples and pinhole plates, alongside investigation of complex scattering distributions and the limitations of heliometric stereo.

\newpage


\chapter{Multiple Detector SHeM}
\label{section:hardware3D}
\section{The \textbeta-SHeM Design}
In Chapter \ref{section:exp_heliometric} a single detector implementation of heliometric stereo was realised, here a new instrument with multiple detectors, designed with 3D surface reconstruction in mind, is detailed. The work presented in Chapter \ref{section:polymer_plates} was critical in the development of the new instrument because it enabled the fabrication of pinhole plates with complex internal geometry. The new microscope, \textbeta-SHeM, was designed by the group and constructed during the work, and has the capacity for multiple physical detectors. \textbeta-SHeM has been designed as the next generation development platform for SHeM that makes significant advances across in comparison to \textalpha-SHeM and is the basis for the ongoing development of the tool. The presented work details the design of the new instrument and how the new features expand upon \textalpha-SHeM's imaging capabilities.
\begin{figure}[H]
    \centering
    \begin{subfigure}{0.9\textwidth}
    \includegraphics[width=\textwidth]{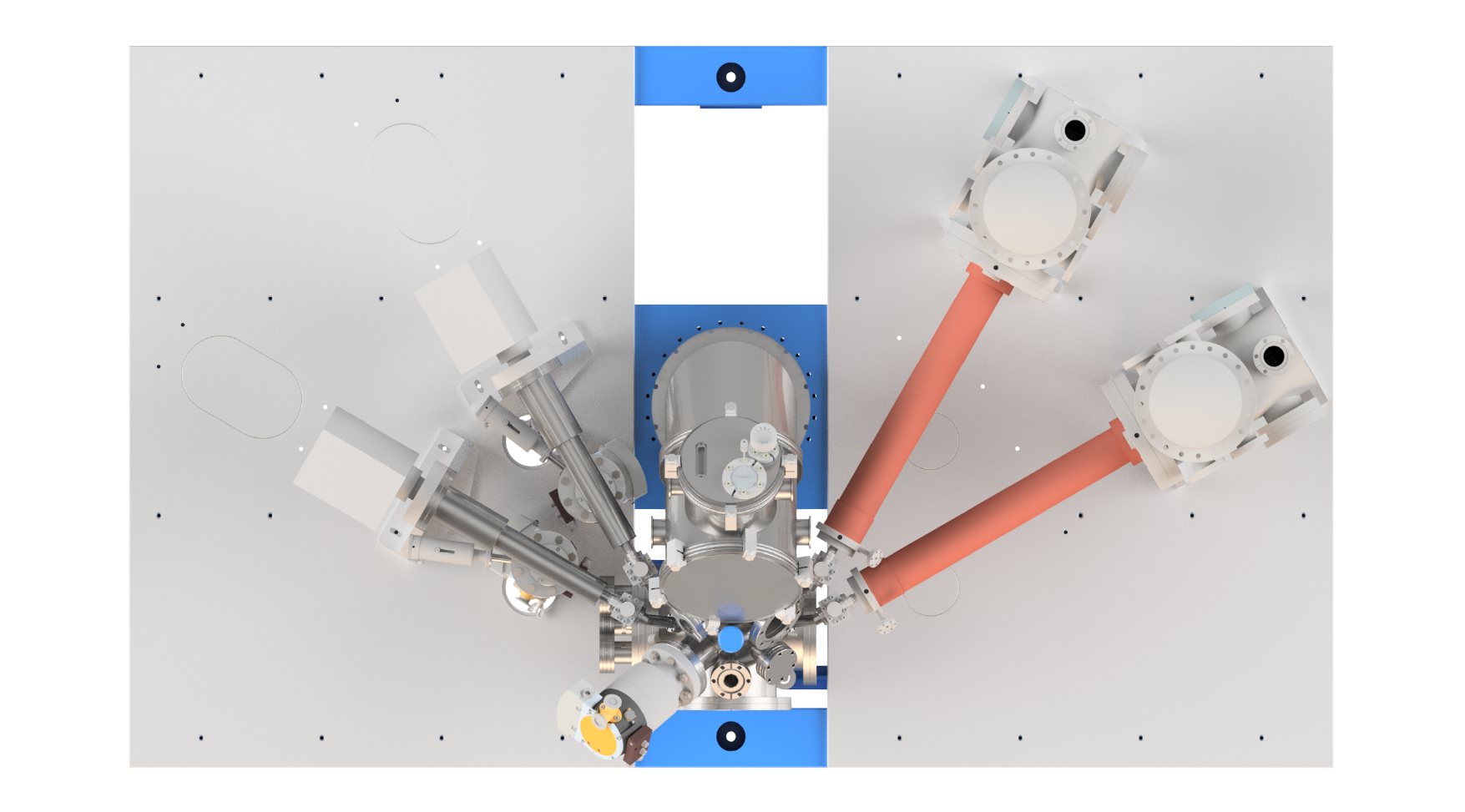}
    \subcaption[a]{Top-down view of \textbeta-SHeM.}
    \label{fig:bshem_overview_a}
    \end{subfigure}
    \begin{subfigure}{0.9\textwidth}
    \includegraphics[width=\textwidth]{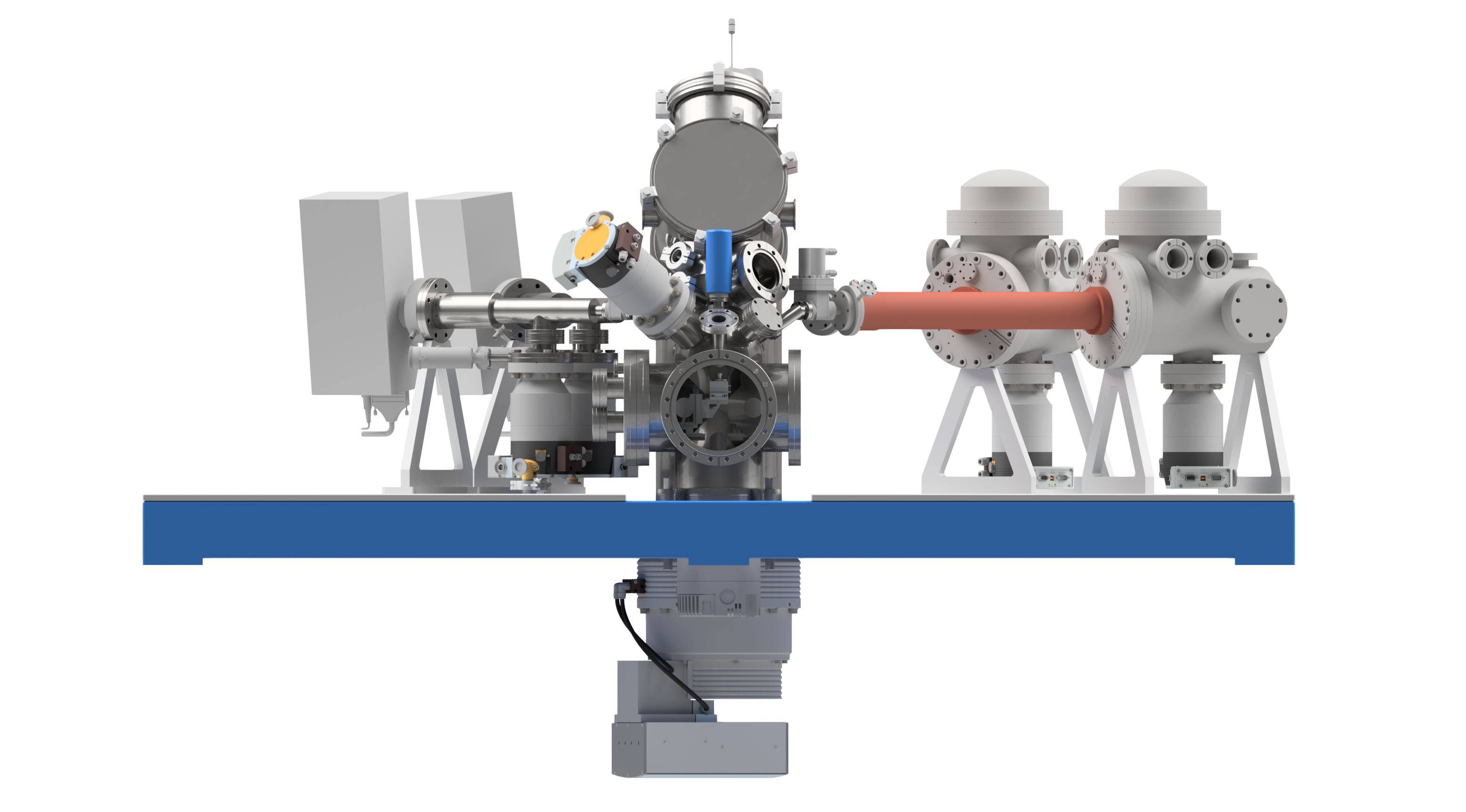}
    \subcaption[b]{Front elevation of \textbeta-SHeM.}
    \label{fig:bshem_overview_b}
    \end{subfigure}
    \caption{Rendered overviews of \textbeta-SHeM. The configuration shown has two quadrupole mass spectrometers (left) and two next generation solenoid detectors that are improvements on the detector design in \textalpha-SHeM. The source mounts into the top-most flange of Panel \ref{fig:bshem_overview_a} with the sample chamber visible in the centre of Panel \ref{fig:bshem_overview_b} with a glass window.}
    \label{fig:bshem_overviews}
\end{figure}
\newpage
\begin{figure}[H]
    \centering
    \begin{subfigure}{0.9\textwidth}
    \includegraphics[width=\textwidth]{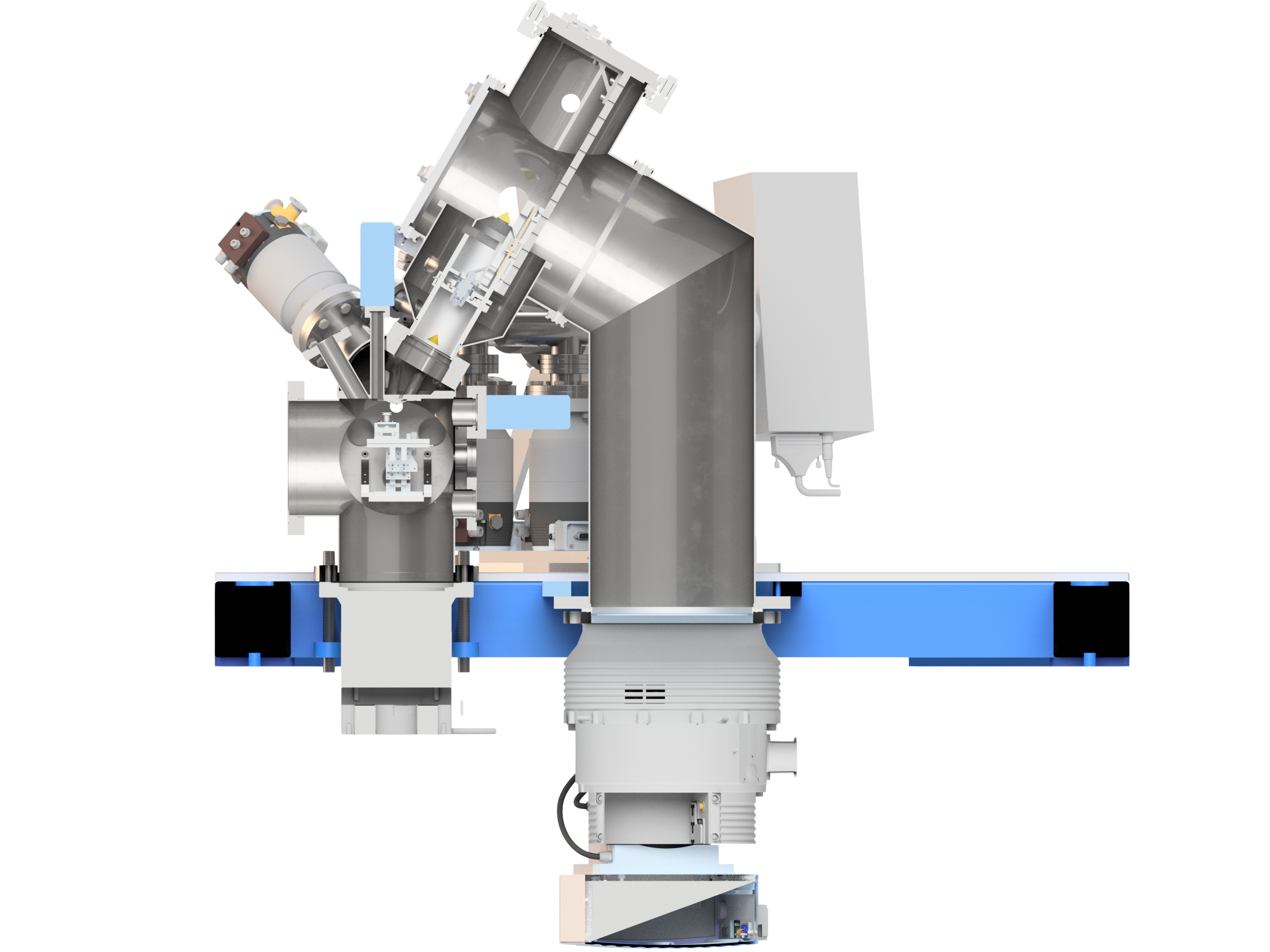}
    \subcaption[a]{Side elevation cross-sectional view of \textbeta-SHeM.}
    \label{fig:bshem_side_a}
    \end{subfigure}
    \begin{subfigure}{0.9\textwidth}
    \includegraphics[width=\textwidth]{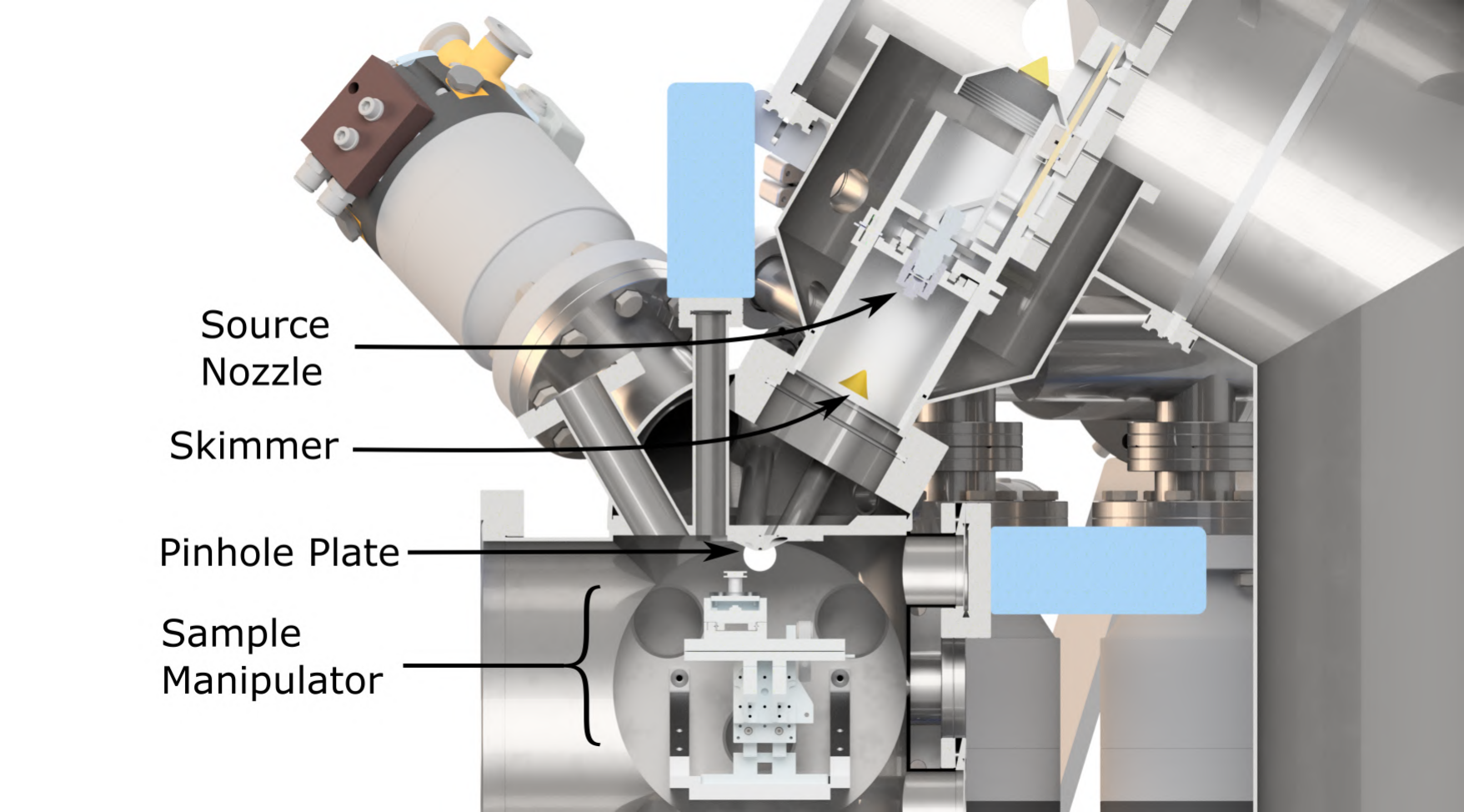}
    \subcaption[b]{Close-up cross-sectional view of sample and source chambers.}
    \label{fig:bshem_side_b}
    \end{subfigure}
    \caption{Side elevation cross-sectional view of \textbeta-SHeM showing the source in the upper-right. The highlighted blue regions are ports where optical microscopes/cameras can be installed.}
    \label{fig:bshem_side}
\end{figure}
\newpage
\begin{flushleft}
Panel \ref{fig:bshem_overview_a} shows the four physical detectors that can be installed on the microscope. Having multiple physical detectors eliminates the need for the azimuthal rotations that were required for the single detector implementation of heliometric stereo, hence \textbeta-SHeM has the capacity for native 3D surface reconstruction. The new instrument can image from up to four azimuthal angles simultaneously, with concurrent 3D reconstruction being a realistic long-term outcome. Due to the versatility of the pinhole plate geometry, using the method detailed in Chapter \ref{section:polymer_plates}, the detector apertures can be arranged in practically any angular configuration relative to each other. Therefore, forward and back-scattered helium can be measured simultaneously so that the scattering distribution can be investigated completely. The current microscope is fitted with two quadrupole mass spectrometers with plans to build four custom solenoid detectors akin to the one fitted to \textalpha-SHeM. Although less sensitive than a custom solenoid detector, the quadrupole mass spectrometers can detect a range of species, making mixed gas beams possible. By mixing a heavier species such as argon in small concentrations with the lighter helium the heavier species becomes super-thermal because it is accelerated to the same mean velocity as the lighter species \cite{1988Aamb}. Therefore, mixed gas beams allow for measurement of the effect of momentum on beam interactions.
\end{flushleft}

\begin{figure}[H]

\begin{subfigure}{0.78\textwidth}
\includegraphics[width=\textwidth]{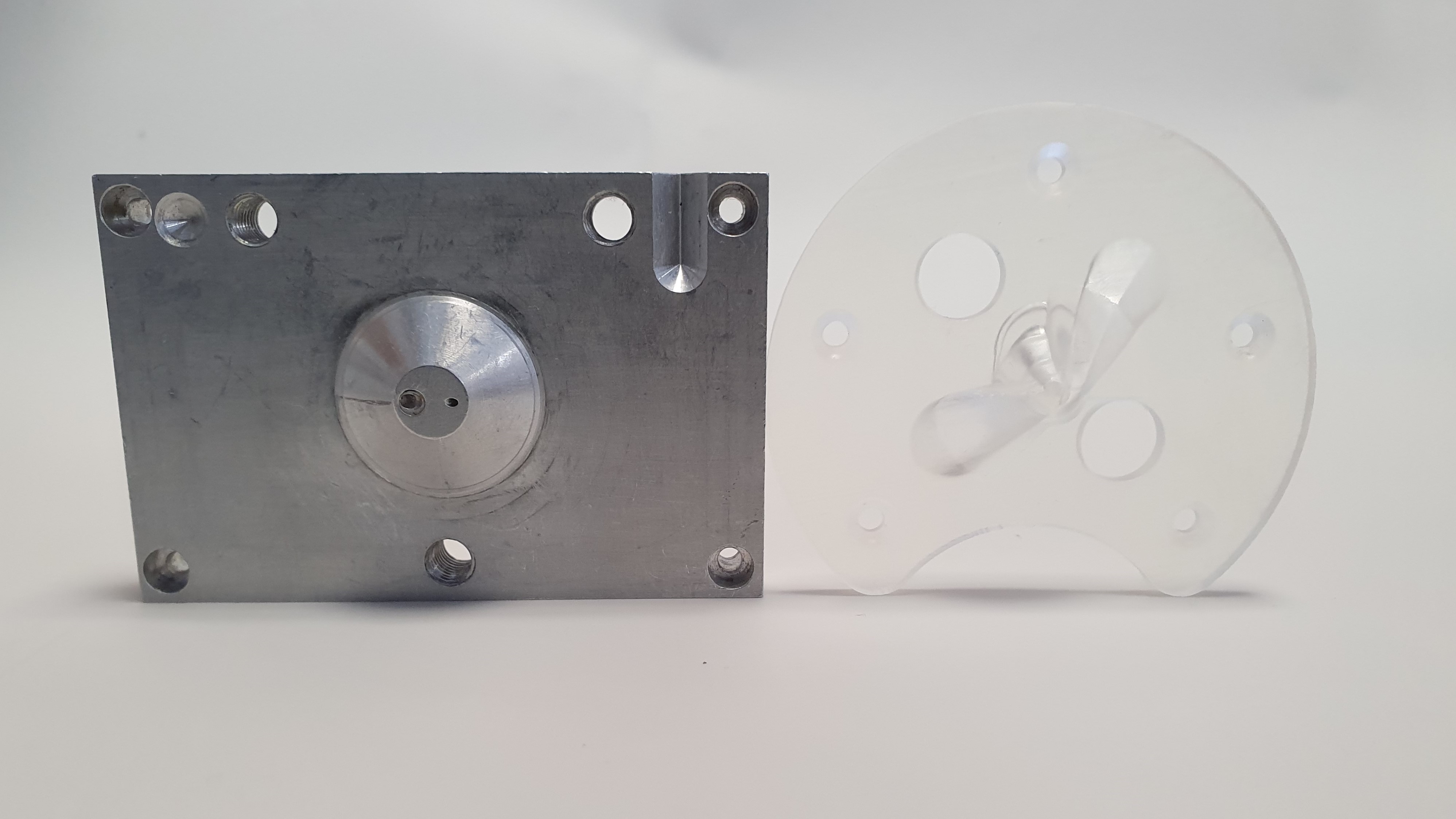}
\end{subfigure}
\begin{subfigure}{0.78\textwidth}
\includegraphics[width=\textwidth]{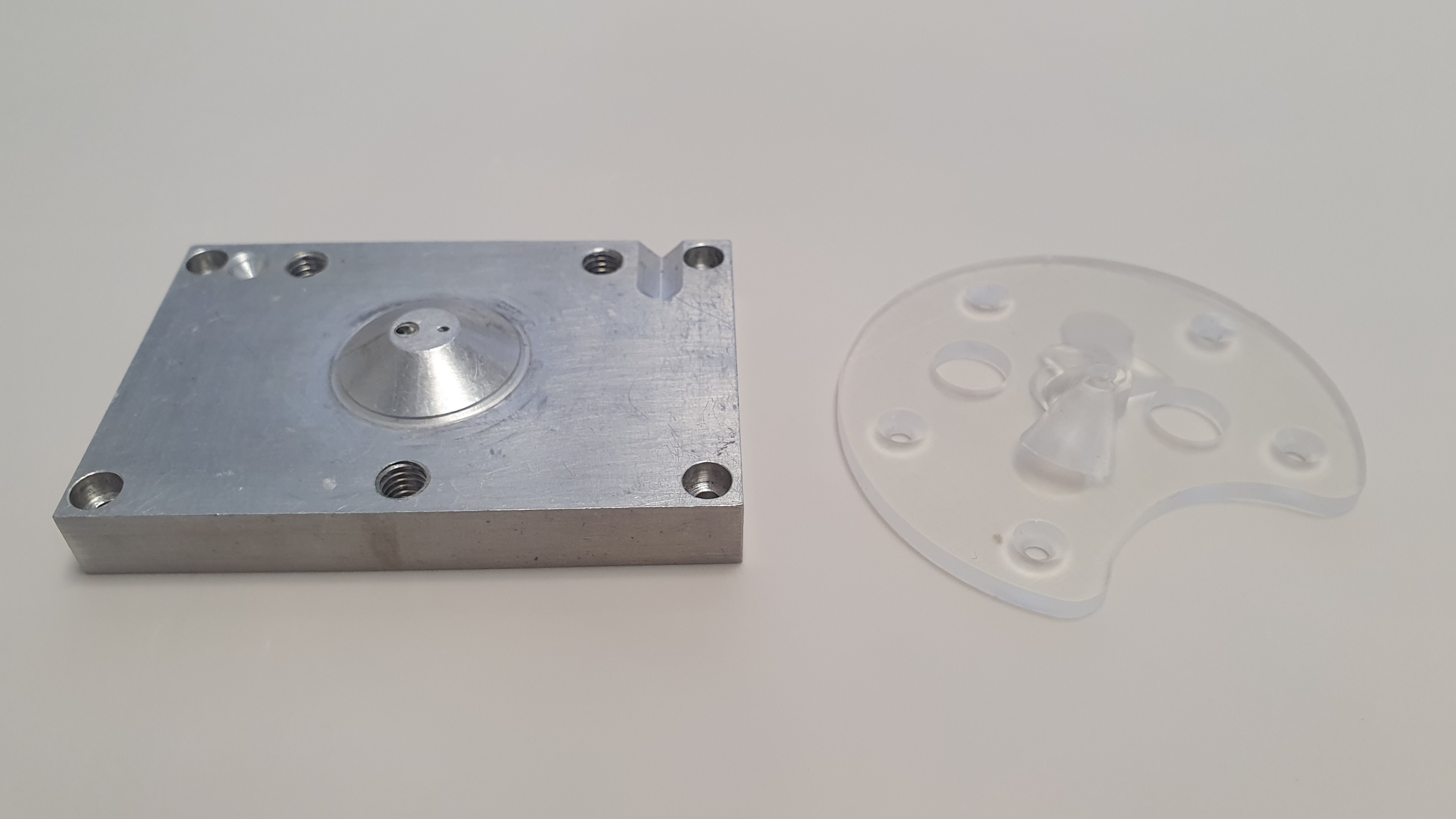}
\end{subfigure}
\begin{subfigure}{0.78\textwidth}
\includegraphics[width=\textwidth]{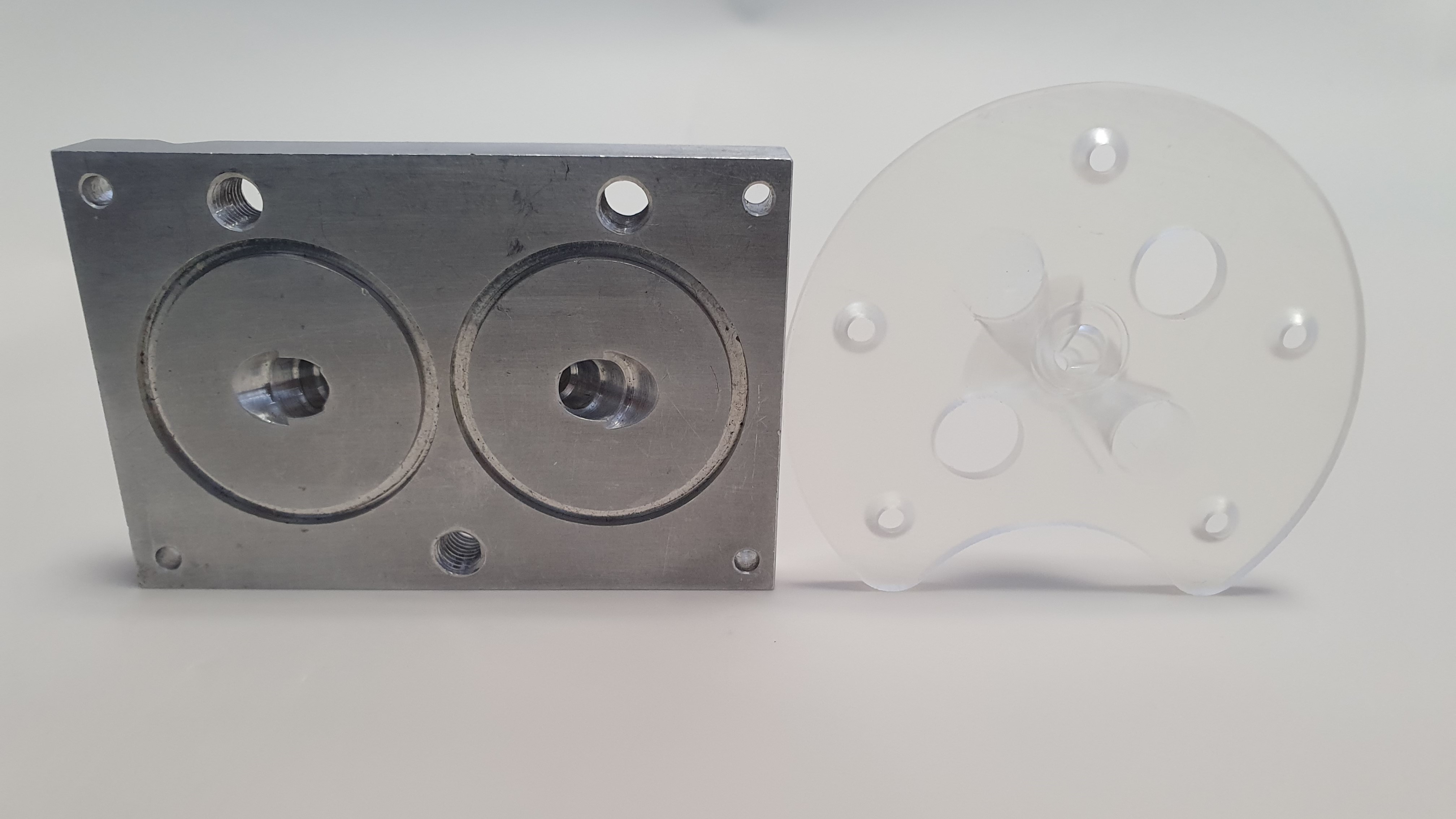}
\end{subfigure}
\caption{Images of \textalpha-SHeM pinhole plate (left) next to \textbeta-SHeM pinhole plate (right). The complex internal geometry was facilitated by work from Chapter \ref{section:polymer_plates}.}
\label{fig:bshem_plates}
\end{figure}

\begin{flushleft}
Panel \ref{fig:bshem_overview_b} gives a clear view of the sample chamber fitted with a glass window. The new sample chamber has been designed with future expansion in mind. In Figure \ref{fig:bshem_side_b} flanges intended for mounting cryogenic cooling to be installed have been included. There are mounting positions for optical microscopes \& other tooling, and sample transfer for in-vacuo cryo-cooled samples. The versatility offered by the new sample environment enables a range of novel measurements using SHeM that were previously impossible, e.g. water ice. The ability to grow and maintain water ice in-situ avoids detector instability due to water degassing, a property of the custom high sensitivity solenoid detectors. In the current detector arrangement which uses Hiden mass spectrometers, such instability is not a problem. Additionally, the sample manipulator has been overhauled to allow horizontal sample mounting, enabling the imaging of liquids for the first time, and simplified pinhole plate and sample change procedures. Previously, the sample manipulator was mounted to the pinhole plate directly meaning a change of pinhole plate required the entire manipulator with the sample to be removed. The sample stages themselves now have longer travel with the option for other in-situ techniques to be installed. The new software environment, in MATLAB and Python, that controls the stages was also developed during the work.
\end{flushleft}
The next generation of SHeM is also more compact than its predecessor, owing to its vertical beam source, making it easier to maintain. Ease-of-use and maintenance has been further improved with more simple pumping and vacuum systems. 
Figure \ref{fig:bshem_side} shows how the source mounts vertically into the microscope, controlled by a 3-axis digital manipulator, visible in Panel \ref{fig:bshem_side_b}. The stages of both the source and sample manipulators have been configured for less cross-talk between their channels to improve stability and accuracy. The potential for a temperature controlled beam nozzle also makes variable beam energy experiments a possibility. The new source is both brighter and easier to configure, and offers a smaller virtual source to demagnify. The beam continues through the skimmer and the pinhole plate, shown in Figure \ref{fig:bshem_plates}, mounted to the top of the sample chamber. The new source configuration also has a smaller beam path which offers higher signal. The distances between nozzle, skimmer and pinhole plate are now more versatile than previously, allowing for further control over the source demagnification. In Chapter \ref{section:exp_heliometric}, images with the default \ang{45} geometry and normal incidence were shown in Figure \ref{fig:45_90_spheres}. \textbeta-SHeM has a beam incident at \ang{30} to the sample surface normal. By installing a goniometer stage, one can also achieve any arbitrary angle of incidence, namely normal incidence in the context of heliometric stereo.

\begin{figure}[ht]
    \centering
    \includegraphics[width=0.7\textwidth]{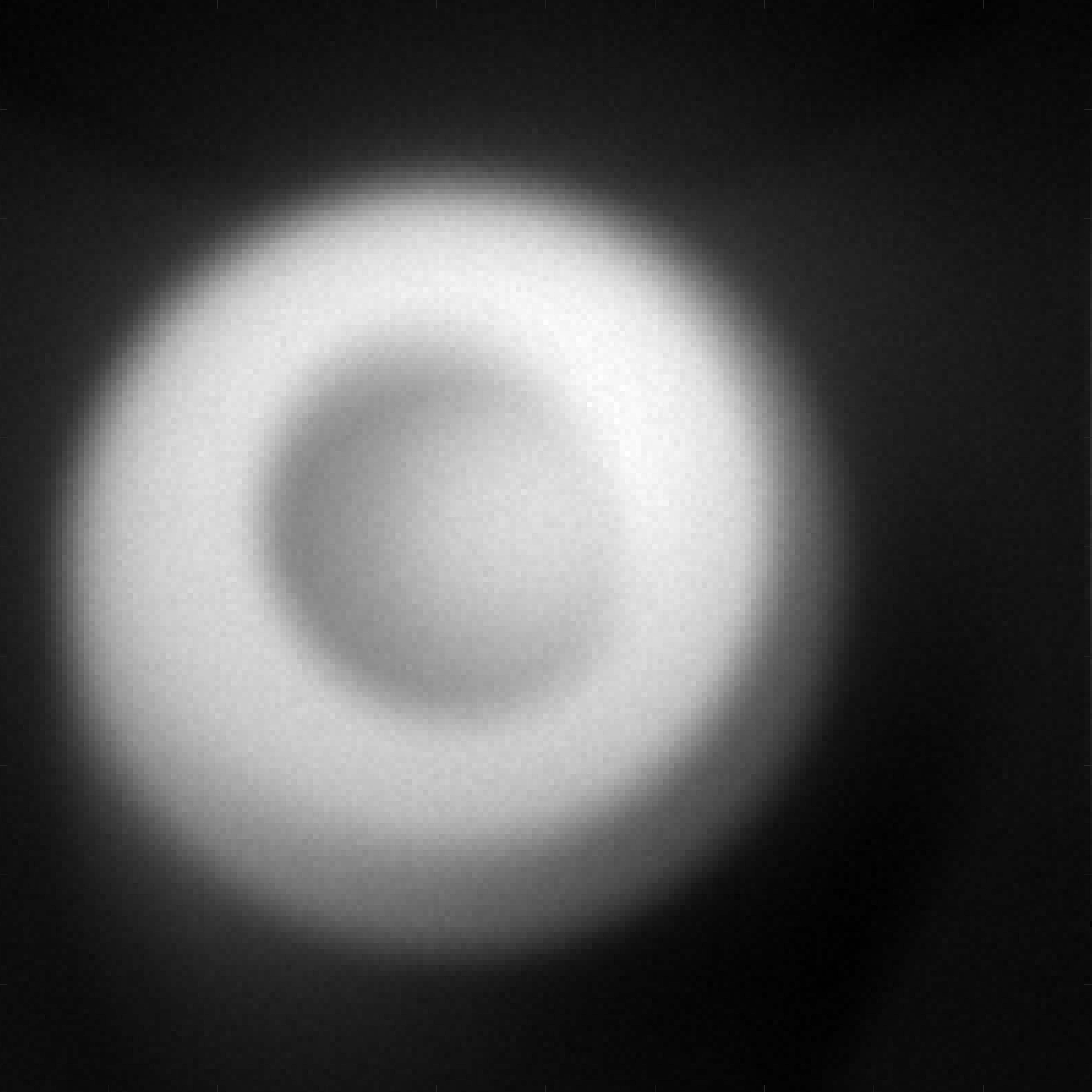}
    \caption{First image taken using \textbeta-SHeM using the plastic pinhole plate shown in Figure \ref{fig:bshem_plates} with no aperture mounted. Image is of the sample mount with no stub or sample fitted.}
    \label{fig:bshem_first}
\end{figure}

\section{Conclusion \& Further Work}
The next generation SHeM platform developed here presents a multitude of new imaging capabilities because of its sophisticated and versatile sample environment to multiple detectors. In the short term there are numerous novel experiments of interest for the new microscope, native heliometric stereo and mixed-species beams being immediately possible. With the addition of in-vacuo sample transfer and cryogenic stage, exciting spatial and diffractive measurements of a new class of samples becomes possible, one particularly interesting candidate is water ice. In the long term, the completion of the microscope with four custom solenoid detectors and zone plate beam focusing \cite{Eder2012}\cite{SalvadorPalau2017}, which is currently in development, in the pinhole plate will allow for down to $10\,nm$ resolution \cite{SalvadorPalau2017} heliometric stereo. $10\,nm$ is well beyond the typical diffraction limits, $\approx200\,nm$, for visible wavelength optical microscopes.
\newpage

\chapter{Conclusion \& Outlook}
\section{Conclusion}
\label{section:conclusion}
In Chapter \ref{section:enhancements_heliometric} a real-space point tracking method was developed to enable true-to-size 3D surface reconstruction using the single detector \textalpha-SHeM. The method has also been shown to be applicable to mosaic spread measurement of multi-faceted polycrystalline materials, a novel imaging mode for the SHeM that could be applied to a range of technological samples, biological crystals and synthetic substrates. The point tracking routine has become a standard operating procedure in general SHeM imaging and the analytical model developed has been used tens of times to calibrate and debug the \textalpha-SHeM sample manipulator. The work then investigated the pumping speeds in complex pinhole plate geometry using CERN's Monte-Carlo based Molflow software. It was determined that the extended source cone in normal incidence pinhole plates, needed for heliometric stereo, would provide sufficient pumping and beam propagation would be unaffected.
\\[12pt]
Chapter \ref{section:polymer_plates} explored the fabrication of complex pinhole plate geometries, where the low-HV to UHV vacuum compatibility of SLA printed FormLabs Clear Resin was investigated and validated. The degassing properties of Clear Resin were investigated and it was found that the primary vacuum contaminant is water, with no evidence to suggest that hydrocarbons evolve from the material. A preparation and baking protocol were developed to achieve the stated vacuum properties. With all pinhole plates now being printed in plastic the protocol has become the standard operating procedure. The work presented also provides strong evidence to suggest that re-wetting under atmospheric conditions only permits the re-absorption of water into the surface layers of the plastic and not into the bulk, on a timescale up to two weeks. Overall, the combination of printing technique and plastic proved to be an excellent choice for cheap and rapid prototyping of small vacuum components whilst allowing fabrication of complex internal structures that were previously near-impossible with traditional machining techniques. 
\\[12pt]
The culmination of the work presented in Chapters \ref{section:enhancements_heliometric} and \ref{section:polymer_plates} enabled the experimental realisation of heliometric stereo using a single detector SHeM. Chapter \ref{section:exp_heliometric} details the successful experimental implementation of heliometric stereo, showing true-to-size accuracy to within 10\%.
\\[12pt]
Motivated by the need to simulate the large data sets for heliometric stereo, the work concludes by evaluating the applicability of GPU parallelisation to the group's in-house ray tracing simulation using Nvidia's CUDA API. The potential speed increase to the ray tracing was investigated using an two analogous problems. The first problem showed a performance increase of 380 times, but also suggested a more significant result in the reduction of the time complexity of the problem from quadratic to linear. The second problem was then designed to have linear time complexity, like the ray tracing itself, when executed serially. Implementing the algorithm on the GPU reduced the time complexity to be constant, confirming that, for ideal problems, the total time complexity is reduced. Changing the time complexity of the problem holds more significance than linear performance improvements because the problem will scale more favourably, so the factor by which the performance improved only increase with the complexity of the simulation. At a critical simulation complexity the GPU execution regresses to the higher order time complexity due to insufficient resources, allowing accurate prediction of the computing resources needed to run a simulation and retain linear time complexity. Resource prediction becomes particularly important when executing the ray tracing on HPC systems.
\\[12pt]
The next generation multiple detector iteration of SHeM, \textbeta-SHeM, which was assembled and tested during the course of the current work, makes significant advances in imaging capabilities over \textalpha-SHeM which was discussed. The new microscope provides a platform for native heliometric stereo and the possibility for completely new classes of samples due to its versatile sample environment. Further than novel imaging capabilities, \textbeta-SHeM allows for the investigation of previously inaccessible core surface science investigations such as the effect of beam momentum on contrast and full scattering distributions simultaneously.
\\[12pt]
The presented work has made significant progress in establishing SHeM as a microscopy tool applicable to a wide range of scientific areas that will undoubtedly underpin fascinating new discoveries.


\section{Outlook}
\label{section:outlook}
The point tracking work presented in Chapter \ref{section:enhancements_heliometric} can be immediately expanded upon by experimentally testing the facet angle measurement routine on a fabricated sample with precisely known geometry so that the accuracy of the method can be quantified. Furthermore, a high resolution, FWHM $\approx 350\,nm$ pinhole plate can be used to provide tighter bounds on the precision of the method.  In addition, the larger data set collected during measurement of multiple facets would give statistically significant results using the further statistical analysis detailed in Chapter \ref{subsection:further_stats}, allowing deeper insight into the nature of \textalpha-SHeM sample manipulator's errors. In the longer term the point tracking should be fully automated to enable SHeM to function as a rapid mosaic spread measuring instrument specialising in materials which would be damaged or chemically altered using XRD rocking curves \cite{Stuckelberger2020}.
\\[12pt]
An important piece of work to conclude that presented in Section \ref{section:polymer_plates} would be to investigate the cause of the delamination observed during the \ang{170}C bake. Exploring slower rates of heating/cooling during the baking process along with a range of peak temperatures starting from \ang{120}C, which is known to be safe, would identify the source of the delamination. In the case that a reasonable high temperature baking procedure cannot be designed for Clear Resin, the experiment could be expanded to other plastics available with the same FormLabs printing technique. The work on characterisation of SLA printed plastics in vacuum could conclude with ToF-SIMS measurements to generate chemical depth profiles of the material \cite{Bailey2015} \cite{Alberici2004}, before and after baking. Such an investigation would be able to confirm that the nature of re-wetting due to exposure to atmosphere is exclusively a surface process.
\\[12pt]
In the short term, the addition of a heavy gas species into the beam, e.g. argon, with helium-4 would allow for the investigation of beam momentum on scattering in \textbeta-SHeM. The long term outlook involves the installation of additional sample environment stages so that cryogenic and in-situ grown samples can be imaged.
\\[12pt]
Building on the preliminary performance results from Chapter \ref{section:cuda} a full implementation of GPU parallelisation should be realised so that ray tracing can be executed efficiently on both desktop and HPC systems.
\newpage

\begin{appendices}
\chapter{Molflow}
\label{appendix:molflow}
\begin{figure}
    \centering
    \includegraphics[width=0.8\textwidth]{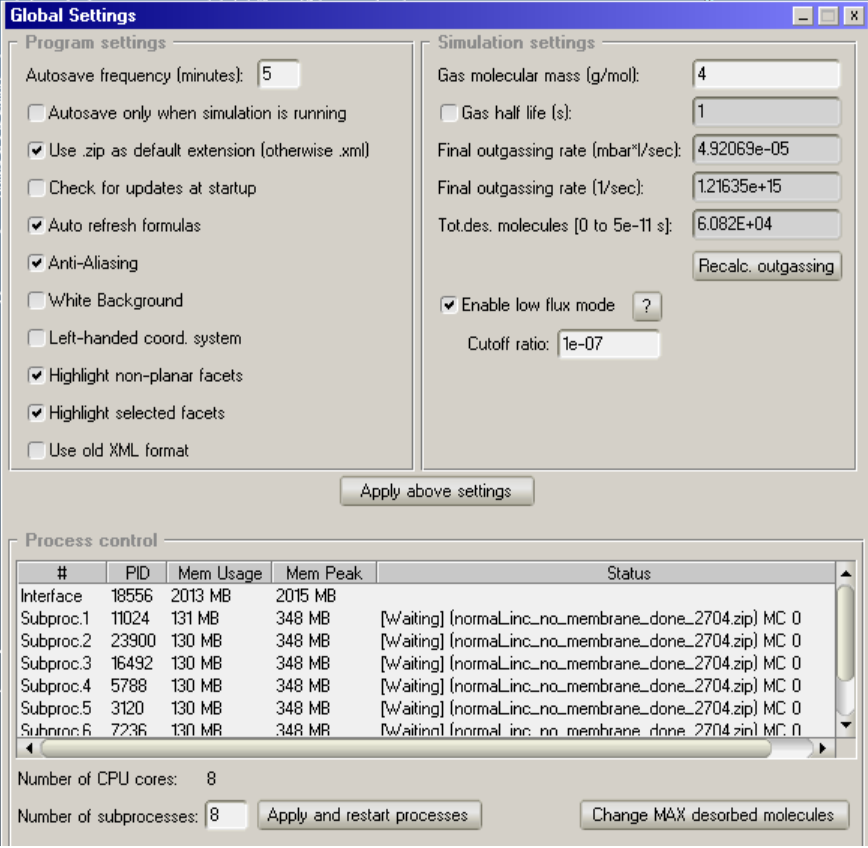}
    \caption{Detailed Molflow parameters used in source cone simulations. An important feature of Molflow is that the molecular mass of the gas being investigated must be specified, the assigned mass is then the mass for all test-particles used in the simulation, therefore calculating a partial pressure.}
    \label{fig:molflow_a}
\end{figure}
\begin{figure}
    \centering
    \includegraphics[width=0.8\textwidth]{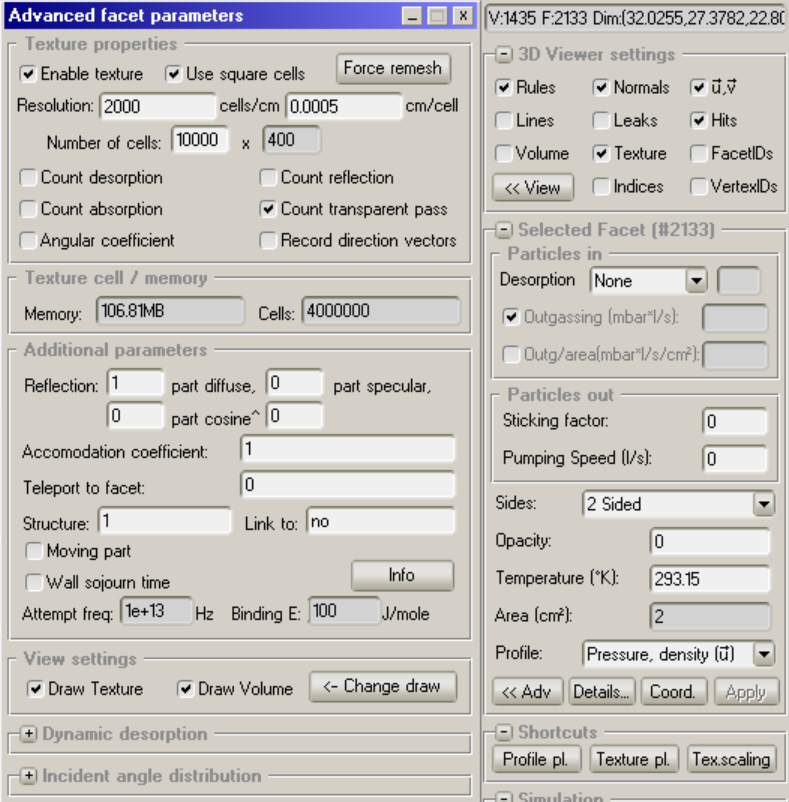}
    \caption{Parameters of transparent facet which records pressure profile down source cone.}
    \label{fig:molflow_b}
\end{figure}

\chapter{Least-Squares Ellipse Fitting Algorithm}
\label{appendix:least_squares_ellipse}
\begin{minted}[
frame=lines,
framesep=2mm,
baselinestretch=0.8,
fontsize=\footnotesize,
linenos
]{matlab}
function ellipse_t = fit_ellipse( x,y,axis_handle )
%https://uk.mathworks.com/matlabcentral/fileexchange/3215-fit_ellipse


% fit_ellipse - finds the best fit to an ellipse for the given set of points.
%
% Format:   ellipse_t = fit_ellipse( x,y,axis_handle )
%
% Input:    x,y         - a set of points in 2 column vectors. AT LEAST 5 points are needed !
%           axis_handle - optional. a handle to an axis, at which the estimated ellipse 
%                         will be drawn along with it's axes
%
% Output:   ellipse_t - structure that defines the best fit to an ellipse
%                       a           - sub axis (radius) of the X axis of the non-tilt ellipse
%                       b           - sub axis (radius) of the Y axis of the non-tilt ellipse
%                       phi         - orientation in radians of the ellipse (tilt)
%                       X0          - center at the X axis of the non-tilt ellipse
%                       Y0          - center at the Y axis of the non-tilt ellipse
%                       X0_in       - center at the X axis of the tilted ellipse
%                       Y0_in       - center at the Y axis of the tilted ellipse
%                       long_axis   - size of the long axis of the ellipse
%                       short_axis  - size of the short axis of the ellipse
%                       status      - status of detection of an ellipse
%
% Note:     if an ellipse was not detected (but a parabola or hyperbola), then
%           an empty structure is returned
% =====================================================================================
%                  Ellipse Fit using Least Squares criterion
% =====================================================================================
% We will try to fit the best ellipse to the given measurements. the mathematical
% representation of use will be the CONIC Equation of the Ellipse which is:
% 
%    Ellipse = a*x^2 + b*x*y + c*y^2 + d*x + e*y + f = 0
%   
% The fit-estimation method of use is the Least Squares method (without any weights)
% The estimator is extracted from the following equations:
%
%    g(x,y;A) := a*x^2 + b*x*y + c*y^2 + d*x + e*y = f
%
%    where:
%       A   - is the vector of parameters to be estimated (a,b,c,d,e)
%       x,y - is a single measurement
%
% We will define the cost function to be:
%
%   Cost(A) := (g_c(x_c,y_c;A)-f_c)'*(g_c(x_c,y_c;A)-f_c)
%            = (X*A+f_c)'*(X*A+f_c) 
%            = A'*X'*X*A + 2*f_c'*X*A + N*f^2
%
%   where:
%       g_c(x_c,y_c;A) - vector function of ALL the measurements
%                        each element of g_c() is g(x,y;A)
%       X              - a matrix of the form: [x_c.^2, x_c.*y_c, y_c.^2, x_c, y_c ]
%       f_c            - is actually defined as ones(length(f),1)*f
%
% Derivation of the Cost function with respect to the vector of parameters "A" yields:
%
%   A'*X'*X = -f_c'*X = -f*ones(1,length(f_c))*X = -f*sum(X)
%
% Which yields the estimator:
%
%       ~~~~~~~~~~~~~~~~~~~~~~~~~~~~~~~~~~~~~~~~~~~~~~~~~~~~~~~~~~~~~~~~~~~~~~~~~~~~
%       |  A_least_squares = -f*sum(X)/(X'*X) ->(normalize by -f) = sum(X)/(X'*X)  |
%       ~~~~~~~~~~~~~~~~~~~~~~~~~~~~~~~~~~~~~~~~~~~~~~~~~~~~~~~~~~~~~~~~~~~~~~~~~~~~
%
% (We will normalize the variables by (-f) since "f" is unknown and can be accounted for later on)
%  
% NOW, all that is left to do is to extract the parameters from the Conic Equation.
% We will deal the vector A into the variables: (A,B,C,D,E) and assume F = -1;
%
%    Recall the conic representation of an ellipse:
% 
%       A*x^2 + B*x*y + C*y^2 + D*x + E*y + F = 0
% 
% We will check if the ellipse has a tilt (=orientation). The orientation is present
% if the coefficient of the term "x*y" is not zero. If so, we first need to remove the
% tilt of the ellipse.
%
% If the parameter "B" is not equal to zero, then we have an orientation (tilt) to the ellipse.
% we will remove the tilt of the ellipse so as to remain with a conic representation of an 
% ellipse without a tilt, for which the math is more simple:
%
% Non tilt conic rep.:  A`*x^2 + C`*y^2 + D`*x + E`*y + F` = 0
%
% We will remove the orientation using the following substitution:
%   
%   Replace x with cx+sy and y with -sx+cy such that the conic representation is:
%   
%   A(cx+sy)^2 + B(cx+sy)(-sx+cy) + C(-sx+cy)^2 + D(cx+sy) + E(-sx+cy) + F = 0
%
%   where:      c = cos(phi)    ,   s = sin(phi)
%
%   and simplify...
%
%       x^2(A*c^2 - Bcs + Cs^2) + xy(2A*cs +(c^2-s^2)B -2Ccs) + ...
%           y^2(As^2 + Bcs + Cc^2) + x(Dc-Es) + y(Ds+Ec) + F = 0
%
%   The orientation is easily found by the condition of (B_new=0) which results in:
% 
%   2A*cs +(c^2-s^2)B -2Ccs = 0  ==> phi = 1/2 * atan( b/(c-a) )
%   
%   Now the constants   c=cos(phi)  and  s=sin(phi)  can be found, and from them
%   all the other constants A`,C`,D`,E` can be found.
%
%   A` = A*c^2 - B*c*s + C*s^2                  D` = D*c-E*s
%   B` = 2*A*c*s +(c^2-s^2)*B -2*C*c*s = 0      E` = D*s+E*c 
%   C` = A*s^2 + B*c*s + C*c^2
%
% Next, we want the representation of the non-tilted ellipse to be as:
%
%       Ellipse = ( (X-X0)/a )^2 + ( (Y-Y0)/b )^2 = 1
%
%       where:  (X0,Y0) is the center of the ellipse
%               a,b     are the ellipse "radiuses" (or sub-axis)
%
% Using a square completion method we will define:
%       
%       F`` = -F` + (D`^2)/(4*A`) + (E`^2)/(4*C`)
%
%       Such that:    a`*(X-X0)^2 = A`(X^2 + X*D`/A` + (D`/(2*A`))^2 )
%                     c`*(Y-Y0)^2 = C`(Y^2 + Y*E`/C` + (E`/(2*C`))^2 )
%
%       which yields the transformations:
%       
%           X0  =   -D`/(2*A`)
%           Y0  =   -E`/(2*C`)
%           a   =   sqrt( abs( F``/A` ) )
%           b   =   sqrt( abs( F``/C` ) )
%
% And finally we can define the remaining parameters:
%
%   long_axis   = 2 * max( a,b )
%   short_axis  = 2 * min( a,b )
%   Orientation = phi
%
%
% initialize
orientation_tolerance = 1e-3;
% empty warning stack
warning( '' );
% prepare vectors, must be column vectors
x = x(:);
y = y(:);
% remove bias of the ellipse - to make matrix inversion more accurate. (will be added later on).
mean_x = mean(x);
mean_y = mean(y);
x = x-mean_x;
y = y-mean_y;
% the estimation for the conic equation of the ellipse
X = [x.^2, x.*y, y.^2, x, y ];
a = sum(X)/(X'*X);
% check for warnings
if ~isempty( lastwarn )
    disp( 'stopped because of a warning regarding matrix inversion' );
    ellipse_t = [];
    return
end
% extract parameters from the conic equation
[a,b,c,d,e] = deal( a(1),a(2),a(3),a(4),a(5) );
% remove the orientation from the ellipse
if ( min(abs(b/a),abs(b/c)) > orientation_tolerance )
    
    orientation_rad = 1/2 * atan( b/(c-a) );
    cos_phi = cos( orientation_rad );
    sin_phi = sin( orientation_rad );
    [a,b,c,d,e] = deal(...
        a*cos_phi^2 - b*cos_phi*sin_phi + c*sin_phi^2,...
        0,...
        a*sin_phi^2 + b*cos_phi*sin_phi + c*cos_phi^2,...
        d*cos_phi - e*sin_phi,...
        d*sin_phi + e*cos_phi );
    [mean_x,mean_y] = deal( ...
        cos_phi*mean_x - sin_phi*mean_y,...
        sin_phi*mean_x + cos_phi*mean_y );
else
    orientation_rad = 0;
    cos_phi = cos( orientation_rad );
    sin_phi = sin( orientation_rad );
end
% check if conic equation represents an ellipse
test = a*c;
switch (1)
case (test>0),  status = '';
case (test==0), status = 'Parabola found';  warning( 'fit_ellipse: Did not locate an ellipse' );
case (test<0),  status = 'Hyperbola found'; warning( 'fit_ellipse: Did not locate an ellipse' );
end
% if we found an ellipse return it's data
if (test>0)
    
    % make sure coefficients are positive as required
    if (a<0), [a,c,d,e] = deal( -a,-c,-d,-e ); end
    
    % final ellipse parameters
    X0          = mean_x - d/2/a;
    Y0          = mean_y - e/2/c;
    F           = 1 + (d^2)/(4*a) + (e^2)/(4*c);
    [a,b]       = deal( sqrt( F/a ),sqrt( F/c ) );    
    long_axis   = 2*max(a,b);
    short_axis  = 2*min(a,b);
    % rotate the axes backwards to find the center point of the original TILTED ellipse
    R           = [ cos_phi sin_phi; -sin_phi cos_phi ];
    P_in        = R * [X0;Y0];
    X0_in       = P_in(1);
    Y0_in       = P_in(2);
    
    % pack ellipse into a structure
    ellipse_t = struct( ...
        'a',a,...
        'b',b,...
        'k',a/b,...
        'phi',orientation_rad,...
        'X0',X0,...
        'Y0',Y0,...
        'X0_in',X0_in,...
        'Y0_in',Y0_in,...
        'long_axis',long_axis,...
        'short_axis',short_axis,...
        'status','' );
else
    % report an empty structure
    ellipse_t = struct( ...
        'a',[],...
        'b',[],...
        'phi',[],...
        'X0',[],...
        'Y0',[],...
        'X0_in',[],...
        'Y0_in',[],...
        'long_axis',[],...
        'short_axis',[],...
        'status',status );
end
% check if we need to plot an ellipse with it's axes.
if (nargin>2) & ~isempty( axis_handle ) & (test>0)
    
    % rotation matrix to rotate the axes with respect to an angle phi
    R = [ cos_phi sin_phi; -sin_phi cos_phi ];
    
    % the axes
    ver_line        = [ [X0 X0]; Y0+b*[-1 1] ];
    horz_line       = [ X0+a*[-1 1]; [Y0 Y0] ];
    new_ver_line    = R*ver_line;
    new_horz_line   = R*horz_line;
    
    % the ellipse
    theta_r         = linspace(0,2*pi);
    ellipse_x_r     = X0 + a*cos( theta_r );
    ellipse_y_r     = Y0 + b*sin( theta_r );
    rotated_ellipse = R * [ellipse_x_r;ellipse_y_r];
    
    % draw
    hold_state = get( axis_handle,'NextPlot' );
    set( axis_handle,'NextPlot','add' );
    plot( new_ver_line(1,:),new_ver_line(2,:),'r' );
    plot( new_horz_line(1,:),new_horz_line(2,:),'r' );
    plot( rotated_ellipse(1,:),rotated_ellipse(2,:),'r' );
    set( axis_handle,'NextPlot',hold_state );
end


\end{minted}

\newpage
\chapter{Brute Force Point Search Algorithm}
\label{appendix:pointsearch}
Brute force point search algorithm for CPU and GPU:
\begin{minted}[
frame=lines,
framesep=2mm,
baselinestretch=0.8,
fontsize=\footnotesize,
linenos
]{cpp}
#include <stdio.h>
#include <iomanip>
#include <iostream>
#include <ctime>
#include <math.h>

#include "cuda_runtime.h"
#include "device_launch_parameters.h"



using namespace std;

__global__ void FindClosestGPU(float3* points, int* indices, int count)
{
	if (count <= 1) return; //If there's 1 point, do nothing

	int idx = threadIdx.x + blockIdx.x * blockDim.x; //Unique thread ID generation

	if (idx < count)
	{
		float3 thisPoint = points[idx];
		float distToClosest = 3.40282e38f;

		for (int i = 0; i < count; i++)
		{
		    //Avoids checking a point against itself resulting in distance being 0
			if (i == idx) continue; 
			

			float dist = sqrt((thisPoint.x - points[i].x) * (thisPoint.x - points[i].x) + 
			(thisPoint.y - points[i].y) * (thisPoint.y - points[i].y) + 
			(thisPoint.z - points[i].z) * (thisPoint.z - points[i].z));

			if (dist < distToClosest) //Update new closest distance
			{
				distToClosest = dist;
				indices[idx] = i;
			}
		}
	}
}


void FindClosestCPU(float3* points, int* indices, int count)
{
if(count <= 1) return;

for(int curPoint = 0; curPoint < count; curPoint++)
{
	float distToClosest = 3.40282e38f;

	for(int i = 0; i<count; i++)
	{

		float dist = ...
		sqrt((points[curPoint].x - points[i].x)*(points[curPoint].x-points[i].x)+
		(points[curPoint].y - points[i].y)*(points[curPoint].y-points[i].y)+
		(points[curPoint].z - points[i].z)*(points[curPoint].z-points[i].z));

			if (dist < distToClosest) //Update new closest distance
			{
				distToClosest = dist;
				indices[curPoint] = i;
			}
		}
	}
}


int main()
{
	//Number of points
	const int count = 10000;

	//Number of iterations
	const double runs = 5;

	//Array of points (CPU)
	int* indexOfClosest = new int[count];
	float3* points = new float3[count]; //float3 contains x,y,z coordinates

	//Arrays of points (GPU)
	int* d_indexOfClosest;
	float3* d_points;
	


	//Create a list of random points
	for (int i = 0; i < count; i++)
	{
		points[i].x = (float)((rand() % 10000) - 5000);
		points[i].y = (float)((rand() % 10000) - 5000);
		points[i].z = (float)((rand() % 10000) - 5000);
	}

	//GPU memory allocation
	cudaMalloc(&d_points, sizeof(float3) * count);
	cudaMemcpy(d_points, points, sizeof(float3) * count, cudaMemcpyHostToDevice);
	cudaMalloc(&d_indexOfClosest, sizeof(int) * count);

	//Variables used to keep track of calculation times
	long fastestTimegpu = 1000000;
	long totalTimegpu = 0;

	//Setting 4 significant figure output:
	cout << setprecision(4);


	cout << "Nearest Point Search GPU:" << endl;

	//Run GPU algorithm q times
	for (int q = 0; q < runs; q++)
	{

		long startTimegpu = clock();

		//Call search function
		FindClosestGPU<<<count/1024 + 1, 64>>>(d_points, d_indexOfClosest, count);
		cudaMemcpy(indexOfClosest, d_indexOfClosest, sizeof(int) * count, cudaMemcpyDeviceToHost);

		long finishTimegpu = clock();

		cout << "Run " << q << " took " <<
		(finishTimegpu - startTimegpu)*1000/(double)CLOCKS_PER_SEC
		<< " ms" << endl;

		totalTimegpu += (finishTimegpu - startTimegpu);


		//If the time was faster than previous, update the fastest time
		if ((finishTimegpu - startTimegpu) < fastestTimegpu)
		{
			fastestTimegpu = (finishTimegpu - startTimegpu);
		}

	}

	//Deallocate GPU memory
	delete[] indexOfClosest;
	delete[] points;
	cudaFree(d_points);
	cudaFree(d_indexOfClosest);

	cudaDeviceReset();

	
	cout << "Fastest run: " << fastestTimegpu/(double)1000<< " ms" << endl;
	cout << "Average run: " << (totalTimegpu /(double)1000)/runs << " ms" << endl;
	cout << "Total run time: " << totalTimegpu/(double)1000 << " ms" << endl;

	//Reset timing variables
	long fastestTime = 1000000;
	long totalTime = 0;

	cout << "\n\nNearest Point Search CPU:" << endl;

	//Run CPU algorithm q times
	for (int q = 0; q < runs; q++)
	{

		long startTime = clock();

		//Call distance function
		FindClosestCPU(points, indexOfClosest, count);
		long finishTime = clock();

		cout << "Run " << q << " took " << (finishTime - startTime)*1000/(double)CLOCKS_PER_SEC
		<< " ms" << endl;

		totalTime += (finishTime - startTime);


		//If the time was faster than previous, update the fastest time
		if ((finishTime - startTime) < fastestTime)
		{
			fastestTime = (finishTime - startTime);
		}

	}

	
	cout << "Fastest run: " << fastestTime/(double)1000<< " ms" << endl;
	cout << "Average run: " << (totalTime /(double)1000)/runs << " ms" << endl;
	cout << "Total run time: " << totalTime/(double)1000 << " ms\n" << endl;

	cout << "Accelerated by factor " <<
	(totalTime/(double)1000)/(totalTimegpu/(double)1000) << endl;

	return 0;
}
\end{minted}

\newpage
\chapter{Vector Addition Algorithm}
\label{appendix:vector_addition}
Vector addition algorithm showing linear time complexity on the CPU:
\begin{minted}[
frame=lines,
framesep=2mm,
baselinestretch=0.8,
fontsize=\footnotesize,
linenos
]{cpp}
#include "cuda_runtime.h"
#include "device_launch_parameters.h"

#include <stdio.h>
#include <time.h>

#define N 10000000 //(1024*1024) was original value - length of vectors
#define M (1000) //10000 original value - number of iterations of full vector addition
#define THREADS_PER_BLOCK 1024

void serial_add(double *a, double *b, double *c, int n, int m)
{
    for(int index=0;index<n;index++)
    {
        for(int j=0;j<m;j++)
        {
            c[index] = a[index]*a[index] + b[index]*b[index];
        }
    }
}

__global__ void vector_add(double *a, double *b, double *c)
{
    int index = blockIdx.x * blockDim.x + threadIdx.x;
        for(int j=0;j<M;j++)
        {
            c[index] = a[index]*a[index] + b[index]*b[index];
        }
}

int main()
{
    clock_t start,end;

    double *a, *b, *c;
    int size = N * sizeof( double );

    a = (double *)malloc( size );
    b = (double *)malloc( size );
    c = (double *)malloc( size );

    for( int i = 0; i < N; i++ )
    {
        a[i] = b[i] = i;
        c[i] = 0;
    }

    start = clock();
    serial_add(a, b, c, N, M);

    printf( "c[0] = %d\n",0,c[0] );
    printf( "c[%d] = %d\n",N-1, c[N-1] );

    end = clock();

    float time1 = ((float)(end-start))/CLOCKS_PER_SEC;
    printf("Serial: %f seconds\n",time1);

    start = clock();
    double *d_a, *d_b, *d_c;


    cudaMalloc( (void **) &d_a, size );
    cudaMalloc( (void **) &d_b, size );
    cudaMalloc( (void **) &d_c, size );


    cudaMemcpy( d_a, a, size, cudaMemcpyHostToDevice );
    cudaMemcpy( d_b, b, size, cudaMemcpyHostToDevice );

    //static threads/blocks allocation:
    vector_add<<< (4096) / THREADS_PER_BLOCK, THREADS_PER_BLOCK >>>( d_a, d_b, d_c ); 



    cudaMemcpy( c, d_c, size, cudaMemcpyDeviceToHost );


    printf( "c[0] = %d\n",0,c[0] );
    printf( "c[%d] = %d\n",N-1, c[N-1] );


    free(a);
    free(b);
    free(c);
    cudaFree( d_a );
    cudaFree( d_b );
    cudaFree( d_c );

    end = clock();
    float time2 = ((float)(end-start))/CLOCKS_PER_SEC;
    printf("CUDA: %f seconds, Speedup: %f\n",time2, time1/time2);

    return 0;
} 
\end{minted}

\end{appendices}

\bibliographystyle{ieeetr}
\bibliography{ref.bib}

\begin{thebibliography}{10}

\bibitem{VanLeeuwenhoek1677}
A.~van Leeuwenhoek, ``{Observations, communicated to the publisher by Mr. Antony van Leewenhoeck, in a dutch letter of the 9th Octob. 1676. here English'd: concerning little animals by him observed in rain-well-sea- and snow water; as also in water wherein pepper had lain infus},'' {\em Philosophical Transactions of the Royal Society of London}, vol.~12, pp.~821--831, mar 1677.

\bibitem{Lane2015}
N.~Lane, ``{The unseen world: reflections on Leeuwenhoek (1677) ‘Concerning little animals'},'' {\em Philosophical Transactions of the Royal Society B: Biological Sciences}, vol.~370, no.~1666, p.~20140344, 2015.

\bibitem{Misri}
R.~Misri, S.~Pande, and U.~Khopkar, ``{Confocal laser microscope},'' {\em PubMed}, 2006.

\bibitem{Bergin2019}
M.~Bergin, {\em {Instrumentation and contrast mechanisms in scanning helium microscopy}}.
\newblock PhD thesis, University of Cambridge, apr 2019.

\bibitem{Bergin2020}
M.~Bergin, S.~M. Lambrick, H.~Sleath, D.~J. Ward, J.~Ellis, and A.~P. Jardine, ``{Observation of diffraction contrast in scanning helium microscopy},'' {\em Scientific Reports}, vol.~10, dec 2020.

\bibitem{Palau2016}
A.~S. Palau, G.~Bracco, and B.~Holst, ``{Theoretical model of the helium pinhole microscope},'' {\em Physical Review A}, vol.~94, dec 2016.

\bibitem{Bracco2013}
G.~Bracco and B.~Holst, {\em {Surface science techniques}}.
\newblock Springer, 2013.

\bibitem{Farias1998}
D.~Far{\'{i}}as and K.~H. Rieder, ``{Atomic beam diffraction from solid surfaces},'' {\em Reports on Progress in Physics}, vol.~61, no.~12, pp.~1575--1664, 1998.

\bibitem{Joy2009}
D.~C. Joy, R.~Ramachandra, and B.~Griffin, ``{Choosing a beam-electrons,protons, he or ga ions?},'' {\em Microscopy and Microanalysis}, vol.~15, no.~SUPPL. 2, pp.~648--649, 2009.

\bibitem{DaveThesis}
D.~J. Ward, {\em A study of spin-echo lineshapes in helium atom scattering from adsorbates}.
\newblock PhD thesis, University of Cambridge, apr 2013.

\bibitem{Koch2008}
M.~Koch, S.~Rehbein, G.~Schmahl, T.~Reisinger, G.~Bracco, W.~E. Ernst, and B.~Holst, ``{Imaging with neutral atoms - A new matter-wave microscope},'' {\em Journal of Microscopy}, vol.~229, pp.~1--5, jan 2008.

\bibitem{Witham2011}
P.~Witham and E.~S{\'{a}}nchez, ``{A simple approach to neutral atom microscopy},'' {\em Review of Scientific Instruments}, vol.~82, oct 2011.

\bibitem{Witham2014}
P.~Witham and E.~Sanchez, ``{Exploring neutral atom microscopy},'' {\em Crystal Research and Technology}, vol.~49, pp.~690--698, sep 2014.

\bibitem{Barr2014}
M.~Barr, A.~Fahy, A.~Jardine, J.~Ellis, D.~Ward, D.~A. MacLaren, W.~Allison, and P.~C. Dastoor, ``{A design for a pinhole scanning helium microscope},'' {\em Elsevier}, vol.~340, pp.~76--80, dec 2014.

\bibitem{1988Aamb}
G.~Scoles, {\em Atomic and molecular beam methods}.
\newblock Oxford: Oxford University Press, 1988.

\bibitem{Reisinger2007}
T.~Reisinger, G.~Braceo, S.~Rehbein, G.~Schmahl, W.~E. Ernst, and B.~Holst, ``{Direct images of the virtual source in a supersonic expansion},'' {\em Journal of Physical Chemistry A}, vol.~111, pp.~12620--12628, dec 2007.

\bibitem{NickThesis}
N.~A. von Jeinsen, {\em {Leveraging spatial and angular resolution in Scanning Helium Microscopy to study technological samples}}.
\newblock PhD thesis, University of Cambridge, August 2021.

\bibitem{Bergin2021}
M.~Bergin, D.~J. Ward, S.~M. Lambrick, N.~A. von Jeinsen, B.~Holst, J.~Ellis, A.~P. Jardine, and W.~Allison, ``{Low-energy electron ionization mass spectrometer for efficient detection of low mass species},'' {\em Review of Scientific Instruments}, vol.~92, p.~073305, jul 2021.

\bibitem{Lambrick2021}
S.~M. Lambrick, A.~S. Palau, P.~E. Hansen, G.~Bracco, J.~Ellis, A.~P. Jardine, and B.~Holst, ``{True-to-size surface mapping with neutral helium atoms},'' {\em Physical Review A}, vol.~103, may 2021.

\bibitem{Bergin2019a}
M.~Bergin, D.~J. Ward, J.~Ellis, and A.~P. Jardine, ``{A method for constrained optimisation of the design of a scanning helium microscope},'' {\em Ultramicroscopy}, vol.~207, dec 2019.

\bibitem{Lambrick2018}
S.~M. Lambrick, M.~Bergin, A.~P. Jardine, and D.~J. Ward, ``{A ray tracing method for predicting contrast in neutral atom beam imaging},'' {\em Micron}, vol.~113, pp.~61--68, oct 2018.

\bibitem{Feres2004}
R.~Feres and G.~Yablonsky, ``{Knudsen's cosine law and random billiards},'' {\em Chemical Engineering Science}, vol.~59, pp.~1541--1556, apr 2004.

\bibitem{Celestini2008}
F.~Celestini and F.~Mortessagne, ``{Cosine law at the atomic scale: Toward realistic simulations of Knudsen diffusion},'' {\em Physical Review E - Statistical, Nonlinear, and Soft Matter Physics}, vol.~77, feb 2008.

\bibitem{OKEEFEDR1971}
{D. R. O'Keefe} and {R. L. Palmer}, ``Atomic and molecular beam scattering from macroscopically rough surfaces,'' {\em JVST-A}, no.~1, pp.~27--30, 1971.

\bibitem{Gontard2016}
L.~C. Gontard, R.~Schierholz, S.~Yu, J.~Cintas, and R.~E. Dunin-Borkowski, ``{Photogrammetry of the three-dimensional shape and texture of a nanoscale particle using scanning electron microscopy and free software},'' {\em Ultramicroscopy}, vol.~169, pp.~80--88, oct 2016.

\bibitem{Postek2013}
M.~T. Postek and A.~E. Vlad{\'{a}}r, ``{Does your SEM Really tell the truth? - How would you know? Part 1},'' {\em Scanning}, vol.~35, pp.~355--361, nov 2013.

\bibitem{AgudoJacome2012}
L.~{Agudo J{\'{a}}come}, G.~Eggeler, and A.~Dlouh{\'{y}}, ``{Advanced scanning transmission stereo electron microscopy of structural and functional engineering materials},'' {\em Ultramicroscopy}, vol.~122, pp.~48--59, nov 2012.

\bibitem{Shen2017}
J.~Shen, D.~Zhang, F.~H. Zhang, and Y.~Gan, ``{AFM tip-sample convolution effects for cylinder protrusions},'' {\em Applied Surface Science}, vol.~422, pp.~482--491, nov 2017.

\bibitem{Vollnhals2018}
F.~Vollnhals and T.~Wirtz, ``{Correlative Microscopy in 3D: Helium Ion Microscopy-Based Photogrammetric Topography Reconstruction Combined with in situ Secondary Ion Mass Spectrometry},'' {\em Analytical Chemistry}, vol.~90, pp.~11989--11995, oct 2018.

\bibitem{Stover2016}
J.~C. Stover, {\em {Optical scattering : measurement and analysis.}}
\newblock Spie Press, 2016.

\bibitem{Woodham1980}
R.~J. Woodham, ``{Photometric Method For Determining Surface Orientation From Multiple Images},'' {\em Optical Engineering}, vol.~19, p.~191139, feb 1980.

\bibitem{Goldman2010}
D.~B. Goldman, B.~Curless, A.~Hertzmann, and S.~M. Seitz, ``{Shape and spatially-varying BRDFs from photometric stereo},'' {\em IEEE Transactions on Pattern Analysis and Machine Intelligence}, vol.~32, no.~6, pp.~1060--1071, 2010.

\bibitem{ISSC23}
{A. Radi\'{c}, N. A. von Jeinsen, S. M. Lambrick, A. P. Jardine, D. J. Ward}, ``{Measuring facet angles using a combined spatially and angularly resolved atom beam instrument},'' {\em poster presentation ISSC-23}, 2021.

\bibitem{vonJeinsen2021}
{N. A. von Jeinsen, S. M. Lambrick, M. Bergin, A. Radi\'{c}, B. Liu, D. Seremet, A. P. Jardine, D. J. Ward}, ``{Selected area helium diffraction for surface science on a spot},'' {\em unpublished}, 2021.

\bibitem{Myles2019}
T.~A. Myles, S.~D. Eder, M.~G. Barr, A.~Fahy, J.~Martens, and P.~C. Dastoor, ``{Taxonomy through the lens of neutral helium microscopy},'' {\em Nature}, vol.~9, no.~1, pp.~1--10, 2019.

\bibitem{Fahy2018}
A.~Fahy, S.~D. Eder, M.~Barr, J.~Martens, T.~A. Myles, and P.~C. Dastoor, ``{Image formation in the scanning helium microscope},'' {\em Ultramicroscopy}, vol.~192, pp.~7--13, sep 2018.

\bibitem{Gal2021}
O.~Gal, ``{Fit Ellipse - File Exchange - MATLAB Central},'' jul 2021.

\bibitem{Oatley1957}
C.~W. Oatley, ``{The flow of gas through composite systems at very low pressures},'' {\em British Journal of Applied Physics}, vol.~8, no.~1, pp.~15--19, 1957.

\bibitem{Mercier2006}
B.~Mercier, ``{Conductance measurement of a conical tube and calculation of the pressure distribution},'' {\em Journal of Vacuum Science \& Technology A: Vacuum, Surfaces, and Films}, vol.~24, p.~529, apr 2006.

\bibitem{Ady}
M.~Ady and R.~Kersevan, ``{MolFlow+ user guide},'' {\em CERN}, 2019.

\bibitem{Kersevan2019}
R.~Kersevan, M.~Ady, and G.~Switzerland, ``{Recent developments of Monte-Carlo codes and MOLFLOW+ and SYNRAD+},'' {\em IPAC-19}, 2019.

\bibitem{1988Aambeqn}
G.~Scoles, {\em Atomic and molecular beam methods}.
\newblock Oxford: Oxford University Press, 1988.

\bibitem{jellis}
J.~Ellis, ``Private communications,'' 2021.

\bibitem{Stuckelberger2020}
M.~E. Stuckelberger, T.~Nietzold, B.~M. West, Y.~Luo, X.~Li, J.~Werner, B.~Niesen, C.~Ballif, V.~Rose, D.~P. Fenning, and M.~I. Bertoni, ``{Effects of X-rays on Perovskite Solar Cells},'' {\em Journal of Physical Chemistry C}, vol.~124, pp.~17949--17956, aug 2020.

\bibitem{Zheng2003}
X.~H. Zheng, H.~Chen, Z.~B. Yan, Y.~J. Han, H.~B. Yu, D.~S. Li, Q.~Huang, and J.~M. Zhou, ``{Determination of twist angle of in-plane mosaic spread of GaN films by high-resolution X-ray diffraction},'' {\em Journal of Crystal Growth}, vol.~255, pp.~63--67, jul 2003.

\bibitem{Zwicker2015}
A.~P. Zwicker, J.~Bloom, R.~Albertson, and S.~Gershman, ``{The suitability of 3D printed plastic parts for laboratory use},'' {\em American Journal of Physics}, vol.~83, pp.~281--285, mar 2015.

\bibitem{Chaneliere2017}
T.~Chaneliere and T.~Chaneli{\`{e}}re, ``{Vacuum compatibility of ABS plastics 3D-printed objects},'' {\em CNRS}, 2017.

\bibitem{Povilus2014}
A.~P. Povilus, C.~J. Wurden, Z.~Vendeiro, M.~Baquero-Ruiz, and J.~Fajans, ``{Vacuum compatibility of 3D-printed materials},'' {\em Journal of Vacuum Science \& Technology A: Vacuum, Surfaces, and Films}, vol.~32, p.~033001, apr 2014.

\bibitem{clear_data}
FormLabs, ``{Clear Resin Safety Data Sheet},'' 2021.

\bibitem{plastic1}
FormLabs, ``{FormLabs Resins Material Properties Datasheet},'' {\em FormLabs}, 2021.

\bibitem{VICTREX2019}
VICTREX, ``{VICTREX™ PEEK 450G Material Properties™},'' 2019.

\bibitem{JB2005}
JB, ``{HumidiProbe Temperature and Humidity Sensor},'' 2005.

\bibitem{LambrickDiffuse}
S.~Lambrick, ``Diffuse scattering observed in scanning helium microscopy,'' {\em in preparation}, 2021.

\bibitem{helios5}
T.~{Fisher Scientific}, ``{Helios 5 CX DualBeam Specifications}.''

\bibitem{cuda}
Nvidia, ``{CUDA Toolkit Documentation},'' 2021.

\bibitem{Owens}
J.~Owens and U.~C. Davis, ``{GPU Architecture Overview},'' {\em SIGGRAPH-07}, 2007.

\bibitem{Liu2008}
W.~Liu, B.~Schmidt, G.~Voss, and W.~M{\"{u}}ller-Wittig, ``{Accelerating molecular dynamics simulations using Graphics Processing Units with CUDA},'' {\em Computer Physics Communications}, vol.~179, pp.~634--641, nov 2008.

\bibitem{Asadchev2012}
A.~Asadchev and M.~S. Gordon, ``{New Multithreaded Hybrid CPU/GPU Approach to Hartree-Fock},'' {\em ACS}, 2012.

\bibitem{Alam2007}
S.~R. Alam, P.~K. Agarwal, M.~C. Smith, J.~S. Vetter, and D.~Caliga, ``{Using FPGA devices to accelerate biomolecular simulations},'' {\em Computer}, vol.~40, pp.~66--73, mar 2007.

\bibitem{Eder2012}
S.~D. Eder, T.~Reisinger, M.~M. Greve, G.~Bracco, and B.~Holst, ``{Focusing of a neutral helium beam below one micron},'' {\em New Journal of Physics}, vol.~14, jul 2012.

\bibitem{SalvadorPalau2017}
A.~{Salvador Palau}, G.~Bracco, and B.~Holst, ``{Theoretical model of the helium zone plate microscope},'' {\em Physical Review A}, vol.~95, jan 2017.

\bibitem{Bailey2015}
J.~Bailey, R.~Havelund, A.~G. Shard, I.~S. Gilmore, M.~R. Alexander, J.~S. Sharp, and D.~J. Scurr, ``{3D ToF-SIMS Imaging of Polymer Multilayer Films Using Argon Cluster Sputter Depth Profiling},'' {\em ACS Applied Materials and Interfaces}, vol.~7, pp.~2654--2659, feb 2015.

\bibitem{Alberici2004}
S.~G. Alberici, R.~Zonca, and B.~Pashmakov, ``{Ti diffusion in chalcogenides: a ToF-SIMS depth profile characterization approach},'' {\em Applied Surface Science}, vol.~231-232, pp.~821--825, jun 2004.

\end{thebibliography}

\end{document}